\documentclass[%
aps,
prd,
amssymb,
amsmath,
amsfonts,
eqsecnum,
showpacs,
nofootinbib,
floatfix,
twocolumn,
twoside,
a4paper,
superscriptaddress,
longbibliography
]{revtex4-1}

\usepackage[T3,T1]{fontenc}
\usepackage{mathrsfs} 
\usepackage{bm} 
\makeatletter
\@ifclassloaded{beamer}
  { 
    \typeout{UsePackages: Detected beamer}
    \usepackage{tgheros}
    
  }
  { 
    \typeout{UsePackages: Did not detect beamer}
    \ifx\asybeamer\undefined 
    \typeout{UsePackages: Detected article}
    \usepackage[varg]{txfonts} 
    \usepackage{tgtermes} 
    \else 
    \typeout{Fonts: Detected asy for beamer}
    \usepackage{tgheros}
    
    \fi
  }
\usepackage{microtype} 
\def\MT@register@subst@font{
  \MT@exp@one@n\MT@in@clist\font@name\MT@font@list
  \ifMT@inlist@\else\xdef\MT@font@list{\MT@font@list\font@name,}\fi}
\makeatother

\usepackage{graphicx} %
\usepackage[x11names,svgnames,rgb]{xcolor} %
\usepackage{xspace}
\usepackage{braket}
\usepackage{accents}
\usepackage{siunitx}
\usepackage{mathtools}
\DeclareSymbolFontAlphabet{\mathrm}{operators}

\definecolor{CiteColor}{rgb}{0.18039, 0.18824, 0.57255}
\definecolor{UrlColor} {rgb}{0.741, 0.173, 0.000}
\definecolor{DarkUrlColor} {rgb}{0.500, 0.110, 0.000}
\definecolor{LinkColor}{rgb}{0.25098, 0.47843, 0.04706}

\makeatletter %
\newcommand{\ShowFont}{%
  \typeout{The main font is \f@encoding \space \f@family \space %
    \f@series \space \f@shape \space at \f@size pt.}%
  \typeout{The math font sizes are \tf@size pt (main), \sf@size pt %
    (script), and \ssf@size pt (scriptscript).}%
  \typeout{The linewidth is \the\linewidth}} %
\makeatother %

\usepackage{xargs}

\usepackage{graphicx}
\usepackage{dcolumn}
\usepackage{bm}
\usepackage{tabularx}
\usepackage{multirow}
\usepackage{capt-of}
\usepackage{color}
\usepackage{url}
\usepackage[utf8]{inputenc}

\usepackage[nolist,nohyperlinks]{acronym}
\usepackage{xspace}
\usepackage[colorlinks]{hyperref}

\usepackage[caption=false]{subfig}

\DeclareMathAlphabet{\mathbfsf}{\encodingdefault}{\sfdefault}{bx}{sl}

\newcommand{\be}{\begin{equation}}
\newcommand{\ee}{\end{equation}}
\newcommand{\bea}{\begin{eqnarray}}
\newcommand{\eea}{\end{eqnarray}}

\newcommand{\phA}{\textsc{IMRPhenomA}\xspace}
\newcommand{\phB}{\textsc{IMRPhenomB}\xspace}
\newcommand{\phC}{\textsc{IMRPhenomC}\xspace}

\newcommand{\phD}{\textsc{IMRPhenomD}\xspace}
\newcommand{\phX}{\textsc{IMRPhenomXAS}\xspace}
\newcommand{\phenX}{\textsc{IMRPhenomX}\xspace}
\newcommand{\phXF}{\textsc{IMRPhenomX}\xspace}
\newcommand{\phP}{\textsc{IMRPhenomP}\xspace}
\newcommand{\phHM}{\textsc{IMRPhenomHM}\xspace}
\newcommand{\phPvtwo}{\textsc{IMRPhenomPv2}\xspace}
\newcommand{\phPvthree}{\textsc{IMRPhenomPv3}\xspace}
\newcommand{\phPvthreehm}{\textsc{IMRPhenomPv3HM}\xspace}
\newcommand{\phXHM}{\textsc{IMRPhenomXHM}\xspace}
\newcommand{\phXP}{\textsc{IMRPhenomXP}\xspace}
\newcommand{\phXPHM}{\textsc{IMRPhenomXPHM}\xspace}
\newcommand{\phT}{\textsc{IMRPhenomT}\xspace}
\newcommand{\phTP}{\textsc{IMRPhenomTP}\xspace}
\newcommand{\phTHM}{\textsc{IMRPhenomTHM}\xspace}
\newcommand{\NRHybSur}{\textsc{NRHybSur3dq8}\xspace}
\newcommand{\NRSur}{\textsc{NRSur7dq4}\xspace}

\newcommand{\seobnrvforhm}{\textsc{SEOBNRv4HM}\xspace}
\newcommand{\seobnrvforrom}{\textsc{SEOBNRv4HM\_ROM}\xspace}
\newcommand{\seobnrvforhmrom}{\textsc{SEOBNRv4\_ROM}\xspace}
\newcommand{\seobnrvforphm}{\textsc{SEOBNRv4PHM}\xspace}

\newcommand{\seobnrvforp}{\textsc{SEOBNRv4P}\xspace}

\newcommand{\Msun}{M_\odot}
\newcommand{\chieff}{\chi_\mathrm{eff}}
\newcommand{\chip}{\chi_\mathrm{p}}

\definecolor{dodgerblue}{HTML}{1E90FF}
\definecolor{viennared}{HTML}{DA0A14}
\definecolor{ctorange}{HTML}{FF6C0C}
\definecolor{granadagreen}{HTML}{078931}
\definecolor{wales}{HTML}{ff0038}
\definecolor{valenciacfred}{HTML}{ee3524}
\definecolor{barcelonafcgold}{HTML}{edbb00}
\definecolor{jam}{HTML}{A50B5E}
\definecolor{austriawien}{HTML}{441678}

\AtBeginDocument{%
  \hypersetup{
    citecolor=dodgerblue,
    linkcolor=dodgerblue,   
    urlcolor=dodgerblue}}

\newcommand{\UIB}{Departament de F\'isica, Universitat de les Illes Balears, IAC3 -- IEEC, Crta. Valldemossa km 7.5, E-07122 Palma, Spain}

\newcommand{\UoB}{School of Physics and Astronomy and Institute for Gravitational Wave Astronomy, University of Birmingham, Edgbaston, Birmingham, B15 9TT, United Kingdom}

\newcommand{\AEI}{Max Planck Institute for Gravitational Physics (Albert Einstein Institute), Am M{\"u}hlenberg 1, Potsdam, 14476, Germany}

\def\MOneSourceCIPhPHM{\ensuremath{28.1_{-4.3}^{+4.8}}\xspace}
\def\MTwoSourceCIPhPHM{\ensuremath{8.8_{-1.1}^{+1.5}}\xspace}
\def\MtotalSourceCIPhPHM{\ensuremath{36.9_{-2.9}^{+3.7}}\xspace}
\def\ChirpMassSourceCIPhPHM{\ensuremath{13.2_{-0.3}^{+0.5}}\xspace}
\def\MassRatioCIPhPHM{\ensuremath{0.31_{-0.07}^{+0.12}}\xspace}
\def\MOneCIPhPHM{\ensuremath{32.3_{-5.2}^{+5.7}}\xspace}
\def\MTwoCIPhPHM{\ensuremath{10.1_{-1.2}^{+1.6}}\xspace}
\def\MtotalCIPhPHM{\ensuremath{42.5_{-3.7}^{+4.4}}\xspace}
\def\ChirpMassCIPhPHM{\ensuremath{15.2_{-0.2}^{+0.3}}\xspace}
\def\ChiEffCIPhPHM{\ensuremath{0.22_{-0.11}^{+0.08}}\xspace}
\def\ChiPCIPhPHM{\ensuremath{0.31_{-0.17}^{+0.24}}\xspace}
\def\SpinMagOneCIPhPHM{\ensuremath{0.41_{-0.24}^{+0.22}}\xspace}

\def\DLCIPhPHM{\ensuremath{740_{-190}^{+150}}\xspace}
\def\RedshiftCIPhPHM{\ensuremath{0.15_{-0.04}^{+0.03}}\xspace}
\def\ThetaJNCIPhPHM{\ensuremath{0.71_{-0.27}^{+0.39}}\xspace}
\def\HSNRCIPhPHM{\ensuremath{9.5_{-0.3}^{+0.2}}\xspace}
\def\LSNRCIPhPHM{\ensuremath{16.1_{-0.3}^{+0.2}}\xspace}
\def\VSNRCIPhPHM{\ensuremath{3.6_{-1.0}^{+0.3}}\xspace}
\def\NetSNRCIPhPHM{\ensuremath{19.0_{-0.3}^{+0.2}}\xspace}
\def\FinalSpinCIPhPHM{\ensuremath{0.67_{-0.07}^{+0.07}}\xspace}
\def\FinalMassCIPhPHM{\ensuremath{35.7_{-3.0}^{+3.8}}\xspace}
\def\MOneSourceCICombined{\ensuremath{29.7_{-5.3}^{+5.0}}\xspace}
\def\MTwoSourceCICombined{\ensuremath{8.4_{-1.0}^{+1.8}}\xspace}
\def\MtotalSourceCICombined{\ensuremath{38.1_{-3.7}^{+4.0}}\xspace}
\def\ChirpMassSourceCICombined{\ensuremath{13.3_{-0.3}^{+0.4}}\xspace}
\def\MassRatioCICombined{\ensuremath{0.28_{-0.06}^{+0.13}}\xspace}
\def\MOneCICombined{\ensuremath{34.2_{-6.5}^{+5.7}}\xspace}
\def\MTwoCICombined{\ensuremath{9.7_{-1.1}^{+1.8}}\xspace}
\def\MtotalCICombined{\ensuremath{43.9_{-4.7}^{+4.6}}\xspace}
\def\ChirpMassCICombined{\ensuremath{15.3_{-0.2}^{+0.2}}\xspace}
\def\ChiEffCICombined{\ensuremath{0.25_{-0.11}^{+0.08}}\xspace}
\def\ChiPCICombined{\ensuremath{0.30_{-0.15}^{+0.19}}\xspace}
\def\SpinMagOneCICombined{\ensuremath{0.43_{-0.26}^{+0.16}}\xspace}

\def\DLCICombined{\ensuremath{730_{-170}^{+140}}\xspace}
\def\RedshiftCICombined{\ensuremath{0.15_{-0.03}^{+0.03}}\xspace}
\def\ThetaJNCICombined{\ensuremath{0.73_{-0.24}^{+0.34}}\xspace}
\def\HSNRCICombined{\ensuremath{9.5_{-0.3}^{+0.1}}\xspace}
\def\LSNRCICombined{\ensuremath{16.2_{-0.3}^{+0.1}}\xspace}
\def\VSNRCICombined{\ensuremath{3.6_{-1.0}^{+0.3}}\xspace}
\def\NetSNRCICombined{\ensuremath{19.1_{-0.3}^{+0.1}}\xspace}
\def\FinalSpinCICombined{\ensuremath{0.67_{-0.07}^{+0.05}}\xspace}
\def\FinalMassCICombined{\ensuremath{37.0_{-3.9}^{+4.1}}\xspace}
\def\MOneSourceCIEOBPHM{\ensuremath{31.7_{-3.5}^{+3.6}}\xspace}
\def\MTwoSourceCIEOBPHM{\ensuremath{8.0_{-0.7}^{+0.9}}\xspace}
\def\MtotalSourceCIEOBPHM{\ensuremath{39.7_{-2.7}^{+3.0}}\xspace}
\def\ChirpMassSourceCIEOBPHM{\ensuremath{13.3_{-0.3}^{+0.3}}\xspace}
\def\MassRatioCIEOBPHM{\ensuremath{0.25_{-0.04}^{+0.06}}\xspace}
\def\MOneCIEOBPHM{\ensuremath{36.5_{-4.2}^{+4.2}}\xspace}
\def\MTwoCIEOBPHM{\ensuremath{9.2_{-0.7}^{+0.9}}\xspace}
\def\MtotalCIEOBPHM{\ensuremath{45.7_{-3.3}^{+3.5}}\xspace}
\def\ChirpMassCIEOBPHM{\ensuremath{15.3_{-0.2}^{+0.1}}\xspace}
\def\ChiEffCIEOBPHM{\ensuremath{0.28_{-0.08}^{+0.06}}\xspace}
\def\ChiPCIEOBPHM{\ensuremath{0.31_{-0.15}^{+0.14}}\xspace}
\def\SpinMagOneCIEOBPHM{\ensuremath{0.46_{-0.15}^{+0.12}}\xspace}

\def\DLCIEOBPHM{\ensuremath{740_{-130}^{+120}}\xspace}
\def\RedshiftCIEOBPHM{\ensuremath{0.15_{-0.02}^{+0.02}}\xspace}
\def\ThetaJNCIEOBPHM{\ensuremath{0.71_{-0.21}^{+0.23}}\xspace}
\def\HSNRCIEOBPHM{\ensuremath{9.5_{-0.2}^{+0.1}}\xspace}
\def\LSNRCIEOBPHM{\ensuremath{16.2_{-0.2}^{+0.1}}\xspace}
\def\VSNRCIEOBPHM{\ensuremath{3.7_{-0.5}^{+0.2}}\xspace}
\def\NetSNRCIEOBPHM{\ensuremath{19.1_{-0.2}^{+0.2}}\xspace}
\def\FinalSpinCIEOBPHM{\ensuremath{0.68_{-0.04}^{+0.04}}\xspace}
\def\FinalMassCIEOBPHM{\ensuremath{38.6_{-2.8}^{+3.1}}\xspace}

\def\MOneSourceCIXPHM{\ensuremath{30.0^{+5.2}_{-4.3}}\xspace}
\def\MTwoSourceCIXPHM{\ensuremath{8.4^{+1.3}_{-1.1}}\xspace}
\def\MtotalSourceCIXPHM{\ensuremath{38.4^{+4.2}_{-3.2}}\xspace}
\def\ChirpMassSourceCIXPHM{\ensuremath{13.3^{+0.5}_{-0.4}}\xspace}
\def\MassRatioCIXPHM{\ensuremath{0.28^{+0.09}_{-0.07}}\xspace}
\def\MOneCIXPHM{\ensuremath{34.4^{+6.2}_{-5.1}}\xspace}
\def\MTwoCIXPHM{\ensuremath{9.6^{+1.4}_{-1.2}}\xspace}
\def\MtotalCIXPHM{\ensuremath{44.1^{+5.0}_{-3.7}}\xspace}
\def\ChirpMassCIXPHM{\ensuremath{15.3^{+0.4}_{-0.2}}\xspace}
\def\ChiEffCIXPHM{\ensuremath{0.25^{+0.1}_{-0.1}}\xspace}
\def\ChiPCIXPHM{\ensuremath{0.23^{+0.20}_{-0.13}}\xspace}
\def\SpinMagOneCIXPHM{\ensuremath{0.39^{+0.16}_{-0.17}}\xspace}

\def\DLCIXPHM{\ensuremath{734^{+161}_{-187}}\xspace}
\def\RedshiftCIXPHM{\ensuremath{0.15_{-0.04}^{+0.03}}\xspace}
\def\ThetaJNCIXPHM{\ensuremath{0.75^{+0.36}_{-0.28}}\xspace}
\def\HSNRCIXPHM{\ensuremath{9.4^{+0.2}_{-0.3}}\xspace}
\def\LSNRCIXPHM{\ensuremath{16.1^{+0.2}_{-0.3}}\xspace}
\def\VSNRCIXPHM{\ensuremath{3.6^{+0.3}_{-0.8}}\xspace}
\def\NetSNRCIXPHM{\ensuremath{18.9_{-0.3}^{+0.2}}\xspace}
\def\FinalSpinCIXPHM{\ensuremath{0.67_{-0.05}^{+0.05}}\xspace}
\def\FinalMassCIXPHM{\ensuremath{37.2_{-3.3}^{+4.3}}\xspace}

\def\MOneSourceCIXPHMSEOB{\ensuremath{30.9^{+4.3}_{-4.4}}\xspace}
\def\MTwoSourceCIXPHMSEOB{\ensuremath{8.2^{+1.2}_{-0.9}}\xspace}
\def\MtotalSourceCIXPHMSEOB{\ensuremath{39.1^{+3.6}_{-3.3}}\xspace}
\def\ChirpMassSourceCIXPHMSEOB{\ensuremath{13.3^{+0.4}_{-0.3}}\xspace}
\def\MassRatioCIXPHMSEOB{\ensuremath{0.27^{+0.09}_{-0.06}}\xspace}
\def\MOneCIXPHMSEOB{\ensuremath{35.5^{+5.2}_{-5.1}}\xspace}
\def\MTwoCIXPHMSEOB{\ensuremath{9.4^{+1.3}_{-0.9}}\xspace}
\def\MtotalCIXPHMSEOB{\ensuremath{44.9^{+4.3}_{-3.8}}\xspace}
\def\ChirpMassCIXPHMSEOB{\ensuremath{15.3^{+0.3}_{-0.2}}\xspace}
\def\ChiEffCIXPHMSEOB{\ensuremath{0.26^{+0.08}_{-0.09}}\xspace}
\def\ChiPCIXPHMSEOB{\ensuremath{0.27^{+0.17}_{-0.15}}\xspace}
\def\SpinMagOneCIXPHMSEOB{\ensuremath{0.43^{+0.14}_{-0.17}}\xspace}

\def\DLCIXPHMSEOB{\ensuremath{737^{+141}_{-159}}\xspace}
\def\RedshiftCIXPHMSEOB{\ensuremath{0.15_{-0.03}^{+0.03}}\xspace}
\def\ThetaJNCIXPHMSEOB{\ensuremath{0.73^{+0.31}_{-0.24}}\xspace}
\def\HSNRCIXPHMSEOB{\ensuremath{9.4^{+0.2}_{-0.3}}\xspace}
\def\LSNRCIXPHMSEOB{\ensuremath{16.2^{+0.2}_{-0.3}}\xspace}
\def\VSNRCIXPHMSEOB{\ensuremath{3.7^{+0.3}_{-0.7}}\xspace}
\def\NetSNRCIXPHMSEOB{\ensuremath{19.0_{-0.4}^{+0.2}}\xspace}
\def\FinalSpinCIXPHMSEOB{\ensuremath{0.67^{+0.05}_{-0.04}}\xspace}
\def\FinalMassCIXPHMSEOB{\ensuremath{38.0^{+3.5}_{-3.3}}\xspace}

\allowdisplaybreaks[1]

\begin{document}

\title[GW190412]
{
Towards the routine use of subdominant harmonics in gravitational-wave inference:\\ re-analysis of GW190412 with generation X waveform models
}


\author{Marta Colleoni}\email{marta.colleoni@uib.es}
\affiliation{\UIB}

\author{Maite Mateu-Lucena}
\affiliation{\UIB}

\author{H\'{e}ctor Estell\'{e}s}
\affiliation{\UIB}

\author{Cecilio Garc{\'i}a-Quir{\'o}s}
\affiliation{\UIB}

\author{David Keitel}\affiliation{\UIB}

\author{Geraint Pratten}
\affiliation{\UoB}

\author{Antoni Ramos-Buades}
\affiliation{\AEI}
\affiliation{\UIB}

\author{Sascha Husa}
\affiliation{\UIB}

\date{\today}

\begin{abstract}
We re-analyse the gravitational-wave event GW190412 with state-of-the-art phenomenological waveform models.
This event, which has been associated with a black hole merger, is interesting due to the significant contribution from subdominant harmonics.
We use both frequency-domain and time-domain waveform models.
The PhenomX waveform models constitute the fourth generation of frequency-domain phenomenological waveforms for black hole binary coalescence; they have more recently been complemented by the time-domain PhenomT models, which open up new strategies to model precession and eccentricity, and to perform tests of general relativity with the phenomenological waveforms approach.
Both PhenomX and PhenomT have been constructed with similar techniques and accuracy goals, and due to their computational efficiency 
this ``generation X'' model family
allows the routine use of subdominant spherical harmonics in Bayesian inference.
We show the good agreement between these and other state-of-the-art waveform models for GW190412, and discuss the improvements over the previous generation of phenomenological waveform models.
We also discuss practical aspects of Bayesian inference such as run convergence, variations of sampling parameters, and computational cost.
\end{abstract}


\maketitle

\section{Introduction}
\label{sec:Introduction}
The analysis of gravitational wave (GW) data from compact binary coalescences (CBCs) has long focused on the dominant quadrupole spherical harmonics, and the use of sub-dominant spherical harmonics has only recently started to play a prominent role for observational results \cite{LIGOScientific:2020stg,GW190814:journal,GW190521:observ,GW190521:properties} with the ground-based detectors Advanced LIGO \cite{TheLIGOScientific:2014jea} and Advanced Virgo \cite{TheVirgo:2014hva}.
The use of waveform models including sub-dominant harmonics can often break degeneracies between source parameters and improve the overall parameter estimation results even if the content of sub-dominant harmonics is weak for the most likely set of parameters, since the additional harmonics help to exclude parameter-space regions which are not consistent with the data under the more complete models.

Here we argue that models including subdominant harmonics should now be used routinely in GW parameter estimation, and how this is facilitated by the computational efficiency and accuracy of the ``generation X'' of phenomenological waveform models:
the frequency-domain \phXF models~\cite{Pratten:2020fqn, Garcia-Quiros:2020qpx,Garcia-Quiros:2020qlt,Pratten:2020ceb}, and the complementary time-domain \phT models~\cite{phenomtp,phenomthm}.
In particular we demonstrate the capability of the generation X waveform family to deliver a suite of Bayesian inference results both quickly, and with relatively low total computational cost.
One interesting application is on measuring the source distance, where breaking the degeneracy with the binary's inclination through inclusion of subdominant harmonics significantly improves accuracy~\cite{PE_PhHM, LIGOScientific:2020stg}.
A quick turnaround on such precise distance measurements is particularly important in the context of electromagnetic counterparts, which have recently become more interesting also
in the context of binary black hole (BBH) mergers, as discussed in connection with the discovery of GW190521 
\cite{GW190521:observ, GW190521:properties}.
Moreover, being able to efficiently compare posteriors for multiple models from the same family including different amount of physics
(aligned spins only vs. precession, dominant modes only vs. inclusion of subdominant modes) allows detailed studies of waveform modelling systematics, increasing the confidence in the final parameter estimates for an event.

To provide a guide to the routine use of subdominant harmonics in practical parameter estimation and to demonstrate the capabilities of the ``generation X'' phenomenological waveform models on a specific example, in this paper we present improved parameter estimation results for the event GW190412 first published in \cite{LIGOScientific:2020stg}.
GW190412 is special among the CBC observations reported by LIGO and Virgo to date \cite{LIGOScientific:2018mvr,Abbott:2020uma,LIGOScientific:2020stg,GW190814:journal,GW190521:observ} as the first BBH signal with clear evidence for significantly unequal component masses.
Since the effect of sub-dominant spherical harmonic modes is stronger in unequal-mass systems, this event also provided the first observational evidence for their presence.
On the other hand, as for previous BBH detections, only limited amounts of information about the spins of the black holes could be extracted in~\cite{LIGOScientific:2020stg}.
The event is of astrophysical interest as the unusual mass ratio hints at a more diverse population of merging BBHs in the Universe than the `vanilla' events previously observed~\cite{LIGOScientific:2018mvr,LIGOScientific:2018jsj}
and starts to provide discriminating evidence between possible formation channels for these systems~\cite{LIGOScientific:2020stg,Mandel:2020lhv,DiCarlo:2020lfa,Olejak:2020oel,Hamers:2020huo,Rodriguez:2020viw,Gerosa:2020bjb,DeLuca:2020qqa,Safarzadeh:2020qrc,Kimball_2020}.

In general, the measurement of CBC source properties through matched filtering and Bayesian inference relies crucially on the quality of the waveform models used as the templates.
Waveform models are typically synthesized from perturbative results (notably the post-Newtonian \cite{Blanchet:2006zz} and effective-one-body \cite{Damour:2001tu, Damour:2012ky, Cotesta:2018fcv} frameworks and results for the ringdown frequencies of Kerr black holes\cite{Berti_2006}), and catalogues of numerical relativity (NR) simulations, such as the large catalogue of waveforms from the SXS collaboration \cite{SXS:catalog, Boyle:2019kee}. 
Several such models of varying complexity have been used in \cite{LIGOScientific:2020stg} to measure the source properties of GW190412.
None of those models includes orbital eccentricity, but two of them include both precession and subdominant harmonics: the frequency-domain model \phPvthreehm \cite{Khan:2019kot} and the time-domain model \seobnrvforphm \cite{Ossokine:2020kjp}.
Both are members of families of models which also include simpler models without precession or subdominant harmonics, which have been used for model comparison to determine the evidence for the presence of these effects in the data.
Systematic errors in such waveform models, in particular regarding precession and sub-dominant harmonics, are not yet well understood, and the high computational cost for both the \phPvthreehm and \seobnrvforphm models is one reason why investigations of potential systematics are challenging.
None of the models used to analyze GW190412 in~\cite{LIGOScientific:2020stg} is calibrated to precessing NR waveforms, and only in \seobnrvforphm have the subdominant harmonics been calibrated to NR, with \phPvthreehm using approximate scalings to include their effects.

Our re-analysis of GW190412 focuses on the \phXF family of inspiral-merger-ringdown waveform models \cite{Pratten:2020fqn,Garcia-Quiros:2020qpx,Garcia-Quiros:2020qlt,Pratten:2020ceb}
which constitute a thorough upgrade of previous versions of the family of frequency-domain phenomenological waveform models  \cite{Husa:2015iqa,Khan:2015jqa,Hannam:2013oca,Bohe:PPv2,London:2017bcn,Khan:2018fmp,Khan:2019kot} that is currently routinely used in CBC data analysis, including the \phPvthreehm model mentioned above.
In addition we study this event with the new \phT family of phenomenological time-domain waveforms (with the versions currently implemented in the LALSuite \cite{lalsuite} package not yet including precession).
An alternative re-analysis focusing on the \NRSur ~\cite{Varma:2019csw} model has been presented in \cite{GW190412:surro}. \NRSur is based on reduced-order-modelling (ROM) \cite{P_rrer_2014} and has been calibrated to numerical relativity waveforms. It does however not span a sufficiently wide frequency range to cover the entire LIGO and Virgo bands for the mass range of GW190412, and   \cite{GW190412:surro} studies variations of the lower cutoff frequency.

None of the \phXF models have yet been calibrated to precessing NR waveforms, but \phXHM, \phXPHM and \phTHM are calibrated to sub-dominant harmonics from NR waveforms.
The modularity and flexibility of the model family allows to compare different approximations for the effects of spin precession and different choices for the final spin of the remnant black hole.
Furthermore the drastically reduced computational cost of the new waveforms allows us to test in detail the impact of varying some of the settings of the Bayesian sampling algorithms we use~\cite{Ashton:2018jfp,Smith:2019ucc,Romero-Shaw:2020owr}.
We find that by replacing \phPvthreehm with our upgraded precessing higher-modes model \phXPHM, the disagreement between the frequency and time-domain models observed in~\cite{LIGOScientific:2020stg} can be reduced significantly, and the uncertainty intervals for key parameters can be tightened.
The agreement between frequency and time-domain models is further improved when assuming that the source does not show significant spin-precession, which we confirm to be consistent with the observational data. 
A summary of our parameter estimation results and a comparison with the results from \cite{LIGOScientific:2020stg} can be found in Table \ref{tab:PEresults} and is discussed below.

The paper is organized as follows.
In Sec.~\ref{sec:preliminaries} we collect preliminaries: remarks on notation, a summary of the results found in \cite{LIGOScientific:2020stg} on the event GW190412 and brief descriptions of the different waveform models and Bayesian inference methods used there and in the present paper.
We then present our main parameter estimation results on GW190412 in Sec.~\ref{sec:PEresults}, further investigations on systematic and sampling errors in Sec.~\ref{sec:extra_checks}, and a summary and our conclusions in Sec.~\ref{sec:conclusions}.
Further comparisons are presented in appendices:
while all our main results are obtained with the parallel Bilby code \cite{Ashton:2018jfp,Smith:2019ucc,Romero-Shaw:2020owr},
in appendix \ref{subsec:LI_vs_pbilby} we also compare these results with comparison runs of the LALInference code \cite{Veitch:2014wba}.
And in appendix \ref{subsec:dist_marg} we compare our main Bilby runs, which have been obtained with marginalisation over distance in order to improve convergence, with runs that do not use this approximation.
Different \phXF implementations of precession and approximations for the spin of the merger remnant are studied in appendix \ref{sec:prec_versions}.
Finally, we compare results in more detail against the \phPvthreehm model in appendix \ref{sec:phenompv3_comparison}
and study the impact of alternative spin priors in appendix~\ref{sec:spin_priors}.

Posterior samples from our preferred run for each waveform model are released in~\cite{datarelease}.

\section{Preliminaries}
\label{sec:preliminaries}

\subsection{Notation and conventions}
\label{sec:notation}

We will report all masses in units of the solar mass $M_\odot$.
Masses are reported both in the detector frame, where they appear redshifted,
and in the source frame, assuming a standard cosmology \cite{Planck2015}
(see Appendix B of \cite{LIGOScientific:2018mvr}).
We will report source-frame masses with a superscript $\mathrm{s}$, as in $m_1^\mathrm{s}$ to denote the mass of the larger black hole in the source frame.
We will drop the superscript to denote masses in the detector frame, and to represent general relations between different mass parameters.
Individual component masses are denoted by $m_i$,
the total mass is $M = m_1 + m_2$, and the chirp mass by $\mathcal M = (m_1 \, m_2)^{3/5} M^{-1/5}$.
The mass ratio is defined as $q = m_2 / m_1 \leq 1$.


We also define two effective spin parameters which are commonly used in waveform modelling and parameter estimation.
First, the parameter $\chi_{\rm{eff}}$~\cite{Damour:2001tu,Ajith:2009bn,Santamaria:2010yb} is defined as
 \begin{equation}\label{def:chieff}
    \chi_{\mathrm{eff}}=\frac{m_1 \chi_1 + m_2 \chi_2}{m_1 + m_2}, 
 \end{equation}
where the $\chi_i$ are the projections of the spin vectors of the individual black holes onto the orbital angular momentum.
Second, the effective spin precession parameter $\chi_p$ \cite{Schmidt:2014iyl} has been designed to capture the dominant effect of precession.
It corresponds to an approximate average over many precession cycles of the spin in the precessing orbital plane, and is defined in terms of the average spin magnitude $S_p$, \cite{Hannam:2013oca,Schmidt:2014iyl}
\begin{align}\label{eq:avNNLOSperp}
    S_p &= \frac{1}{2} \left( A_1 S_{1,\perp} + A_2 S_{2,\perp} + | A_{1} S_{1,\perp} - A_2 S_{2,\perp} |\right) , \\
    &= {\rm{max}} \left( A_1 S_{1,\perp} , A_2 S_{2,\perp} \right),
\end{align}
where $A_1 = 2 + 3 / (2 q)$, and $\chi_p$
is then defined as
\begin{align}\label{def:chip}
    \chi_p &= \frac{S_p}{A_1 m^2_1} .
\end{align}
Both  $\chi_{\mathrm{eff}}$ and $\chi_p$ are dimensionless and thus independent of the frame (source or detector).

Throughout this work we will employ waveforms with several multipoles beyond the quadrupolar contribution.
Unless otherwise stated, we will consider pairs of both positive and negative modes when referring to a particular multipole.
For example, to refer to a set of multipoles $(l,m)=(2,\pm2), (2,\pm1)$ we will use the simplified notation $(l,|m|)=(2,2), (2,1)$ or simply $(2,2),(2,1)$.

\subsection{Summary of the event GW190412}
\label{sec:whathappenedbefore}

\begin{table*}
\caption{Inferred parameter values for GW190412 and
their 90\% credible intervals, obtained using precessing models including higher
multipoles.
Columns 2--4 correspond to the results from the LVC analyses~\cite{LIGOScientific:2020stg},
the fifth column gives the new results from our preferred inference run, which is run 26 from Table~\ref{tab:tabRuns} with the precessing higher-modes model \phXPHM, and the last column combines the LVC \seobnrvforphm results with \phXPHM in the same way that they were combined with \phPvthreehm in~\cite{LIGOScientific:2020stg}.}
\label{tab:PEresults}
\begin{ruledtabular}

\begin{tabular}{lrrrrr} 
 parameter
 & \seobnrvforphm & \phPvthreehm & LVC Combined & \phXPHM & Combined \\ \hline
$m_1^\mathrm{s} / M_\odot$ & \MOneSourceCIEOBPHM & \MOneSourceCIPhPHM &
\MOneSourceCICombined  & \MOneSourceCIXPHM & \MOneSourceCIXPHMSEOB\\
$m_2^\mathrm{s} / M_\odot$ & \MTwoSourceCIEOBPHM & \MTwoSourceCIPhPHM &
\MTwoSourceCICombined  &\MTwoSourceCIXPHM  & \MTwoSourceCIXPHMSEOB\\
$M^\mathrm{s} / M_\odot$ & \MtotalSourceCIEOBPHM & \MtotalSourceCIPhPHM &
\MtotalSourceCICombined  & \MtotalSourceCIXPHM & \MtotalSourceCIXPHMSEOB\\
$\mathcal M^\mathrm{s} / M_\odot$ & \ChirpMassSourceCIEOBPHM & \ChirpMassSourceCIPhPHM &
\ChirpMassSourceCICombined   & \ChirpMassSourceCIXPHM & \ChirpMassSourceCIXPHMSEOB \\
$q$ & \MassRatioCIEOBPHM & \MassRatioCIPhPHM &
\MassRatioCICombined  &\MassRatioCIXPHM & \MassRatioCIXPHMSEOB \\[6pt]
$M_\mathrm{f} / M_\odot$ & \FinalMassCIEOBPHM & \FinalMassCIPhPHM &
\FinalMassCICombined  &\FinalMassCIXPHM & \FinalMassCIXPHMSEOB\\
$\chi_\mathrm{f}$ & \FinalSpinCIEOBPHM & \FinalSpinCIPhPHM &
\FinalSpinCICombined  & \FinalSpinCIXPHM& \FinalSpinCIXPHMSEOB \\[6pt]
$m_1/ M_\odot$ & \MOneCIEOBPHM & \MOneCIPhPHM &
\MOneCICombined  &  \MOneCIXPHM& \MOneCIXPHMSEOB \\
$m_2 / M_\odot$ & \MTwoCIEOBPHM & \MTwoCIPhPHM &
\MTwoCICombined  &\MTwoCIXPHM & \MTwoCIXPHMSEOB \\
$M / M_\odot$ & \MtotalCIEOBPHM & \MtotalCIPhPHM &
\MtotalCICombined  & \MtotalCIXPHM &  \MtotalCIXPHMSEOB \\
$\mathcal M / M_\odot$ & \ChirpMassCIEOBPHM & \ChirpMassCIPhPHM &
\ChirpMassCICombined   &\ChirpMassCIXPHM & \ChirpMassCIXPHMSEOB\\[6pt]
$\chi_{\rm eff}$ &  \ChiEffCIEOBPHM & \ChiEffCIPhPHM &
\ChiEffCICombined  & \ChiEffCIXPHM & \ChiEffCIXPHMSEOB\\
$\chi_\mathrm{p}$ &  \ChiPCIEOBPHM & \ChiPCIPhPHM &
\ChiPCICombined  & \ChiPCIXPHM & \ChiPCIXPHMSEOB \\
$\chi_1$ & \SpinMagOneCIEOBPHM & \SpinMagOneCIPhPHM &
\SpinMagOneCICombined &\SpinMagOneCIXPHM & \SpinMagOneCIXPHMSEOB
\\[6pt]
$D_\mathrm{L} / \textrm{Mpc}$ &  \DLCIEOBPHM & \DLCIPhPHM &
\DLCICombined & \DLCIXPHM & \DLCIXPHMSEOB\\
$z$ &  \RedshiftCIEOBPHM & \RedshiftCIPhPHM &
\RedshiftCICombined&\RedshiftCIXPHM & \RedshiftCIXPHMSEOB  \\
$\hat \theta_{JN}$ &  \ThetaJNCIEOBPHM & \ThetaJNCIPhPHM &
\ThetaJNCICombined & \ThetaJNCIXPHM & \ThetaJNCIXPHMSEOB \\[6pt]
$\rho_\mathrm{H}$  &  \HSNRCIEOBPHM & \HSNRCIPhPHM &
\HSNRCICombined & \HSNRCIXPHM & \HSNRCIXPHMSEOB \\
$\rho_\mathrm{L}$  &  \LSNRCIEOBPHM & \LSNRCIPhPHM &
\LSNRCICombined & \LSNRCIXPHM & \LSNRCIXPHMSEOB \\
$\rho_\mathrm{V}$  &  \VSNRCIEOBPHM & \VSNRCIPhPHM &
\VSNRCICombined & \VSNRCIXPHM & \VSNRCIXPHMSEOB \\
$\rho_\mathrm{HLV}$  &  \NetSNRCIEOBPHM & \NetSNRCIPhPHM &
\NetSNRCICombined & \NetSNRCIXPHM & \NetSNRCIXPHMSEOB
\end{tabular}
\end{ruledtabular}
\end{table*}

GW190412 is the first BBH detection from the O3 observing run published~\cite{LIGOScientific:2020stg} by the LIGO-Virgo Collaboration (LVC).
It was observed by all three detectors, with a network signal-to-noise ratio (SNR) of $19.1_{-0.3}^{+0.1}$ from the final coherent Bayesian analysis reported in~\cite{LIGOScientific:2020stg}.
In briefly summarizing the event properties,
we concentrate here on the results from Bayesian inference
using the Bilby~\cite{Ashton:2018jfp,Smith:2019ucc,Romero-Shaw:2020owr},
LALInference~\cite{Veitch:2014wba}
and RIFT~\cite{Lange:2018pyp,Wysocki:2019grj}
packages
and the \seobnrvforphm~\cite{Ossokine:2020kjp}
and \phPvthreehm~\cite{Khan:2019kot} waveform models.
These LVC estimates of the properties of GW190412's source are listed in Table \ref{tab:PEresults} together with our own results.

In summary, GW190412 came from a BBH with individually unremarkable source-frame masses
$m_1^\mathrm{s} \approx30\Msun$ and $m_2^\mathrm{s}\approx8\Msun$,
but the mass ratio $q=m_2/m_1\approx0.3$ ($<0.5$ at 99\% probability)
is much lower than inferred for any previous detection.
The system's effective spin parameter appears to be low,
but significantly different from zero: $\chieff=0.25_{-0.11}^{+0.08}$ when combining results from both waveforms.
The effective precession parameter has also been constrained, but more weakly, to
$\chip=0.31_{-0.16}^{+0.19}$.
While there is some information gain compared to the prior on this quantity,
and the posterior is peaked away from zero, it prefers lower values than the prior.
The results from \seobnrvforphm and \phPvthreehm agree within 90\% uncertainties
for all quantities reported in~\cite{LIGOScientific:2020stg}, but do show some qualitative differences --
for example,

the \phPvthreehm posteriors tend to less unequal $q$ together with lower $\chieff$.

Additional inference runs were also reported in \cite{LIGOScientific:2020stg} for several waveform models with reduced physics content
(see their Table I and references therein),
with the goal of estimating the evidence for the presence of precession and HMs in the observed signal.
Clear and robust evidence was found for the presence of HMs,
with $\log_{10}$ Bayes factors between 3.0 and 4.1 in favour of models including  HMs over those only including $\ell=2$ modes,
depending on the waveform model family, sampling method and whether precession was also included at the same time.
On the other hand, there was no clear evidence for or against precession, with the obtained Bayes factors for that hypothesis test remaining within systematic uncertainties for each waveform model family.
We summarise Bayes factors listed in the LVC publication, together with those obtained from our own analysis, in Table~\ref{tab:tabBF}.

The main astrophysical conclusions that \cite{LIGOScientific:2020stg} drew from GW190412 include that the event's properties,
especially its mass ratio,
are somewhat unexpected for draws from a BBH population as inferred from the previous two observing runs~\cite{LIGOScientific:2018mvr,LIGOScientific:2018jsj},
but not in clear tension with it.
Furthermore, the formation of the source system challenges some astrophysical models that mostly predict mergers with near-equal masses, but GW190412 is still compatible with most versions of both isolated binary evolution and dynamical assembly~\cite{Postnov:2014lrr,Benacquista:2013lrr}.
Implications for specific formation scenarios have since been the topic of many studies~\cite{Mandel:2020lhv,DiCarlo:2020lfa,Olejak:2020oel,Hamers:2020huo,Rodriguez:2020viw,Gerosa:2020bjb,DeLuca:2020qqa,Safarzadeh:2020qrc},
and more accurate and robust inference of the system's mass and spin parameters can be crucial in further constraining these channels.

Additional PE analyses of GW190412 have also been reported
by~\cite{Zevin:2020gxf}, focusing on the impact of alternative spin priors,
and recently by~\cite{GW190412:surro},
focusing on the \NRSur model and the effect of varying the lower cutoff frequency.

\subsection{Waveform models used}
\label{sec:systematics}

\begin{table*}[htpb]
 \caption{Waveform models used in this paper.
 We indicate which models include precession and which multipoles are included for each model.
 For precessing models, the multipoles correspond to those in the co-precessing frame.
 \label{tab:models}
 }
 \begin{ruledtabular}
 \begin{tabular}{llccc}
  \textbf{family} &  \textbf{full name} & \textbf{precession} & \textbf{multipoles $(\ell,\,|m|)$} & \textbf{ref.}\\
\hline
\multirow{4}{*}{EOBNR} 
  &  \seobnrvforrom & $\times$ &  (2, 2) & \cite{Bohe:2016gbl} \\
  &  \seobnrvforhmrom &  $\times$ &  (2, 2), (2,1), (3, 3), (4, 4), (5,5) & \cite{Cotesta:2018fcv, Cotesta:2020qhw} \\
  &  \seobnrvforp &  \checkmark &  (2, 2), (2, 1) & \cite{Ossokine:2020kjp,Babak:2016tgq,Pan:2013rra}\\
  &  \seobnrvforphm &  \checkmark & (2, 2), (2, 1), (3, 3), (4, 4), (5,5) & \cite{Ossokine:2020kjp,Babak:2016tgq,Pan:2013rra}
\\[6pt]
\multirow{4}{*} {Phenom - Gen. 3}
 &  \phD & $\times$ & (2, 2) & \cite{Husa:2015iqa, Khan:2015jqa} \\
 &  \phHM    &  $\times$ & (2, 2), (2, 1), (3, 3), (3, 2), (4,4), (4, 3) & \cite{London:2017bcn} \\
 &  \phPvtwo   &  \checkmark & (2, 2) & \cite{Hannam:2013oca, Bohe:PPv2} \\
 &  IMRPhenomPv3   &  \checkmark & (2, 2) & \cite{Khan:2018fmp} \\
 &  \phPvthreehm &  \checkmark & (2, 2), (2, 1), (3, 3), (3, 2),(4, 4), (4, 3) & \cite{Khan:2019kot}
\\[6pt]
 NR surrogate & \NRHybSur &  $\times$ & $\ell \leq 4$, (5, 5) but
not (4, 0), (4, 1)&\cite{Varma:2018mmi}\\[6pt]
\multirow{4}{*}{PhenomX} &  \phX & $\times$ & (2, 2) &
\cite{Pratten:2020fqn}
\\
  &  \phXHM &  $\times$ & (2, 2), (2, 1), (3, 3), (3, 2), (4,4)  & \cite{Garcia-Quiros:2020qpx,Garcia-Quiros:2020qlt} \\
  &  \phXP
  &  \checkmark & (2, 2) & \cite{Pratten:2020ceb} \\
  &  \phXPHM &  \checkmark & (2, 2), (2, 1), (3, 3), (3, 2),(4, 4) & \cite{Pratten:2020ceb}\\[6pt]
\multirow{2}{*}{PhenomT}  
  &  \phT &  $\times$ & (2, 2)  & \cite{phenomtp} \\
  &  \phTHM
  &  $\times$ &  (2, 2), (2, 1), (3, 3), (4,4), (5,5) & \cite{phenomthm}\\
 \end{tabular}
 \end{ruledtabular}
\end{table*}

CBC parameter estimation currently mostly uses waveform models from three different families:
\begin{itemize}
    \item Models constructed within the effective-one-body (EOB) framework \cite{Damour:2001tu, Damour:2012ky, Pan:2013rra, Cotesta:2018fcv}. In a first stage these are constructed as time-domain models, where Hamiltonians and GW fluxes are calibrated to NR simulations, and the ordinary differential equations resulting from the Hamiltonian equations are solved numerically for the inspiral, and carried through the merger and ringdown with phenomenological models.
    One of two models used in 
    \cite{LIGOScientific:2020stg} and describing both precession and subdominant harmonics, \seobnrvforphm \cite{Ossokine:2020kjp}, and its restriction to the dominant quadrupole content, \seobnrvforp, belong to this family.
    These models are typically computationally expensive,
    thus it is common to produce reduced-order-models (ROMs)
    to accelerate the evaluation of the waveforms.
    Two such models have been used in \cite{LIGOScientific:2020stg}:
    \seobnrvforhmrom, which describes non-precessing systems including HMs, and \seobnrvforrom, which corresponds to the $l=\vert m\vert =2$ content.
    Several generations of these models have been built, and
    \cite{LIGOScientific:2020stg} uses the fourth generation (``v4'').
    \item Phenomenological models, which are constructed as piecewise closed-form expressions that are calibrated to post-Newtonian or EOB inspiral descriptions and NR waveforms, and which can be evaluated very rapidly. As for the EOB models, several generations of such models have been built, with the generation used in \cite{LIGOScientific:2020stg} all constructed from the baseline \phD model for the $l=\vert m\vert =2$ modes of non-precessing binaries.
    Analytical approximate maps are used to model the HM content and precession.
    We will refer to the phenomenological models used  in \cite{LIGOScientific:2020stg} as the {\em third generation} (counting \phA \cite{Ajith_2007} as the first generation, and \phB \cite{Ajith:2009bn}, \phC \cite{Santamaria:2010yb} and \phP \cite{Hannam:2013oca} as the second generation).
    The \phXF family constitutes the next generation and current state of the art for such models, and corresponds to an update of essentially all aspects of the model.
    We will provide further details on how it relates to the third generation below.
    While previous phenomenological waveform models have all been constructed in the frequency domain, \phenX is complemented by the new 
    \phT time-domain models \cite{phenomtp,phenomthm}.
    Currently, only non-precessing versions of \phT have been implemented in LALSuite \cite{lalsuite} and can be used for the analysis presented here (see however \cite{phenomtp} for a Mathematica implementation lacking sub-dominant harmonics). 
    \item Finally, ROMs have also been successfully applied directly to interpolate between NR or hybrid waveforms in the time domain.
    Hybrids are built from appropriately ``gluing'' NR waveforms to an early inspiral described by an EOB model.
    The latest models of this kind are the non-precessing \NRHybSur \cite{Varma_2019}, which has been used in \cite{LIGOScientific:2020stg}, and the precessing \NRSur~\cite{Varma:2019csw}, which has not been used in that original discovery paper, since it does not span a sufficiently wide frequency range to cover the entire LIGO and Virgo bands for the mass range of GW190412.
    See however \cite{GW190412:surro} for results obtained more recently with \NRSur, and studies on varying the lower cutoff frequency.
\end{itemize}

For a complete list of all the waveform models used in \cite{LIGOScientific:2020stg} and the 
present paper see Table \ref{tab:models}. We now turn to describing the new \phXF and \phT families in more detail.

The starting point of the \phXF family is  \phX \cite{Pratten:2020fqn}, which models the $(\ell, \vert m\vert) = (2,2)$ modes
of signals from non-precessing BBHs, and has been extended to include the 
$(\ell, \vert m\vert) = (2,1), (3,3), (3,2), (4,4)$ modes by \phXHM.
All these modes
have been calibrated to around 500 numerical waveforms, as reported in \cite{Garcia-Quiros:2020qlt},
whereas for the third generation of Phenom waveforms only the dominant $(\ell, \vert m\vert) = (2,2)$ modes had 
been calibrated to 19 NR waveforms in 2015 \cite{Husa:2015iqa,Khan:2015jqa}.
\phXPHM~\cite{Pratten:2020ceb} is our most complete state-of-the-art phenomenological waveform model for quasi-circular precessing signals including the same modes as \phXHM, and \phXP is the corresponding precessing version of the dominant-mode \phX.

Following the paradigm \cite{Schmidt:2010it,Schmidt:2012rh,Hannam:2013oca} 
adopted by previous phenomenological models, the precessing versions are obtained by extending the non-precessing waveforms by an approximate map, which identifies the non-precessing spherical
harmonic modes with the precessing modes in a non-inertial co-precessing frame. We refer to this procedure as ``twisting up''.
For a detailed discussion of the procedure and the conventions employed to describe precessing
waveforms in the frequency domain see \cite{Pratten:2020ceb}.
The twisting construction relies on a 
prescription for the three frequency-dependent Euler angles which rotate the modes in the non-inertial frame into
the intertial frame, which is used for gravitational wave data analysis.
These angles encode the amplitude and phase modulations determined by the precession dynamics and are by  default computed following the double spin multiple-scale-analysis (MSA) prescription \cite{Chatziioannou:2017tdw},
although the model has the option to instead evolve them at next-to-next-to-leading order (NNLO) in the post-Newtonian approximation as a function of an effective single spin.
The MSA prescription is similar to that used in \phPvthree, and the NNLO prescription to what is used in \phPvtwo.
The mode content of \phXPHM in the co-precessing frame can be freely specified, with the full set comprising the modes  $(\ell, \vert m\vert) = (2,2), (2,1), (3,3), (3,2), (4,4)$.
Note also that the modelling of the co-precessing $(3,|2|)$ mode incorporates mode-mixing effects.
Another notable feature of \phXPHM is its use of the multibanding interpolation method introduced by \cite{Vinciguerra:2017ngf}, implemented in an improved way as described in \cite{Garcia-Quiros:2020qlt}.
The algorithm is applied to the evaluation of both the aligned-spin waveforms and the precession angles, with default thresholds that can be further relaxed by the user to allow for even quicker exploratory runs.
In Sec. \ref{subsec:cost}, we will discuss the benefits of multibanding for parameter estimation studies.

The ``twisting-up'' procedures adopted by \phPvthreehm and \phXPHM are almost equivalent, as both models adopt by default the MSA prescription.
However, \phXPHM by default falls back to NNLO angles when MSA failures are encountered.
This feature is advantageous in parameter estimation studies as, with no fallback in place, samplers may get stuck in regions of parameter space where the MSA equations become numerically unstable.
\phXPHM offers several options to set the precession prescription, with the default version (also called version 223) being the one just described. Another relevant version for this work will be version 102, which instead enforces the use of NNLO angles. 
\phXF also allows to choose among four final-spin formulas, with the default version using a precession-averaged equation inspired by the MSA formalism (``FS version 3''), see \cite{Pratten:2020ceb} for details.
Alternative versions attach the in-plane spins to the larger mass, either relying on the usual effective precession spin $\chi_p$ (version 0, which is adopted by all third generation Phenom models), or by taking the norm of the in-plane spin vectors at the reference frequency (version 2). We will discuss the impact of these settings on parameter estimation in
 appendix \ref{sec:prec_versions}.

A further important element distinguishing \phXPHM from its predecessor \phPvthreehm is the calibration of the underlying aligned-spin higher modes (HM) to NR waveforms.
It has been already shown in \cite{Garcia-Quiros:2020qlt} that this results in an increased faithfulness both with respect to NR and to \NRSur.
We will study the impact of this HM calibration on parameter estimation in appendix \ref{sec:phenompv3_comparison}.

Thanks to the modular way that \phXHM and \phXPHM are constructed from the baseline dominant-mode model \phX,
their LALSuite implementations when called with the dominant $(2,2)$ modes only will reproduce the \phX and \phXP models respectively.
This approach has the added advantage of providing multibanding speedup,
and we used it for most of the dominant-mode PE runs included in this paper.
For clarity those runs are usually referred to as ``\textsc{IMRPhenomX(P)HM (2,2)}'' (or shortened versions of this) in the following.

Recently, \phenX models have been complemented by the new \phT time-domain models \cite{phenomtp,phenomthm}.
\phT and \phTHM model the $(\ell, \vert m\vert) = (2,2), (2,1), (3,3), (4,4), (5,5)$ modes of non-precessing signals, while \phTP describes the $\ell=2$ sector of precessing signals applying the ``twisting-up'' procedures to the dominant mode $(\ell, \vert m\vert) = (2,2)$ described by \phT.
Like the corresponding non-precessing models of the \phXF family, \phT and \phTHM have been calibrated to around 500 numerical waveforms.
Model construction follows a similar design as in other phenomenological models, describing the inspiral region of the signal through a post-Newtonian quasi-circular approximant, TaylorT3~\cite{Buonanno_2009},
extended to higher pseudo-PN orders calibrated with numerical waveforms, providing phenomenological descriptions of the merger signal and including a calibrated ringdown description based on the quasinormal mode expansion of the signal \cite{Damour_2014}.

\subsection{Methodology for our parameter estimation analysis}\label{sec:PEsetup}

For this reanalysis, we use v2 of the strain data~\cite{GW190412:gwosc-v2} for GW190412 released through the Gravitational Wave Open Science Center~(GWOSC) \cite{Vallisneri:2014vxa,Abbott:2019ebz}, with a default sampling rate of 16384\,Hz, for consistency with the official LVC study.
This version has non-linear subtraction~\cite{Vajente:2019ycy} of 60\,Hz power lines applied to it.
We also use the power spectral densities (PSDs)~\cite{Littenberg:2014oda,Chatziioannou:2019zvs} and calibration uncertainties~\cite{Sun:2020wke} included in v11 of the posterior sample release~\cite{GW190412:dcc} for the event.
We analyse 8\,s of strain data from each of the Hanford, Livingston and Virgo detectors around the trigger time of the event, as reported in GraceDB~\cite{GraceDB}.

We have carried out most of our Bayesian parameter estimation runs using the python-based package pBilby \cite{10.1093/mnras/staa2483,Ashton:2018jfp,Smith:2016qas} with static nested sampling \cite{Skilling:2004ns} as implemented in the dynesty Python code \cite{10.1093/mnras/staa278}. We use the default parameters of the dynesty implementation in Bilby, apart from the following more important parameters:
We vary the number of nested sampling live points ($n_\mathrm{live}$) and the minimum length of the chain, in terms of multiples ($n_\mathrm{act}$) of its auto-correlation length. We also set a minimal ($walks$) and maximal ($maxmcmc$) number of Markov-Chain Monte Carlo (MCMC) steps. All parallel Bilby runs are carried out with four statistically independent sampling runs, usually referred to as ``seeds''.

The lower and upper cutoff frequencies for the likelihood integration were taken to be 20\,Hz and 2048\,Hz respectively. 
We adopted the same ``tight'' priors as employed for the LVC pBilby runs, with narrow bounds based on initial exploratory runs.
Explicit prior settings are listed in appendix \ref{sec:spin_priors}.
We have also examined the impact of alternative spin priors, as explained in appendix \ref{sec:spin_priors}.  
Besides custom post-processing scripts, the PESummary~\cite{Hoy:2020vys} package was also used for comparisons of multiple runs.

In order to claim confidence in our results, we need to check how posteriors change when refining our sampler settings.
To this end we have varied the number of live points $n_\mathrm{live}$ and the number $n_\mathrm{act}$ of autocorrelation times to use before accepting a point.
Our main series of runs, summarized in Table~\ref{tab:tabRuns}, uses marginalisation over the distance parameter and then a reconstruction of the distance posterior as discussed in \cite{Romero-Shaw:2020owr,Thrane:2018qnx} to accelerate the convergence of nested sampling\footnote{We do not use time-marginalization, due to potential issues in the sky-localization posteriors.
We also do not use phase-marginalization since it is not fully compatible with precessing and higher-modes waveforms.
}.
In Sec.~\ref{subsec:dist_marg}
we compare the distance-marginalised results with a series of simulations that do not use distance marginalisation (listed in Table \ref{tab:tab_oldRuns}).

Some additional comparison runs as listed in Table\,\ref{tab:LI_runs} were performed with the C library LALInference (LI) \cite{Veitch:2014wba}, using both MCMC and nested sampling, since the parallel Bilby code is still relatively new. The LALInference runs do not use distance (or other types of) marginalisation and they are compared with our default series of runs in appendix~\ref{subsec:LI_vs_pbilby}.

\begin{table*}[hptb]
\begin{center}
\begin{tabular}{|c|c|c|cccc|c|cc|r|c|c|}
\hline 

Approximant &  Run $nº$  &         Modes (l,|m|) & PV &   FS &    PMB & MB & Prior &  $n_\mathrm{live}$  &  $n_\mathrm{act}$ &  CPU h & L. eval. & Cost/L. eval. [ms]\\
\hline \hline 
\multirow{8}{*}{\phXHM} & 1 & (2,2) & - & - & - & D &   Aligned spin  &   2048 &  10 &   1186 & $1.10 \times 10^{8}$ & 38.96 \\ 
& 2 & (2,2) & - & - & - & D &   Aligned spin  &   2048 &  50 &  5819  & $5.48 \times 10^{8}$ & 38.26 \\ 
& 3 & (2,2) & - & - & - & $10^{-1}$ &   Aligned spin  &   2048 &  10 &  1171  & $1.09 \times 10^{8}$ & 38.83 \\ 
&4 & (2,2) & - & - & - & $10^{-2}$ &   Aligned spin  &   2048 &  10 &  1166 & $1.09 \times 10^{8}$ & 38.43  \\ 

&5& D & - & - & - & D &   Aligned spin  &   2048 &  10 &  1666 & $1.29 \times 10^{8}$ & 46.39  \\ 
&6& D & - & - & - & D &   Aligned spin  &   2048 &  50 & 7918 & $6.38 \times 10^{8}$ & 44.71 \\ 
&7& D & - & - & - & $10^{-1}$ &   Aligned spin  &   2048 &  10 &  1501 & $1.29 \times 10^{8}$ & 41.78 \\ 
&8& D & - & - & - & $10^{-2}$ &   Aligned spin  &   2048 &  10 &   1494 & $1.28 \times 10^{8}$ & 42.15 \\ 

\hline 
\multirow{4}{*}{\phTHM} &9& (2,2) & - & - & - & - &   Aligned spin  &   1024 &  10 & 874 & $5.06 \times 10^{7}$ & 62.22 \\ 
&10& (2,2) & - & - & - & - &   Aligned spin  &   2048 &  10 &  1649 & $1.02 \times 10^{8}$ & 58.00 \\ 
&11& D & - & - & - & - &   Aligned spin  &   1024 &  10 & 1226  & $6.19 \times 10^{7}$ & 71.27  \\ 
&12& D & - & - & - & - &   Aligned spin  &   2048 &  10 & 2332  & $1.22 \times 10^{8}$ & 68.69  \\ 

\hline 

\multirow{26}{*}{\phXPHM} &13& (2,2) & D & D & D & D &   Precessing  &   2048 &  10 & 1375 & $1.11 \times 10^{8}$ & 44.30 \\ 
&14& (2,2) & D & D & D & D &   Precessing  &   2048 &  50 & 6681 & $5.54 \times 10^{8}$ & 43.42 \\ 
&15& (2,2) & D & D & $10^{-1}$ & $10^{-1}$ &   Precessing  &   2048 &  10 & 1319 & $1.11 \times 10^{8}$ & 42.40 \\ 
&16& (2,2) & D & D & $10^{-1}$ & $10^{-2}$ &   Precessing  &   2048 &  10 & 1320 &$1.13 \times 10^8$ & 41.89\\ 
&17& (2,2) & D & D & $10^{-2}$ & $10^{-1}$ &   Precessing  &   2048 &  10 & 1297 &$1.11 \times 10^8$  & 42.00\\ 
&18& (2,2) & D & D & $10^{-2}$ & $10^{-2}$ &   Precessing  &   2048 &  10 & 1296 & $1.31 \times 10^8$ & 41.54\\ 
&19& D & D & D & D & D &   Precessing  &   512 &  10 & 679 &  $3.02 \times 10^7$ & 80.84 \\
&20& D & D & D & D & D &   Precessing  &   512 &  50 & 3002 & $1.36 \times 10^8$ & 79.47 \\
&21& D & D & D & D & D &   Precessing  &   1024 &  10 & 1318 & $6.11 \times 10^7$ & 77.62 \\
&22& D & D & D & D & D &   Precessing  &   1024 &  50 & 6129 & $3.01 \times 10^8$ & 73.20 \\
&\pmb{23}&  \textbf{D} & \textbf{D} & \textbf{D} & \textbf{D} & \textbf{D} &   \textbf{Precessing}  &   \textbf{2048} &  \textbf{10} & \textbf{2670} & $\mathbf{1.32 \times 10^8}$ & \textbf{72.95} \\
&24&  (2,2),(2,1) & D & D & D & D &   Precessing  &   2048 &  10 & 1580  & $1.19 \times 10^8$ & 47.77 \\
&25&  (2,2),(2,1),(3,2),(3,3) & D & D & D & D &   Precessing  &   2048 &  10 & 2325 & $1.04 \times 10^8$  & 80.44 \\
&\pmb{26}& \textbf{D} & \textbf{D} & \textbf{D} & \textbf{D} & \textbf{D} &  \textbf{Precessing}  &   \textbf{2048} &  \textbf{50} & \textbf{13530} & $\mathbf{6.61 \times 10^8}$ & \textbf{73.64} \\
&27& D & D & D & D & D &   Precessing  &   4096 &  10 & 5718 & $2.79 \times 10^8$ & 73.79 \\
&28& D & D & D & $10^{-1}$ & $10^{-1}$ &   Precessing  &   2048 &  10 & 1900 & $1.31 \times 10^8$ & 52.01 \\ 
&29& D & D & D & $10^{-1}$ & $10^{-2}$ &   Precessing  &   2048 &  10 &  1984 & $1.33 \times 10^8$ & 53.70 \\ 
&30& D & D & D & $10^{-2}$ & $10^{-1}$ &   Precessing  &   2048 &  10 &  1989 & $1.32 \times 10^8$ & 54.19 \\
&31& D & D & D & $10^{-2}$ & $10^{-2}$ &   Precessing  &   2048 &  10 & 2001 & $1.32 \times 10^8$  & 54.63 \\ 
&32& D & 223 & 0 & D & D &   Precessing  &   2048 &  10 & 2679 & $1.31 \times 10^8$ & 73.80 \\
&33& D & 223 & 1 & D & D &   Precessing  &   2048 &  10 & 2711 & $1.03 \times 10^8$ & 94.83 \\
&34& D & 223 & 2 & D & D &   Precessing  &   2048 &  10 & 2647 & $1.31 \times 10^8$ & 72.52 
\\
&35& D & 102 & 0 & D & D &   Precessing  &   2048 &  10 & 2217 & $1.29 \times 10^8$ & 61.49 \\
&36& D & 102 & 1 & D & D &   Precessing  &   2048 &  10 & 2160 & $1.28 \times 10^8$ & 60.79 \\
&37& D & 102 & 2 & D & D &   Precessing  &   2048 &  10 & 2230 & $1.30 \times 10^8$ & 61.68 \\

\hline 
\end{tabular}
\end{center}
\caption{
\label{tab:tabRuns}
This table lists all the runs we have performed with parallel Bilby and distance marginalisation on open data for GW190412.
For each run, we indicate the LALSuite waveform approximant called along with various waveform settings:
the mode content, precessing prescription (PV), final spin version (FS), and multibanding thresholds applied to the evaluation of the Euler angles (PMB) and of the aligned-spin modes (MB);
as well as the prior used and the chosen sampler settings ($n_\mathrm{live}$ and $n_\mathrm{act}$).
We also provide the computational cost of each run, the number of likelihood evaluations and the mean cost of each evaluation.
Two runs are highlighted in boldface:
run 26 is our preferred \phXPHM run for astrophysical results,
while the cheaper but essentially equally accurate run 23 serves as a comparison baseline for alternative model choices and sampler settings.}

\end{table*}

To quantify the differences between results which use different sampler settings, we also compute their Jensen-Shannon (JS) divergence \cite{Majtey2005} and analyze the results in Sec.~\ref{subsec:cost}.
This quantity measures the distance between two probability distributions as a value in [0,1], where 0 means that both distributions are exactly alike and 1 means maximal divergence.
The JS divergence between two distributions $p(x)$ and $q(x)$ is defined as
\begin{equation}
    D_\mathrm{JS}(p|q) = \frac{1}{2}\bigg[D_\mathrm{KL}\bigg(p\bigg\rvert\frac{1}{2}(p+q)\bigg) + D_\mathrm{KL}\bigg(\frac{1}{2}(p+q)\bigg\rvert q\bigg)\bigg],
\end{equation}
where the Kullback-Leibler divergence $D_\mathrm{KL}$  \cite{kullback1951} is defined as
\begin{equation}
    D_\mathrm{KL}(p|q) = \int p(x)\log_2\left(\frac{p(x)}{q(x)}\right)dx.
\end{equation}

\section{PE results for GW190412}
\label{sec:PEresults}

\subsection{Analysis with \phXPHM}\label{subsec:XPHMruns}

\begin{figure*}[htpb]
\begin{center}
\includegraphics[width=0.8\columnwidth]{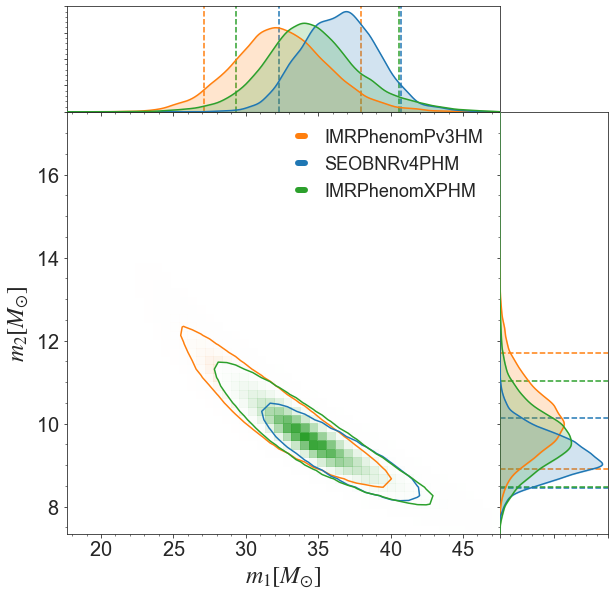}
\includegraphics[width=0.825\columnwidth]{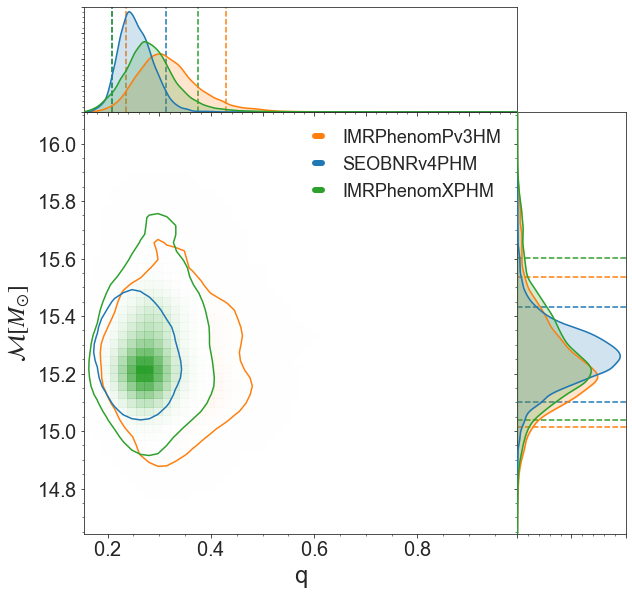}\\
\includegraphics[width=0.8\columnwidth]{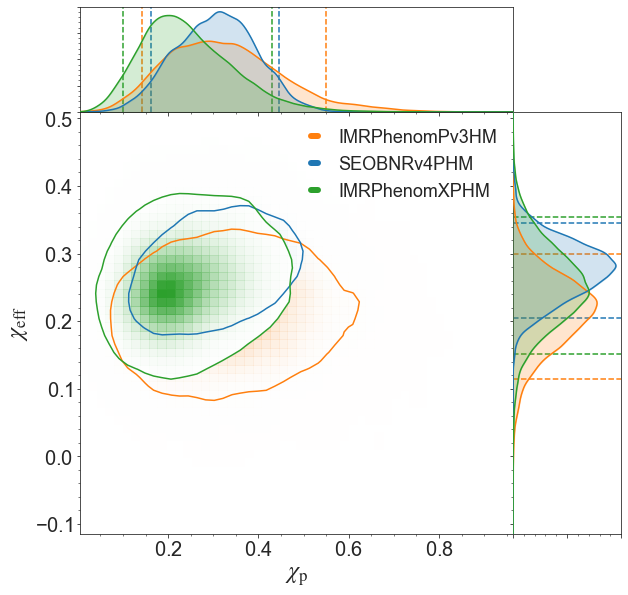}
\includegraphics[width=0.825\columnwidth]{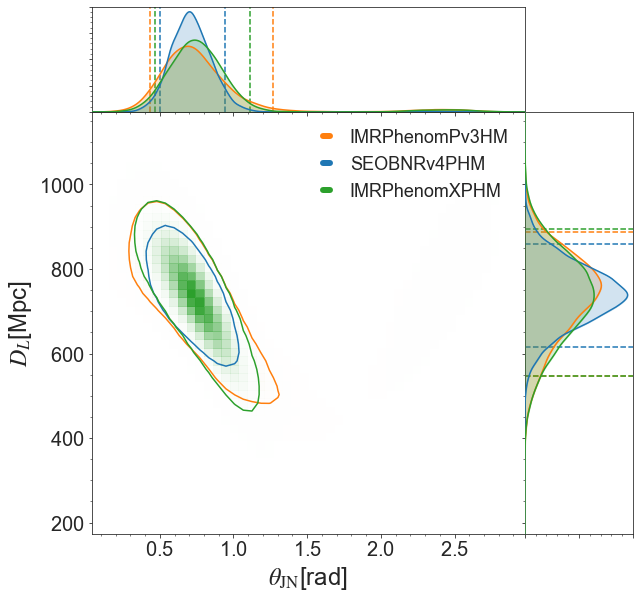}
\caption{Posterior distributions for the masses, effective spin parameters, distance and inclination of GW190412, as estimated with \seobnrvforphm (blue), \phPvthreehm (orange) and \phXPHM (green) (using run 26 from Table \,\ref{tab:tabRuns}).
For each pair of parameters, the square central panel shows the 2D joint posteriors with contours marking 90\% credible regions; while the top and right side-panels show one-dimensional distributions for the individual parameters, with dashed lines indicating 90\% credible intervals.
\label{fig:precessing_posteriors}
}
\end{center}
\end{figure*}

We first compare posterior distributions for the two precessing models used in the LVC paper \cite{LIGOScientific:2020stg}, \seobnrvforphm and \phPvthreehm, with our \phXPHM model in Fig.~\ref{fig:precessing_posteriors}.
The posterior results for \seobnrvforphm and \phPvthreehm are taken from the official LVC release samples, while those shown for \phXPHM correspond to our preferred run (run 26, shown in bold, in Table\,\ref{tab:tabRuns}),
which uses $n_{\mathrm{live}}=2048$ and $n_\mathrm{act}=50$.

\begin{figure}[htbp]
\begin{center}
\includegraphics[width=\columnwidth]{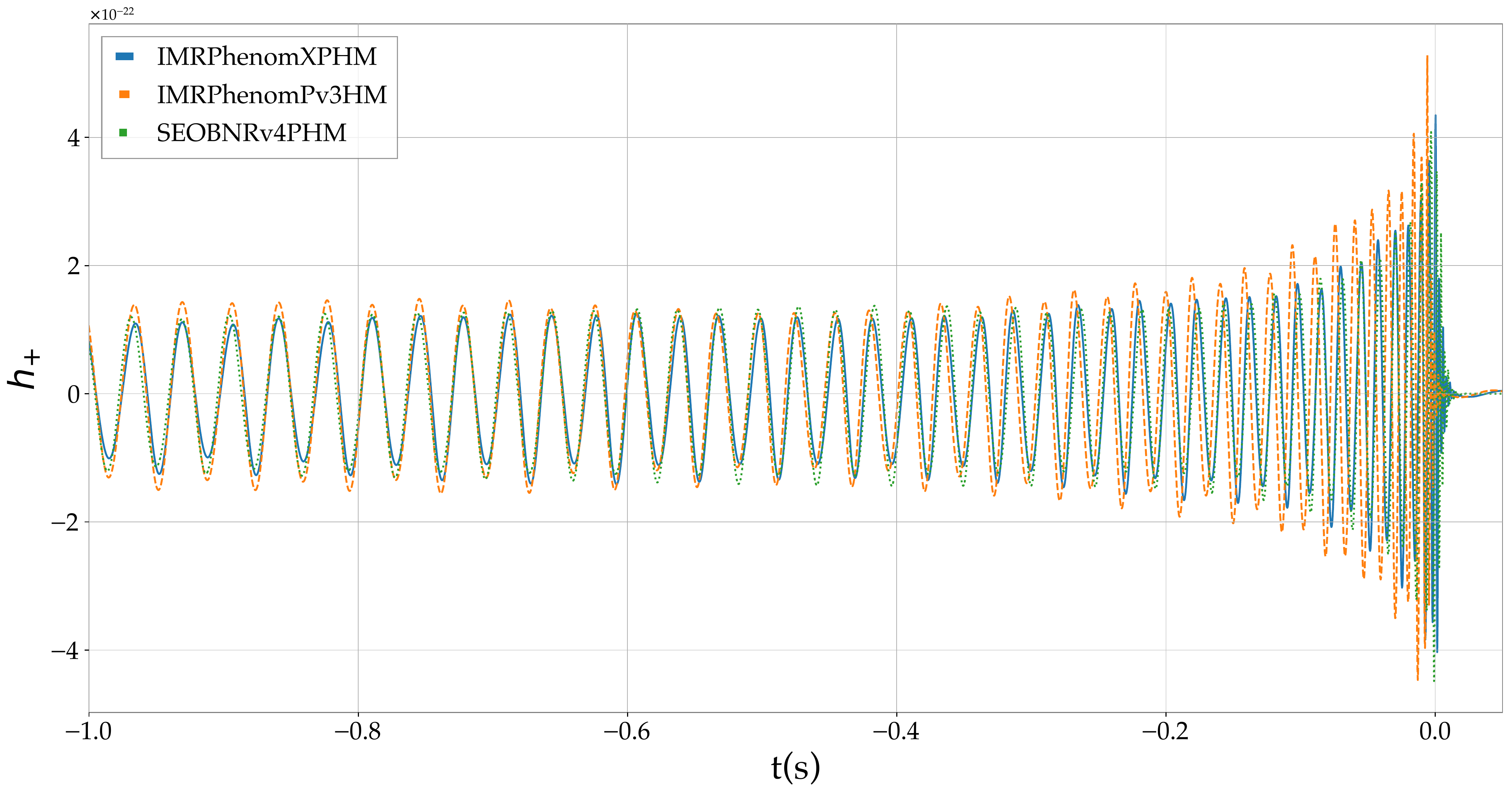}
\caption{Waveforms ($h_+$ polarization as a function of time) corresponding to the maximum-likelihood samples for \phXPHM (blue), \phPvthreehm (orange) and \seobnrvforphm (green).
The models show good agreement during the inspiral, while larger differences can be appreciated around merger. The maximum-likelihood sample for \phXPHM is taken from run 26.
\label{fig:maxLwaveforms}
}
\end{center}
\end{figure}

The parameters recovered with the three models considered here are broadly consistent, as can be seen by inspecting the joint posterior distributions for some of the key source properties (Fig.\,\ref{fig:precessing_posteriors})
as well as the corresponding maximum-likelihood waveforms (Fig.\,\ref{fig:maxLwaveforms}).

While all these models agree fairly well on the estimated distance and inclination of the source, differences are clearly visible in the mass and spin parameters. The component masses estimated with \phXPHM lie in between those estimated with \seobnrvforphm and \phPvthreehm.
We notice that \phPvthreehm tends to return broader posteriors than the other models, which appear to be somewhat more consistent with each other. We also see that \seobnrvforphm is favouring more asymmetric masses and higher $\chi_{\mathrm{eff}}$ and $\chi_{\mathrm{p}}$, while \phXPHM prefers lower values of $\chi_p$, consistently with our analysis of Bayes factors, which will be presented in the next section.

Given the better agreement between the \seobnrvforphm and \phXPHM models, we compute their combined posterior as our best estimate of the source parameters, and list the estimated source parameters and error estimates in Table \ref{tab:PEresults}. These improvements of the LVC estimates which combine
\seobnrvforphm and \phPvthreehm  posteriors constitute our main astrophysical result. 

\begin{figure*}[htpb]
\begin{center}
\includegraphics[width=0.8\columnwidth]{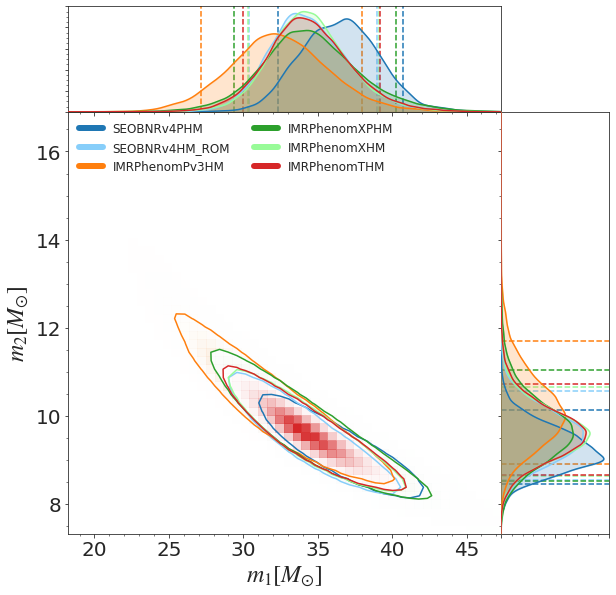}
\includegraphics[width=0.825\columnwidth]{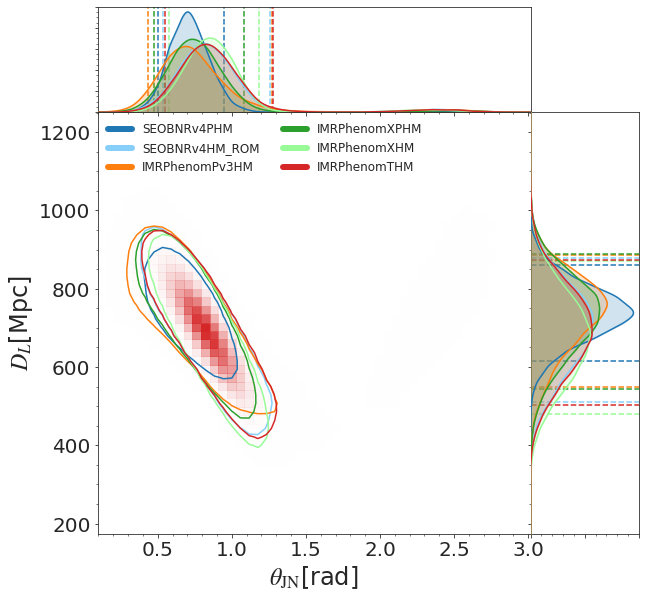}\\

\caption{Comparison of posterior distributions for the component masses,  distance and inclination of GW190412, using \seobnrvforphm (dark blue), \seobnrvforhmrom (light blue), \phPvthreehm (orange), \phXPHM (dark green), \phXHM (light green) and \phTHM (red).}
\label{fig:aliged_vs_prec_HM_models}
\end{center}
\end{figure*}

\subsection{Comparison with models of reduced content}\label{sec:reduced_comparison}

To investigate support in the GW190412 data for the presence of subdominant modes and precession, we compare a reference run using the full \phXPHM model (23 in Table~\ref{tab:tabRuns})
against additional runs with equal sampler settings but using the models \phT, \phX, \phXP which drop either one or both of these additional aspects.
As discussed above, in fact we called \phXHM and \phXPHM with $(2,2)$-modes only instead of the named \textsc{XAS} / \textsc{XP} LALSuite approximants.

Our main tool to compare how well each model fits the data are Bayes factors, i.e. ratios of the marginal likelihoods of the models to be compared. If we indicate by $\mathcal{Z}_{A}$ the marginal likelihood of model A and by $\mathcal{Z}_{B}$ that of model B, the Bayes factor will take the form
\begin{align}
\mathcal{B_{B/A}}=\frac{\mathcal{Z}_{B}}{\mathcal{Z}_{A}},    
\end{align}
with $\log_{10}{\mathcal{B}_{B/A}}>1$ indicating that model B is strongly preferred over model A.  
In Table~\ref{tab:tabBF} we list $\log_{10}{\mathcal{B}_{B/A}}$ for two sets of hypotheses, namely whether the signal is best described 1) by an aligned-spin or precessing model and 2) by a quadrupole-only or higher-mode model.
Error bars are computed using error propagation, from the uncertainties associated to the evidence integral for each run.

We find strong evidence ($\log_{10}\mathcal{B}>3$)
for the presence of higher modes,
consistent with the results from~\cite{LIGOScientific:2020stg}. 
On the other hand, there is no actual evidence from the data to
infer the presence of precession effects over aligned-only spins,
neither when limiting to dominant multipoles or when including the subdominant harmonics.
This second result is also consistent with~\cite{LIGOScientific:2020stg}
where no numerical $\log_{10}\mathcal{B}$ were reported for this comparison,
since they were within numerical and systematic uncertainties --
matching our results which are very close to zero.

\begin{figure*}[htpb]
\begin{center}
\includegraphics[width=0.8\columnwidth]{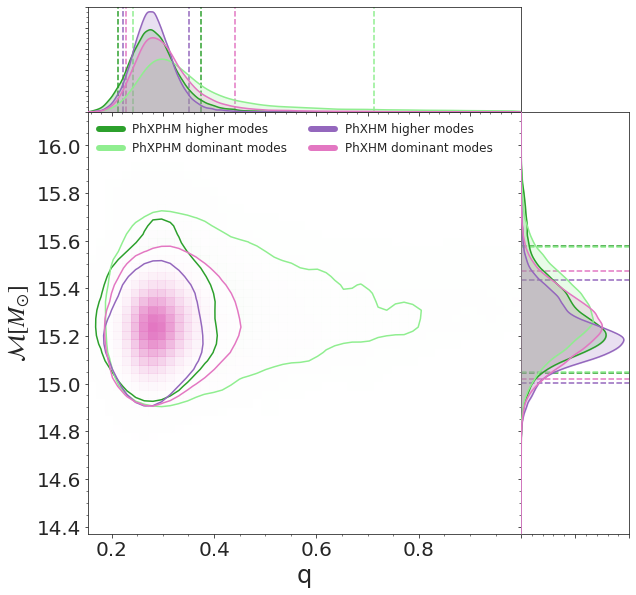}
\includegraphics[width=0.825\columnwidth]{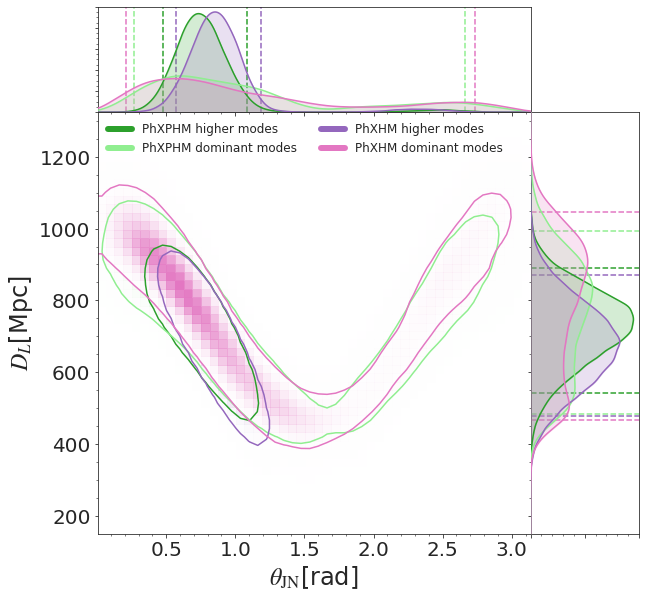}\\
\caption{Comparison of some estimated source parameters, using aligned (pink and purple) and precessing (light and dark green) waveform models from the \phXF family.
Light green and pink correspond to runs where only the quadrupole mode has been activated (note that, for the precessing model, this corresponds to the mode-content in the co-precessing frame).
The inclusion of higher multipoles helps to break the degeneracy between distance and inclination, as seen in the plot on the right.
\label{fig:HM_vs_22_models}
}
\end{center}
\end{figure*}

\begin{figure*}[htpb]
\begin{center}
\includegraphics[width=0.8\columnwidth]{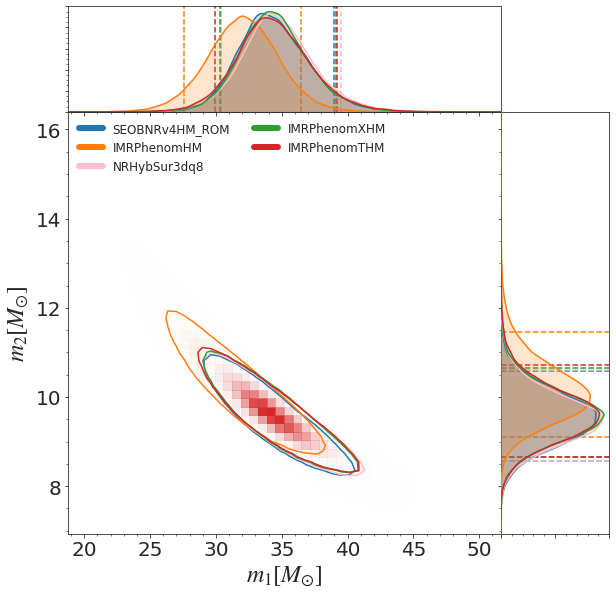}
\includegraphics[width=0.8\columnwidth]{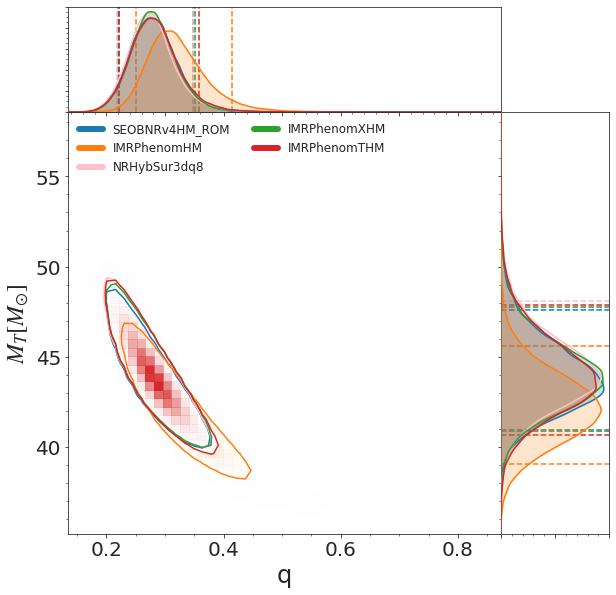}\\
\includegraphics[width=0.8\columnwidth]{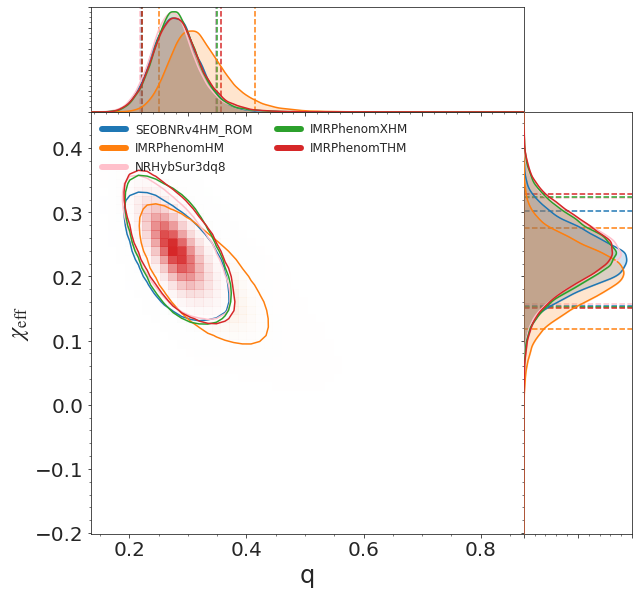}
\includegraphics[width=0.8\columnwidth]{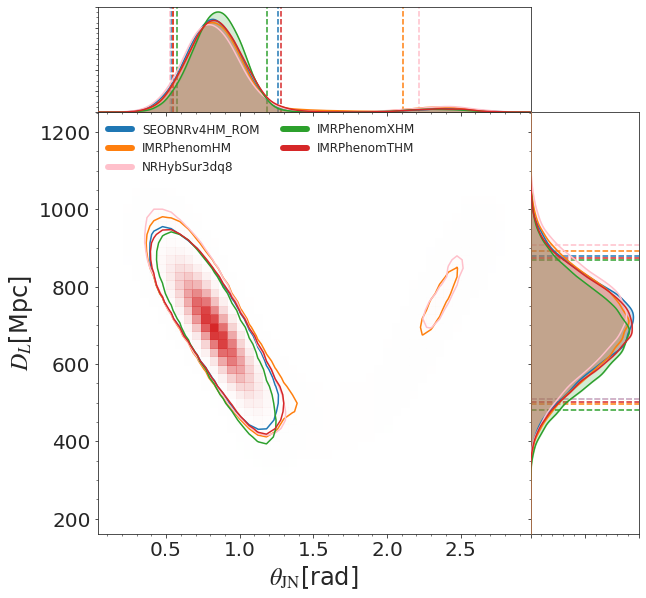}
\caption{
Posterior distributions for the masses, spin parameters, distance and inclination of GW190412,
as estimated with aligned-spin waveform models (\seobnrvforhmrom: blue, \NRHybSur: pink, \phHM: orange, \phXHM: green, \phTHM: red).
Models with multipoles individually calibrated to NR (\seobnrvforhmrom, \phXHM and \phTHM) are in excellent agreement with the numerical relativity surrogate \NRHybSur,
while \phHM returns visibly shifted posteriors.
\label{fig:posteriors_nonprec}
}
\end{center}
\end{figure*}

To explore in more detail the differences between the parameters inferred using these models, we compare some of the estimated source parameters for aligned-spin vs precessing models in Fig.~\ref{fig:aliged_vs_prec_HM_models} and for dominant-mode-only vs. higher-mode models in Fig.~\ref{fig:HM_vs_22_models}.
One can see here that the inclusion of precession alone does not significantly affect the posterior distributions, while much tighter constraints follow from including higher multipoles, in line with previous studies \cite{PhysRevD.101.124054,PhysRevD.99.124005,Usman_2019}. 
In Fig.~\ref{fig:posteriors_nonprec} we also show comparisons of our own aligned-spin results against those from \seobnrvforhmrom, \NRHybSur and \phHM, finding excellent agreement with the first two, while \phHM is an outlier due to its higher modes not being directly calibrated to NR.

\begin{table*}[hptb]
\begin{center}
\begin{tabular}{|c|c|cccc|}
\hline
Hypotheses & Model properties & EOB & Phenom & PhenomX & PhenomT   \\
\hline \hline 
\multirow{2}{*}{HM vs $\ell=2=\vert m\vert $} & aligned  &  3.5 & 3.4 & $3.5\pm 0.1$ & $3.4\pm 0.1$ \\
\cline{2-6}
 & precessing &  4.1 & 3.6  &  $3.3\pm0.1$ & - \\
 \hline
\multirow{2}{*}{prec. vs aligned} & dominant multipoles &  - & - & $0.0\pm0.1$ & -\\
 \cline{2-6}
& higher multipoles &  - & - & $0.3\pm0.1$ & -\\
\hline
\end{tabular}
\end{center}
\caption{Comparison of Bayes factors $\log_{10}\mathcal{B}$ between EOBNR, Phenom, PhenomX and PhenomT families for two hypothesis tests:
(i) whether the signal contains higher or only dominant multipoles,
(ii) whether spin precession is present or not.
The multipoles test is performed for both aligned and precessing waveforms,
and the test for precession both when including only dominant multipoles or also the higher multipoles.
No $\log_{10}\mathcal{B}$ numbers are quoted for ``prec. vs aligned`` from EOBNR and Phenom because in~\cite{LIGOScientific:2020stg} these were only noted to be not significant in comparison with statistical and systematic error bars.
}
\label{tab:tabBF}
\end{table*}

\subsection{Contributions from subdominant harmonics}

In order to quantify the strength of higher multipoles in GW190412, 
we consider the posterior samples obtained by analysing the event with \phXHM\footnote{Note that, in the twisting-up approximation, one can cleanly separate different mode-contributions in the inertial frame only in the aligned-spin limit, and that is why we carry out this analysis with \phXHM.}
(using run number 5 in Table~\ref{tab:tabRuns}) and compute the SNR distribution corresponding to each mode. 
The optimal SNR for one mode is defined as 
\begin{equation}
    \mathrm{\rho^{lm}_{opt}} = \sqrt{(h^{lm} | h^{lm})},
\end{equation}
where $(\:|\:)$ refers to the usual noise-weighted inner product
\begin{equation}
    (a|b) := \mathrm{Re} \int_0^{f_{max}} \frac{\tilde{a}(f)\tilde{b}^\ast(f)}{S_n(f)}df
\end{equation}
and $h^{lm}$ 
denotes the contribution of a given mode to the strain measured by the detector:
\begin{align}
    h^{lm} &= F_+ h_+^{lm} + F_\times h_\times^{lm},\\
    h_+^{lm} -i h_\times^{lm} &= h_{lm} \,_{-2}Y_{lm} + h_{l-m} \,_{-2}Y_{l-m}.
\end{align}

\begin{figure}[htpb]
    \centering
    \includegraphics[width=0.9\columnwidth]{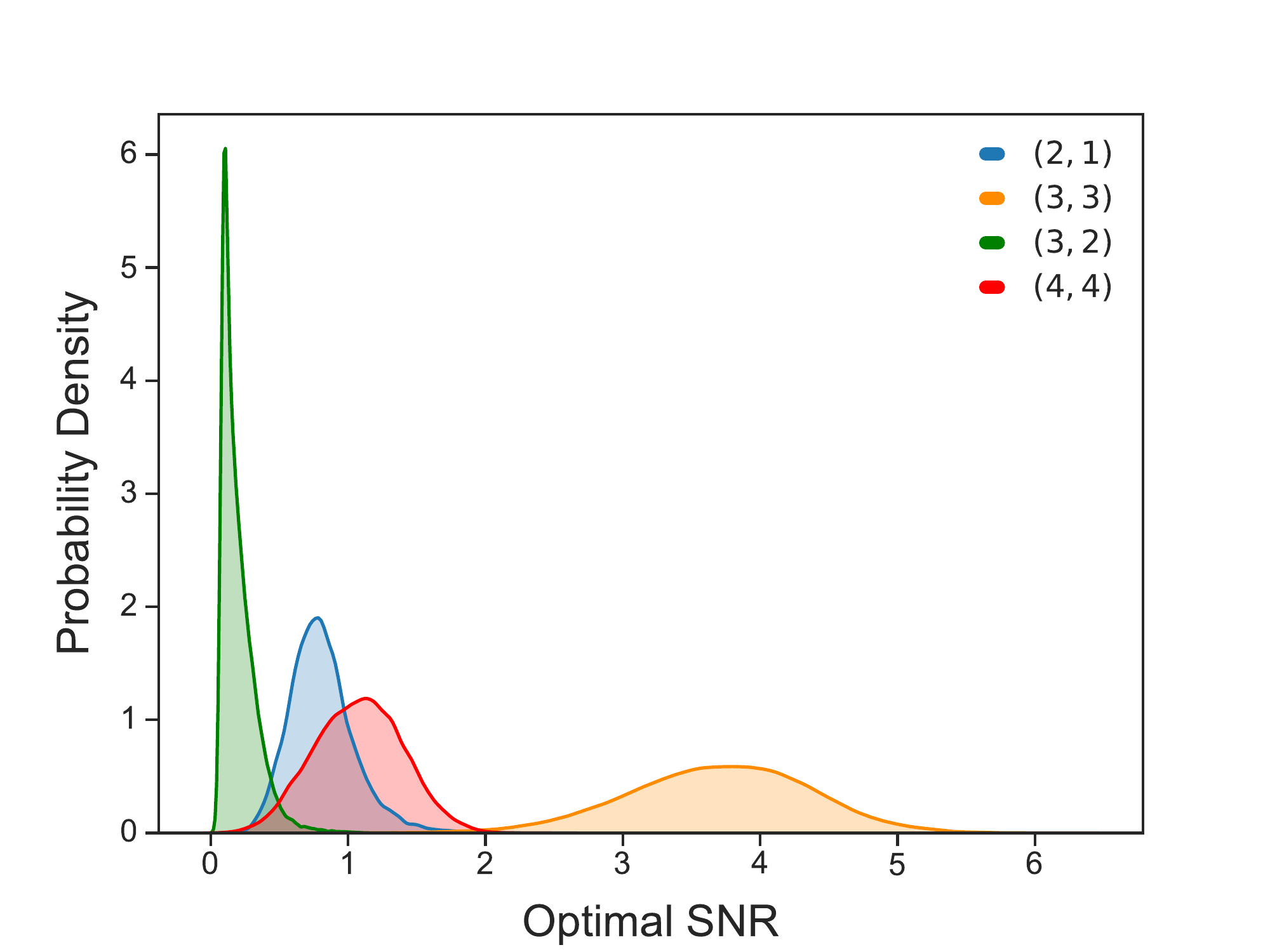}
    \includegraphics[width=0.9\columnwidth]{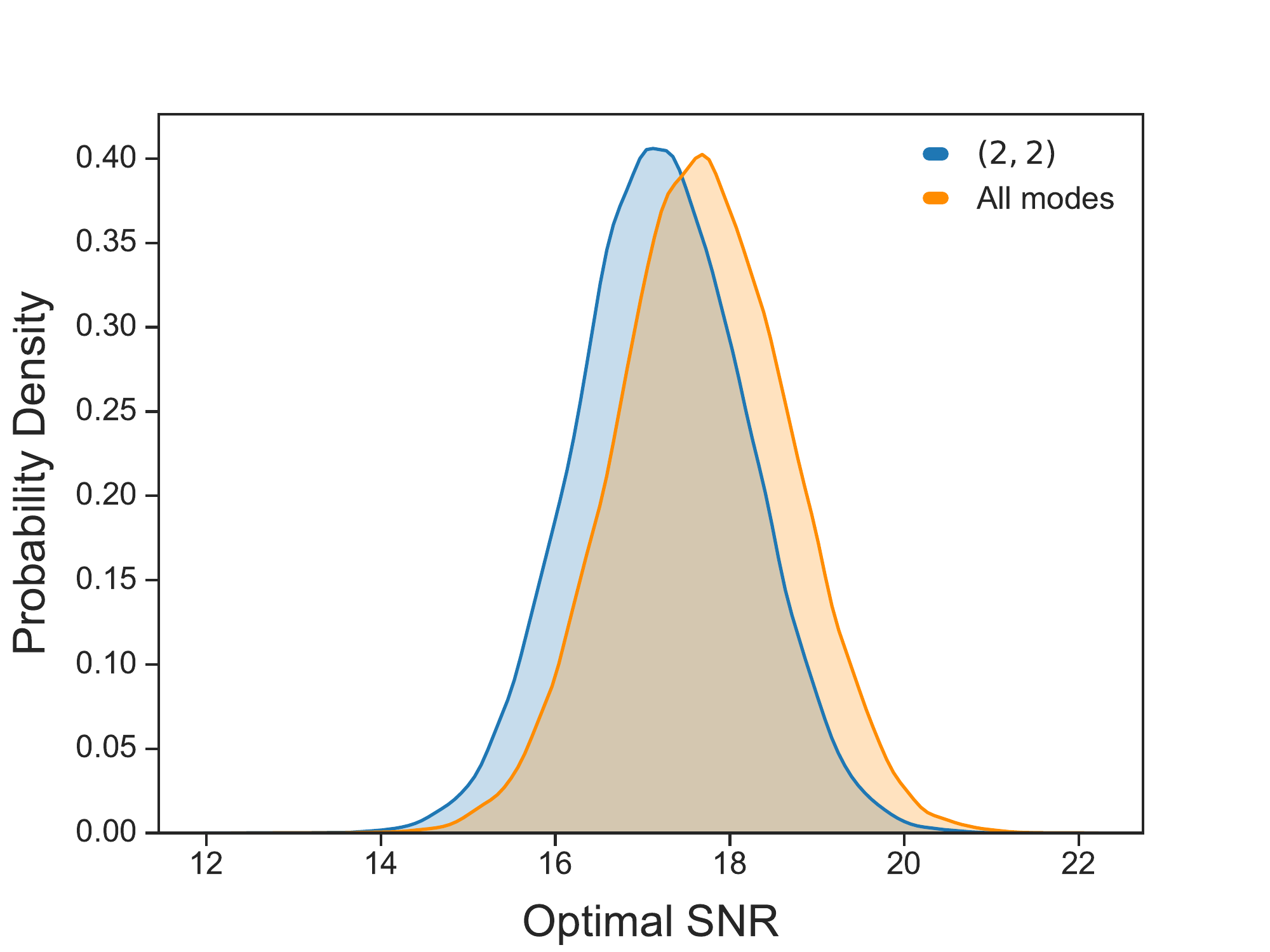}
    \caption{Top panel: Optimal SNR distributions for higher multipoles, when recovering the signal with \phXHM. The waveform used to compute the SNR contains just one single mode. Bottom panel: SNR distribution for the dominant $(2,2)$ mode only waveform and the whole multimode waveform.
    \label{fig:snr_modes}
    }
\end{figure}

\begin{figure}[htpb]
    \centering
    \includegraphics[width=0.9\columnwidth]{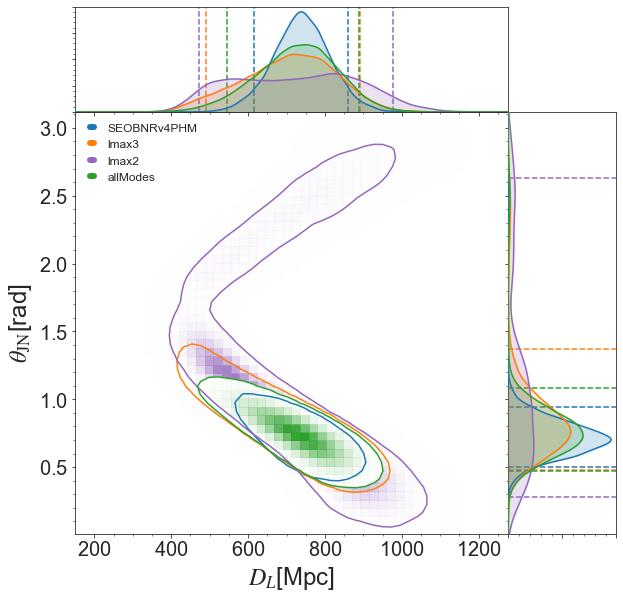}
    \caption{Joint posterior distributions for luminosity distance and inclination of GW190412, as estimated with \seobnrvforphm (blue), \phXPHM with its full mode-content (green) and \phXPHM with only the $\ell=2$ (purple) or $\ell\leq3$ modes activated.
    \label{fig:lum_inc_modes}
    }
\end{figure}

Fig.~\ref{fig:snr_modes} summarizes our results. In the top panel, we compare the SNR distributions for the subdominant multipoles available in \phXHM, and find that the strongest contribution comes from the $(3,3)$ mode, in line with \cite{LIGOScientific:2020stg} \footnote{We have verified that the single-mode SNR contributions calculated here are consistent with those obtained in \cite{LIGOScientific:2020stg}, where the SNR of each subdominant mode is first orthogonalised with respect to that of the $(2,2)$ mode.}. In the lower panel, we compare the full optimal SNR distribution to that of the quadrupole mode only. As expected, the $(2,2)$ is by far the strongest mode, however, the contribution of subdominant harmonics is clearly non-negligible. This is also illustrated in Fig.\,\ref{fig:lum_inc_modes}, where we show the estimated luminosity distance and inclinations obtained when activating only the $\ell=2$ and $\ell\leq3$ modes and compare them with the results of a standard run, for which $\ell_{\mathrm{max}}=4$. One can clearly see that increasing the mode content of the model results in tighter constraints on the source location. We note also the addition of the $(4,4)$ mode alone, which is the second-strongest subdominant mode, has a visible effect on the posteriors.

\subsection{Convergence, computational efficiency and cost}\label{subsec:cost}

One of the main advantages of \phXPHM is its computational efficiency, which makes it a potential workhorse for large scale parameter estimation studies, and an ideal tool for convergence tests and systematic studies of sampler settings. This is particularly important considering that future sensitivity improvements of the LIGO-Virgo detector network will result in an increased detection rate~\cite{Aasi:2013wya}.
The modularity of \phXPHM, and in particular the possibility to relax the multibanding thresholds, naturally lends itself to a hierarchical PE workflow, where computationally cheap runs can be set up to find optimal choices of priors, while more expensive runs are reserved for final data releases. 

Our strategy here has been to select a default configuration with 2048 nested sampling live points ($n_\mathrm{live}$) and ($n_\mathrm{act}=10$), corresponding to run 23 in Table~\ref{tab:tabRuns}.\footnote{Run 26 with $n_\mathrm{act}=50$ has been used for our main astrophysical results instead, but run 23 with $n_\mathrm{act}=10$ is the baseline for efficiently comparing to alternative settings and models.}
As we report below, from all our tests this configuration gives robust results when compared both with more expensive settings (increasing $n_\mathrm{live}$ and $n_\mathrm{act}$), and with computationally cheaper settings.
This justifies to use our default setting for comparisons of different choices of precession treatment in appendices~\ref{sec:prec_versions} and \ref{sec:phenompv3_comparison},
and for comparisons with runs that reduce the physics content in the model to non-precessing spins  or the dominant quadrupole in Sec.~\ref{sec:reduced_comparison}. 
We will also show here that configurations with drastically reduced computational cost still reproduce the posteriors of more accurate runs rather well, and are sufficient for fast exploratory runs,
e.g. to determine prior settings,
or to obtain rapid accurate distance measurements that take advantage of precession and higher harmonics.

Table~\ref{tab:tabRuns} summarizes all the runs performed 
with parallel Bilby and distance marginalisation switched on, with different models of the \phXF family as well as \phTHM.
Additional parallel Bilby runs without distance marginalisation are listed in Table\ \ref{tab:tab_oldRuns}.
Most of our runs were performed on IntelXeon Platinum CPUs with 2.1GHz clock-rate (BSC MareNostrum), except for some runs in Table\ \ref{tab:tab_oldRuns}, which 
have been performed on Intel Xeon E5-2670 CPUs with 2.6GHz clock-rate (Picasso machine). All the runs of Table \ref{tab:tabRuns} used the master branch of LALsuite, with git hash f253e1307b9c19b0fa974fe627651db483f38170, compiled with gcc version 5.4.0 (BSC MareNostrum) and 4.9.4  (Picasso). 
The cost of a full higher-mode precessing run with $n_{\mathrm{live}}=2048, n_\mathrm{act}=10$ can be as low as $\approx$ 1300 cpu hrs per seed, using aggressive multibanding thresholds and distance marginalization, which translates into a total sampling time of roughly 13 hrs when using 96 CPU-cores. The cost can be further decreased by lowering the number of live points: the fastest run in our analysis, using $n_\mathrm{live}=512, n_\mathrm{act}=10$ had a computational cost of less than 700 CPU h per seed. We should stress the fact that the computational costs reported here correspond to the analysis of 8 s of data and will be therefore even lower for BBH events with high total masses, for which segment lengths of 4 s are appropriate.

\begin{figure*}[htpb]
\begin{center}
\includegraphics[width=0.8\columnwidth]{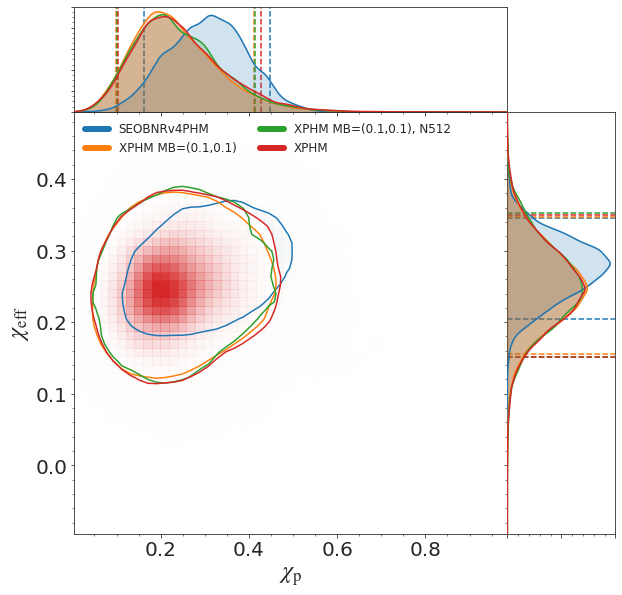}
\includegraphics[width=0.825\columnwidth]{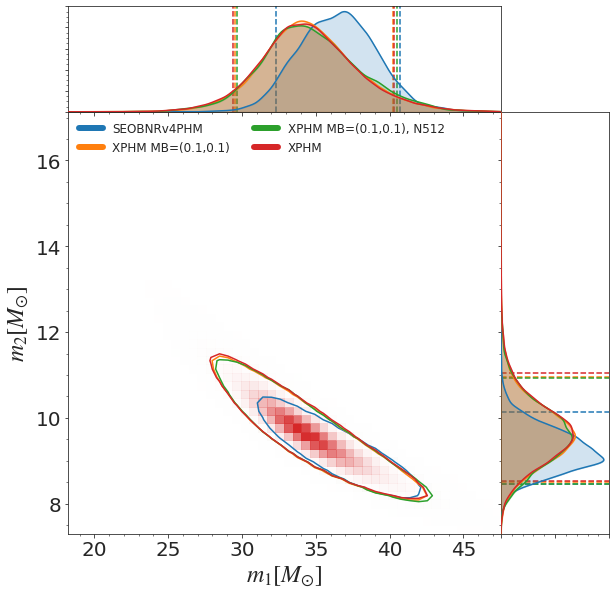}\\
\includegraphics[width=0.8\columnwidth]{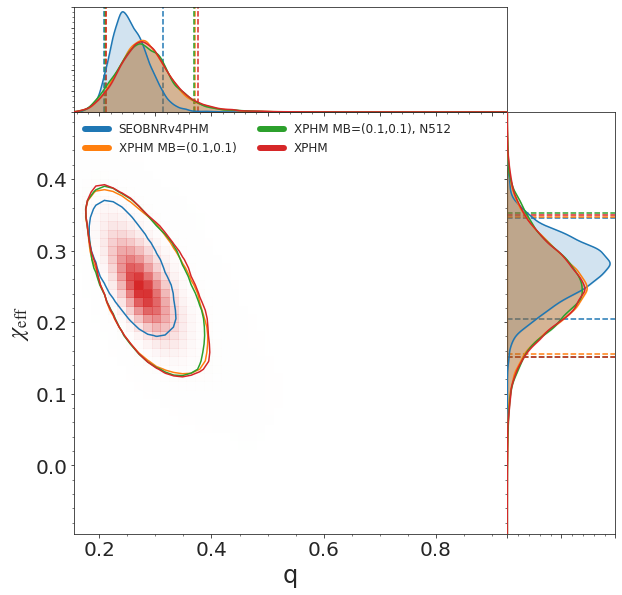}
\includegraphics[width=0.825\columnwidth]{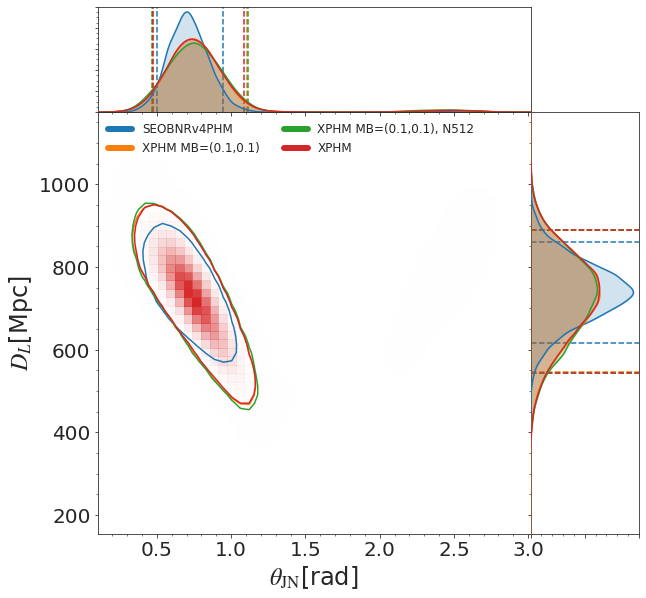}
\caption{Comparison of some estimated source parameters obtained with \seobnrvforphm (blue) and \phXPHM,
with different combinations of multibanding thresholds and sampler settings
(red for the default version, orange with aggressive multibanding thresholds for both the aligned-spin waveforms and precession angles, green with aggressive multibanding and $n_{\mathrm{live}}=512$).
\label{fig:multibanding_posteriors}
}
\end{center}
\end{figure*}

Fig.~\ref{fig:multibanding_posteriors} compares some estimated source parameters obtained with \phXPHM, for different combinations of multibanding thresholds and sampler settings. Even for the most aggressive settings, with moderately high multibanding thresholds and $n_{\mathrm{live}}=512$, one cannot appreciate large differences with respect to our default run.

To quantify the robustness of results under changes of sampler settings, we have computed the JS divergences (see definition in Sec.~\ref{sec:PEsetup}) between \phXPHM runs employing different numbers of $n_\mathrm{live}$ and/or $n_\mathrm{act}$.
Our results are summarised in Table\,\ref{tab:tabJS_settings}.
For comparison Table~\ref{tab:tabJS_models} lists JS divergences between runs with different models.

Among the divergences of runs with different sampler settings but the same \phXPHM model, statistically significant discrepancies (defined as $JS\geq0.002$, according to the same criterion adopted in \cite{Romero-Shaw:2020owr}), occur more frequently for the cheapest configurations ($n_{\mathrm{live}}=512$), as expected.
We also observe that $n_{\mathrm{live}}=2048, n_\mathrm{act}=10$ gives already entirely satisfactory results when compared with more expensive settings and could be therefore considered a safe configuration for productions runs on this event.

\begin{table}[hptb]
\begin{center}
\begin{tabular}{|*{5}{c|}}

\cline{1-2}
 XAS & -  \\ \cline{1-3}
 XHM & $JS_{\phi_c}=0.2929$ & - \\ \cline{1-4}
 XP & $JS_{t_1}=0.6689$ & $JS_{t_1}=0.6863$ & -  \\ \cline{1-5}
 XPHM &  $JS_{t_1}=0.6691$  &   $JS_{t_1}=0.68791$& $JS_{\psi}=0.2659$  & - \\ \cline{1-5}
 & XAS & XHM  & XP &XPHM  \\
\hline
\end{tabular}
\end{center}
\caption{Maximum Jensen-Shannon (JS) divergence values between the posterior distributions of GW190412 estimated with \phX, \phXHM, \phXP and \phXPHM.
\label{tab:tabJS_models}
}
\end{table}

\begin{table*}[hptb]
\begin{center}
\begin{tabular}{|*{8}{c|}}

\cline{1-2}
  N512 NA10 & -  \\ \cline{1-3}
  N512 NA50 &$JS_{t_1}=0.0010$ & -  \\\cline{1-4}
  N1024 NA10 & $JS_{t_2}=0.0032$ & $JS_{t_2}=0.0023$ & - \\ \cline{1-5}
  N1024 NA50 & $JS_{t_1}=0.0008$ & $JS_{ra}=0.0007$ & $JS_{ra}=0.0019$ & -\\ \cline{1-6}
  N2048 NA10 & $JS_{ra}=0.0020$ & $JS_{ra}=0.0020$ &$JS_{t_2}=0.0010$ & $JS_{ra}=0.0017$ & -  \\ \cline{1-7}
   N2048 NA50 & $JS_{\phi_{JL}}=0.0015$ & $JS_{\phi_{JL}}=0.0016$ &$JS_{t_2}=0.0014$ & $JS_{a_2}=0.0017$ & $JS_{\phi_{JL}}=0.0013$ & - \\ \cline{1-8}
  N4096 NA10 & $JS_{ra}=0.0023$ & $JS_{ra}=0.0027$ & $JS_{\theta_{JN}}=0.0030$ & $JS_{\theta_{JN}}=0.0018$ & $JS_{\theta_{JN}}=0.0033$ &  $JS_{\theta_{JN}}=0.0020$ & - \\ \cline{1-8}
 
  & N512 NA10  & N512 NA50 & N1024 NA10 & N1024 NA50 & N2048 NA10 & N2048 NA50 & N4096 NA10\\ 
\hline
\end{tabular}
\end{center}
\caption{Maximum Jensen-Shannon (JS) divergence values between the posterior distributions of GW190412 estimated with \phXPHM using different sampler settings.
In this table, N is short for $n_\mathrm{live}$ and NA is short for $n_\mathrm{act}$.
We report here only values larger than 0.001.
\label{tab:tabJS_settings}
}
\end{table*}

\section{Further waveform systematics studies}\label{sec:extra_checks}

\subsection{Matches for low total-mass binaries}\label{subsec:matches}

Waveform systematics can be further investigated by computing noise weighted frequency-domain overlaps between model and signal waveforms:
\begin{equation}
\Braket{h_\mathrm{M}, h_\mathrm{S}} = 4 \Re \int_{f_{\min}}^{f_{\max}} \frac{\tilde{h}_\mathrm{M}(f) \:\tilde{h}^*_\mathrm{S}(f)}{S_\mathrm{n}(f)},
\end{equation}
where $S_\mathrm{n}(f)$ is the one-sided PSD of the detector noise. In what follows, we will take \phXPHM as the model and assume that the signal is given either by a NR or \seobnrvforphm waveform. 
We will quantify the agreement between model and signal by means of the match function:
\begin{equation}
\mathcal{M}(h_\mathrm{M},h_\mathrm{S}) = \max_{t_c, \phi_0, \psi_0} \frac{\Braket{h_\mathrm{M}, h_\mathrm{S}}}{\sqrt{\Braket{h_\mathrm{M}, h_\mathrm{M}}}\sqrt{\Braket{h_\mathrm{S}, h_\mathrm{S}}}},
\end{equation}
where we optimize the overlap between normalized waveforms over polarization angle $\psi_0$, coalescence time $t_c$ and reference phase $\phi_0$. In general, the definition of the spin configuration at the reference frequency will not be the same for the model and the signal. To account for this, we also optimize the match over rotations of the initial in-plane spins. We use the Advanced-LIGO~\cite{TheLIGOScientific:2014jea} design sensitivity Zero-Detuned-High-Power PSD \cite{adligopsd} with $f_{\mathrm{min}}=$20 Hz and $f_{\mathrm{max}}=$2048 Hz. 

Match calculations are far cheaper than a fully fledged Bayesian analysis and can be leveraged for extensive explorations in parameter space. We extended our previous results \cite{Pratten:2020ceb} and specifically targeted low total-mass binaries similar to GW190412. 

\begin{figure}[htpb]
\begin{center}
\includegraphics[width=0.95\columnwidth]{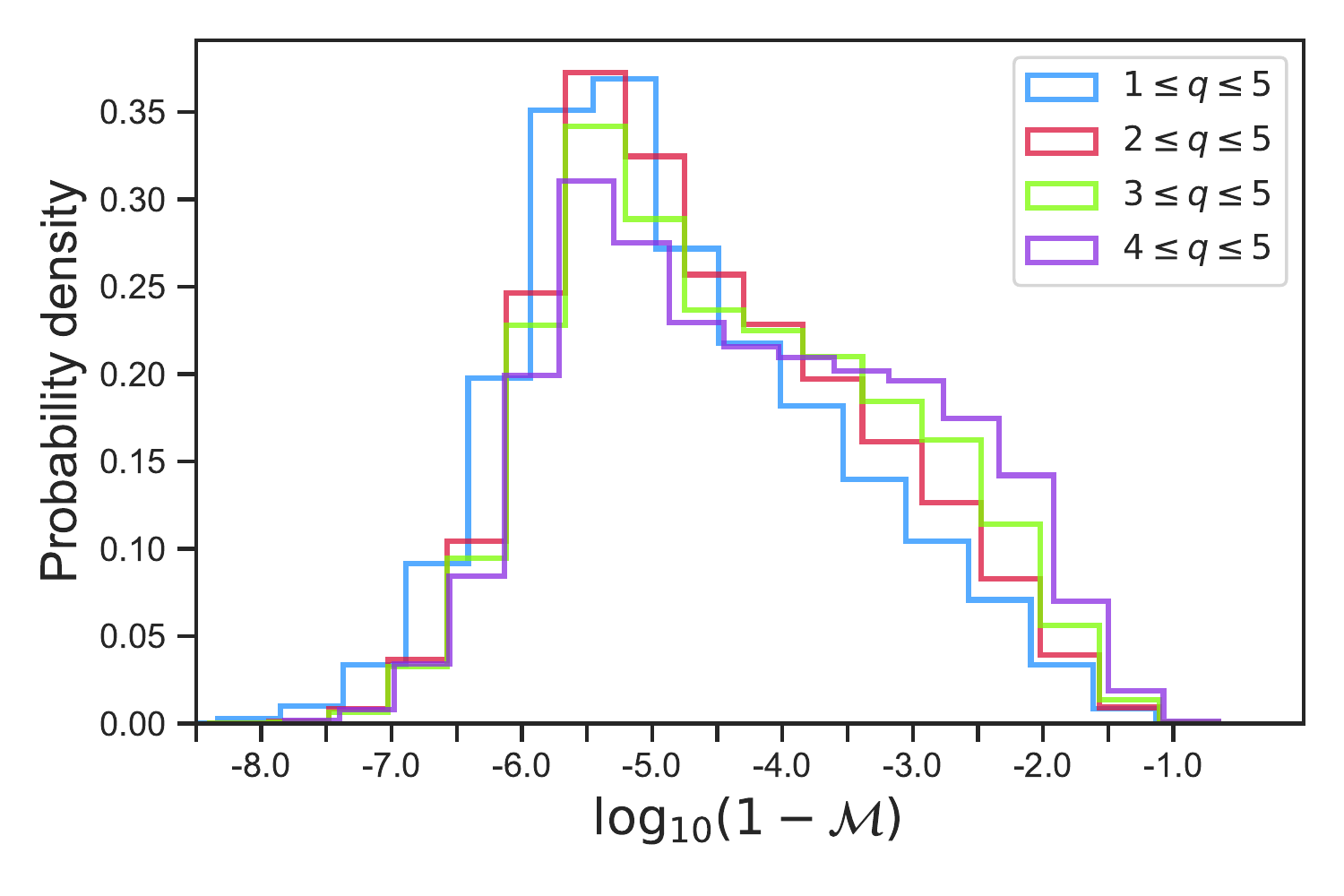}\\\includegraphics[width=0.95\columnwidth]{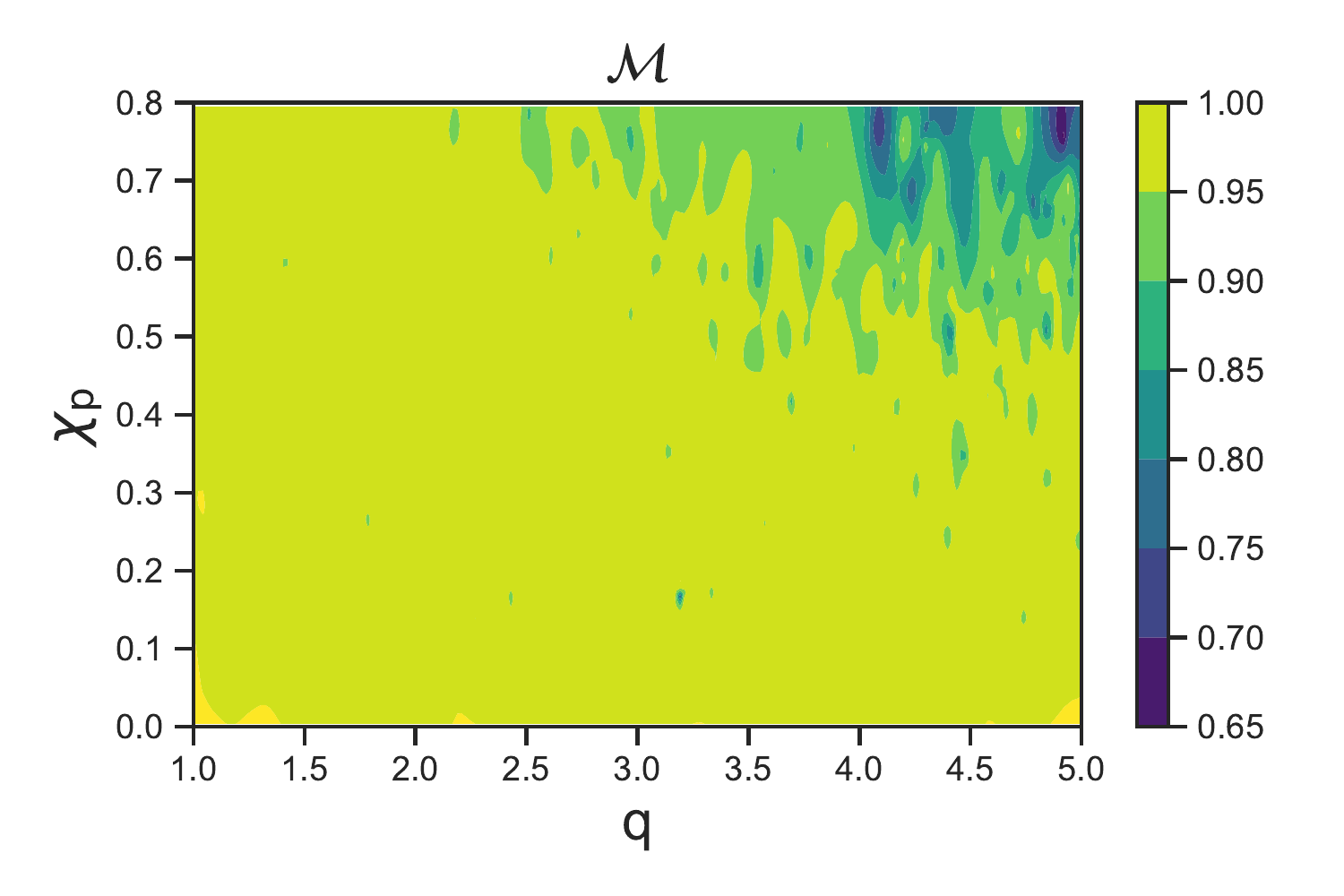}
\caption{Matches between \phXPHM and \seobnrvforphm waveforms.
The top panel shows the distribution of mismatches for different ranges in mass-ratio $q$.
The lower panel shows the match as a function of the mass ratio $q$ and effective precession spin $\chi_\mathrm{p}$.
Light green (dark blue) areas indicate good (poor) agreement between the two models.
The worst matches correspond to strongly precessing, high mass-ratio binaries.
\label{fig:matches_seob}
}
\end{center}
\end{figure}

We computed the match between \seobnrvforphm and \phXPHM waveforms for 20000 random configurations with mass ratios uniformly distributed in the interval $q\in\left[1,5\right]$, spins isotropically distributed on the unit sphere and total mass varying between 30$\Msun$ and 50$\Msun$ in bins of 5$\Msun$ width. Our results are illustrated in Fig.\,\ref{fig:matches_seob}. The top panel shows the normalized probability distribution of $\log(1-\mathcal{M})$. Overall, the two models agree very well, with $\approx$ 90\% of the configurations achieving $\mathcal{M}\geq 95\%$; however, it can be seen that matches tend to degrade for higher mass ratios. 
The lower panel shows instead the match $\mathcal{M}$ as a function of the mass ratio $q$ and the effective precession spin $\chi_\mathrm p$. We notice that the two models can show fairly large disagreement when it comes to strongly precessing, high mass ratio systems.  

\begin{figure}[htbp!]
\begin{center}
\includegraphics[width=0.95\columnwidth]{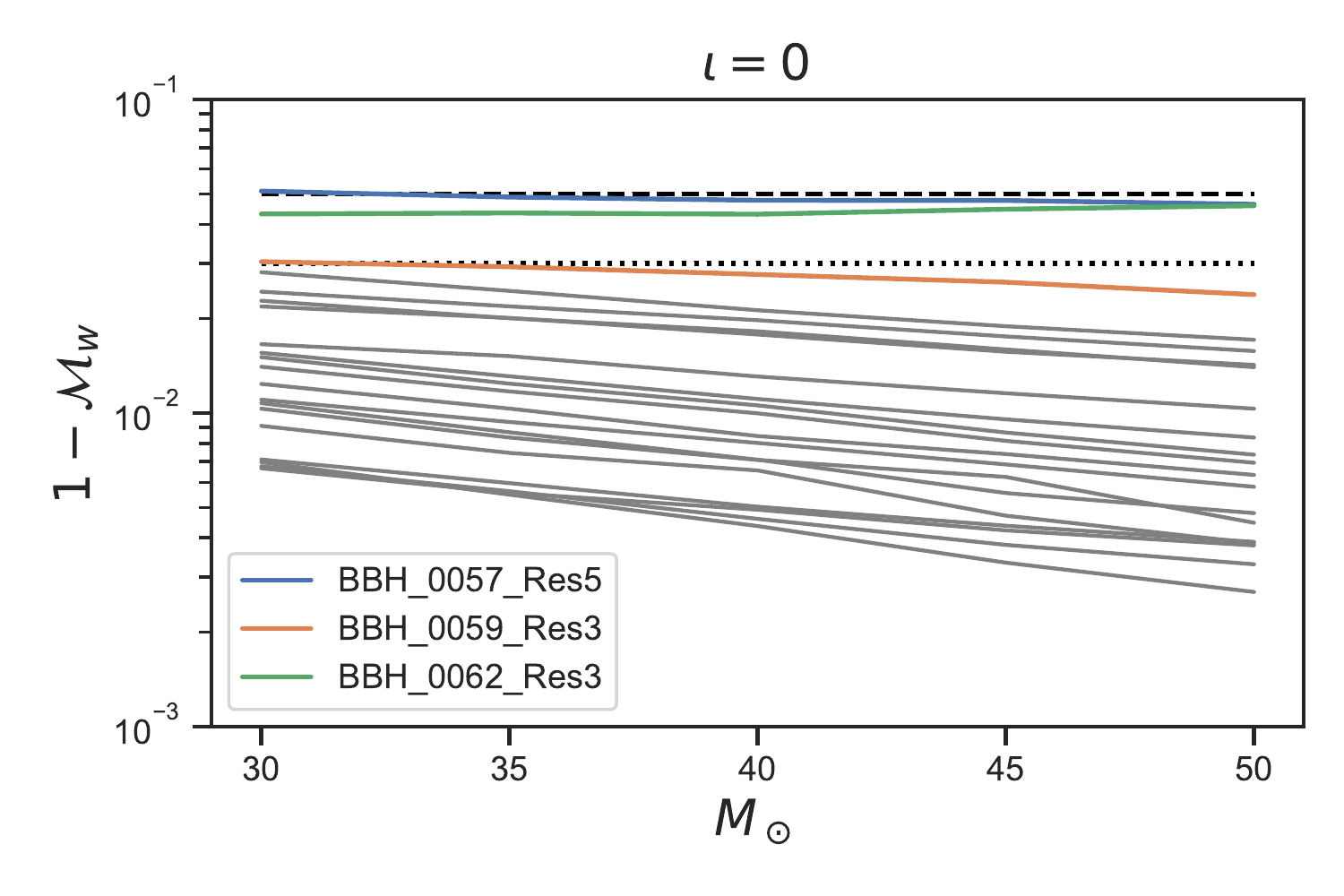}\\\includegraphics[width=0.95\columnwidth]{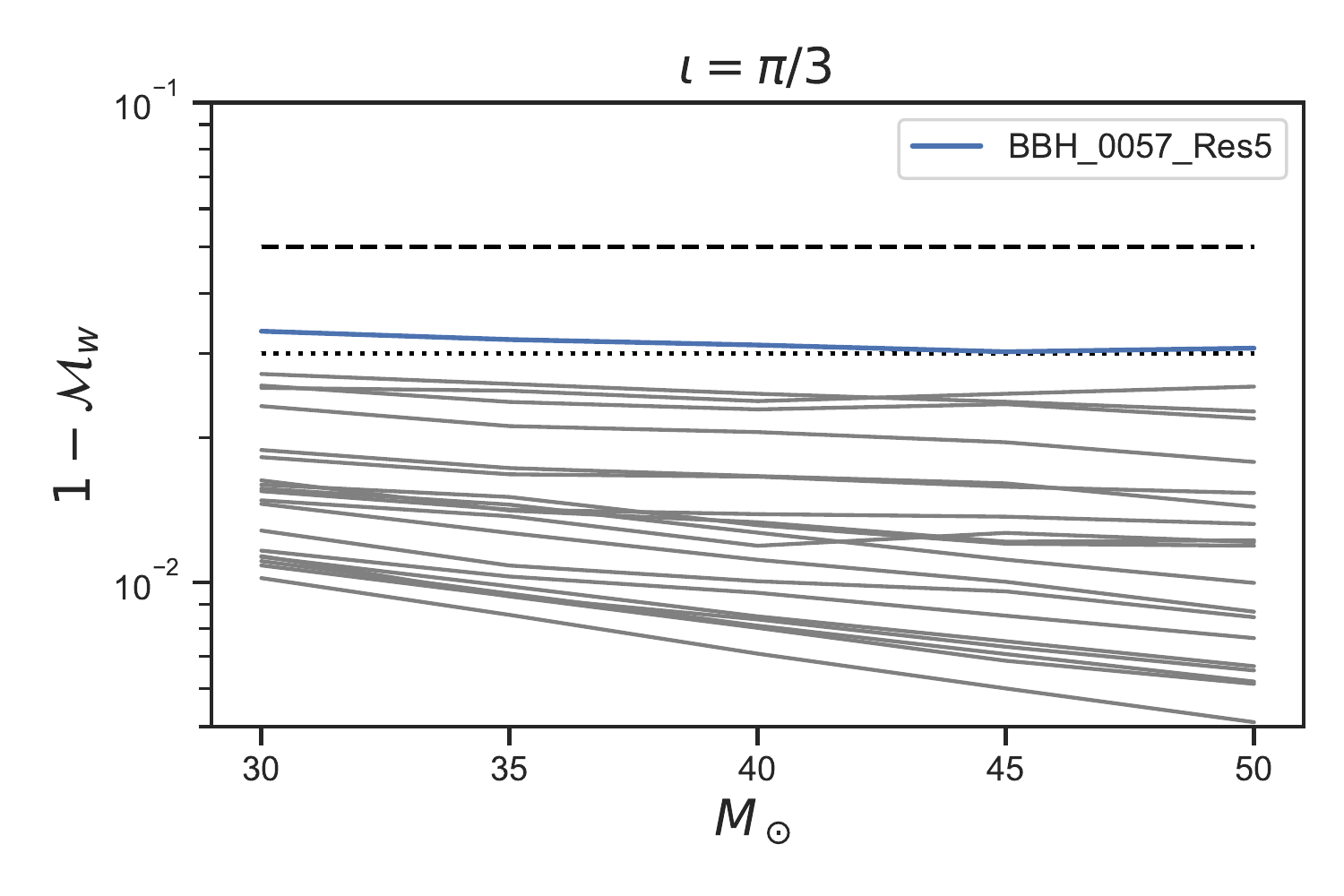}\\\includegraphics[width=0.95\columnwidth]{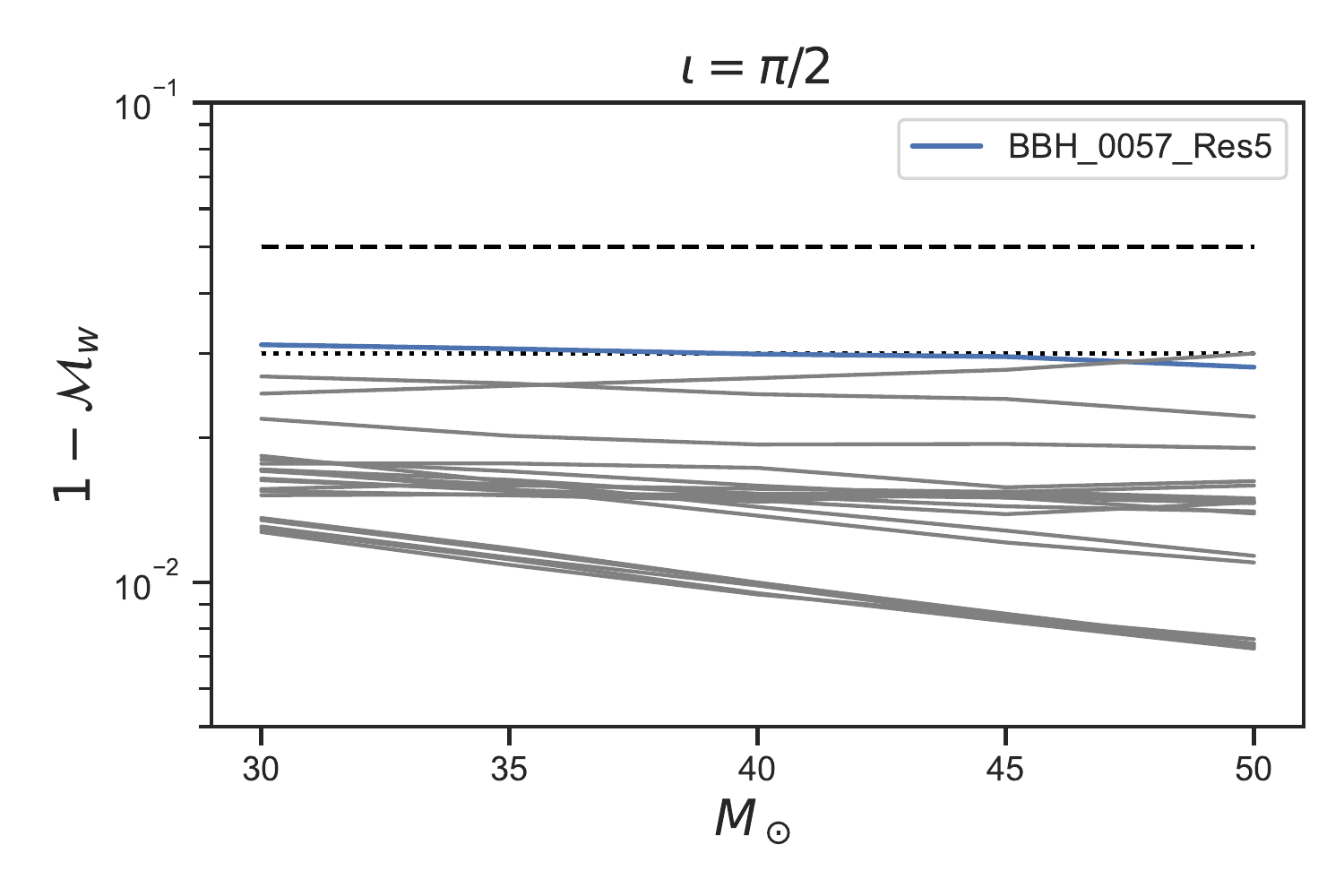}
\caption{SNR-weighted mismatches between \phXPHM and precessing SXS waveforms with $q\in[2,5]$.
The dotted and dashed lines in each panel denote thresholds of 3\% and 5\%,  respectively.
Cases for which the maximum mismatch exceeds the 3\% threshold are shown with coloured lines and listed in the legends.
\label{fig:matches_NR}
}
\end{center}
\end{figure}

We have also checked the agreement between the model and NR for low total mass binaries. The NR simulations considered here are all public SXS precessing waveforms available in the \texttt{lvcnr} catalog \cite{Schmidt:2017btt}. For simplicity, we focused on systems with $q\in[2,5]$, as all the precessing higher mode models considered here tend to exclude mass ratios outside of this range for GW190412. In this case we quantify the agreement between model and signal (NR) through the SNR-weighted match $\mathcal{M}_{\mathrm{w}}$ ~\cite{Harry:2016aa}
\begin{equation}
\mathcal{M}_{\mathrm{w}}=\left(\frac{\sum_i\mathcal{M}_i^3 \Braket{h_{i,\mathrm{NR}}, h_{i,\mathrm{NR}}}^{3/2}}{\sum_i{\Braket{h_{i,\mathrm{NR}},h_{i,\mathrm{NR}}}^{3/2}}}\right)^{1/3},
\end{equation}
where the subscript $i$ goes over different polarization and reference phases of the signal. Fig.\,\ref{fig:matches_NR} shows our results: coloured lines correspond to simulations for which the maximum mismatch $1-\mathcal{M_\mathrm{w}}$ is above 3\% in the mass range considered. The three outliers with mismatches above this threshold are \textsc{SXS:0057}, \textsc{SXS:0059} and \textsc{SXS:0062}, which all correspond to $q=5$ binaries. These result confirm the conclusions drawn in \cite{Pratten:2020ceb}. Overall, \seobnrvforphm and \phXPHM show good agreement over most of the parameter space and are expected to deliver comparable estimates for typical BBH events. However, exceptional events with very asymmetric masses and/or strong precession might require a more in-depth analysis, which we leave for future work. Further insight on waveform systematics can be found in previous works on the \phenX family models \cite{Pratten:2020fqn, Garcia-Quiros:2020qpx, Garcia-Quiros:2020qlt, Pratten:2020ceb} where extensive comparisons through matches with previous waveform models and NR simulations are carried out.

\subsection{Injection study}\label{susec:injections}
In order to assess the accuracy of our measurements, in particular of the precessing spin, we injected into a Hanford-Livingston-Virgo detector network three synthetic signals and recovered their parameters with \phXPHM. Two of these signals are based on NR simulations with moderately high $\chi_p$, while the third one makes use of a \seobnrvforphm waveform, corresponding to the maximum-likelihood sample of the LVC release. We injected the signals in zero noise, but we fed into the likelihood estimation the PSDs used to analyse GW190412. For the NR signals, the total mass and luminosity distance were chosen so as to achieve a network SNR comparable to that of GW190412. In all cases, we set the reference frequency to be the same as the minimum frequency, i.e. $f_{\mathrm{ref}}=f_{\mathrm{low}}$ in our setup. 

\begin{figure*}[htpb]
\includegraphics[width=0.8\columnwidth]{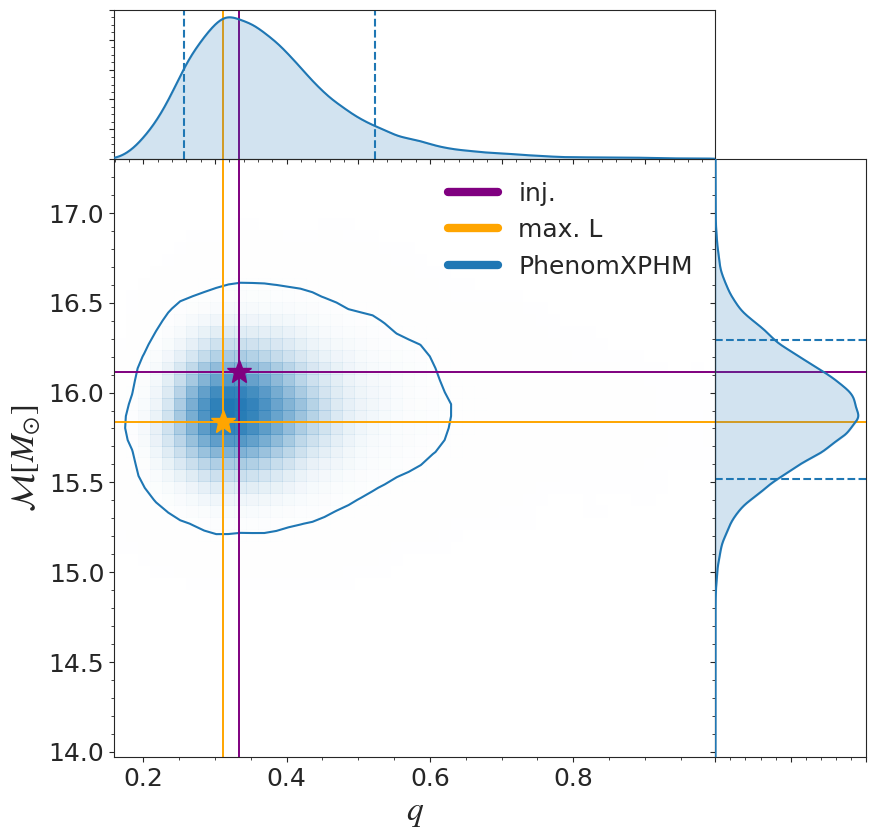}
\includegraphics[width=0.8\columnwidth]{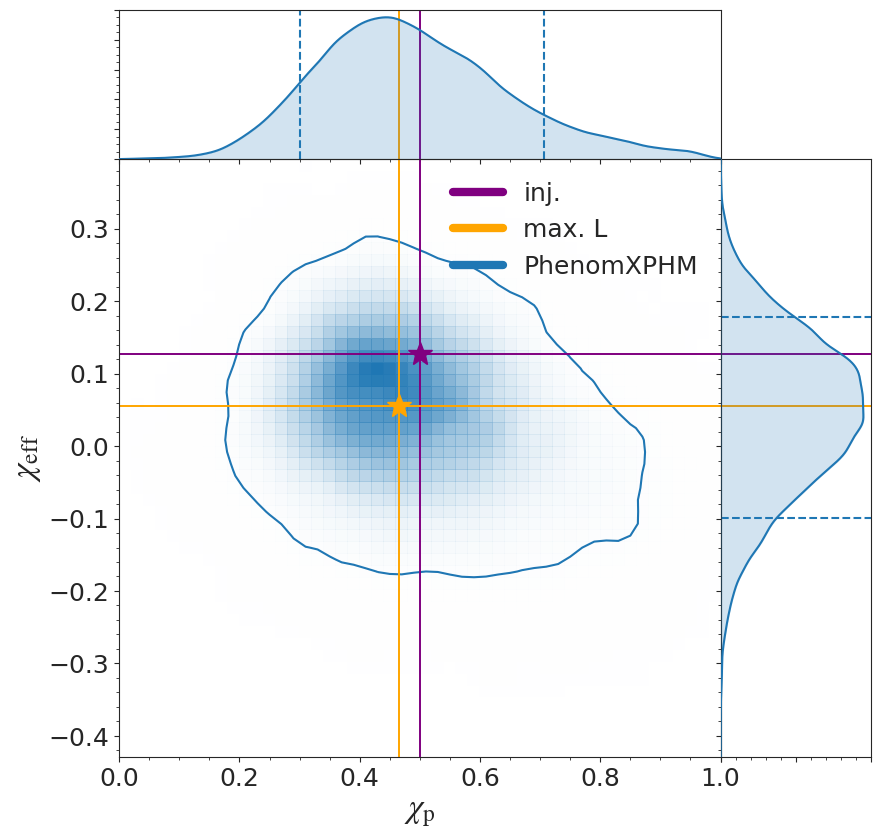}\\
\includegraphics[width=0.8\columnwidth]{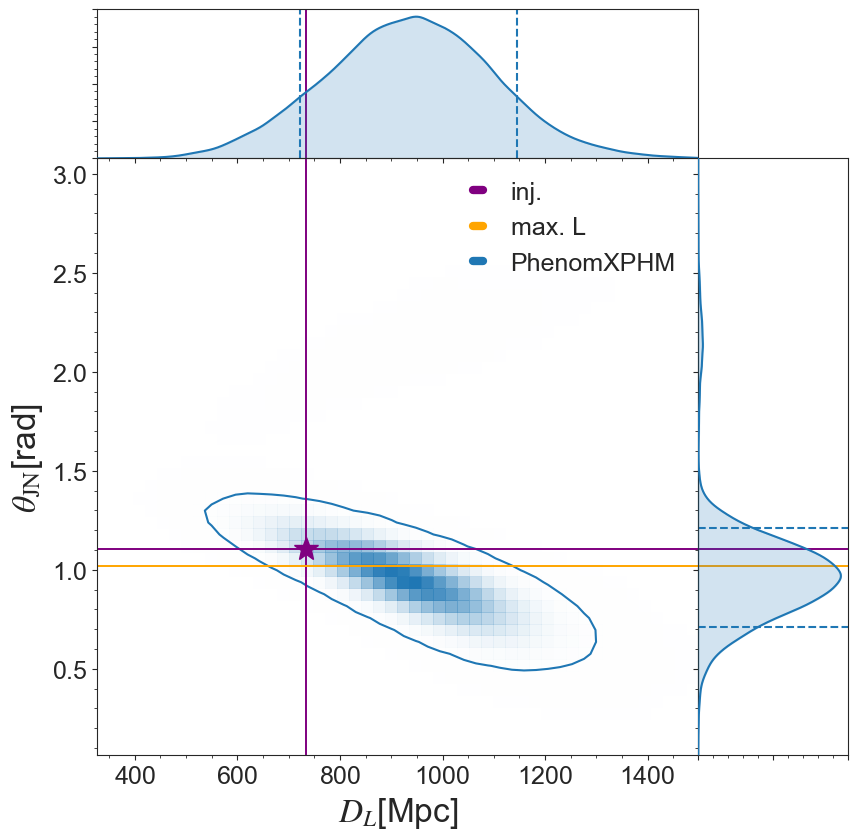}
\includegraphics[width=0.8\columnwidth]{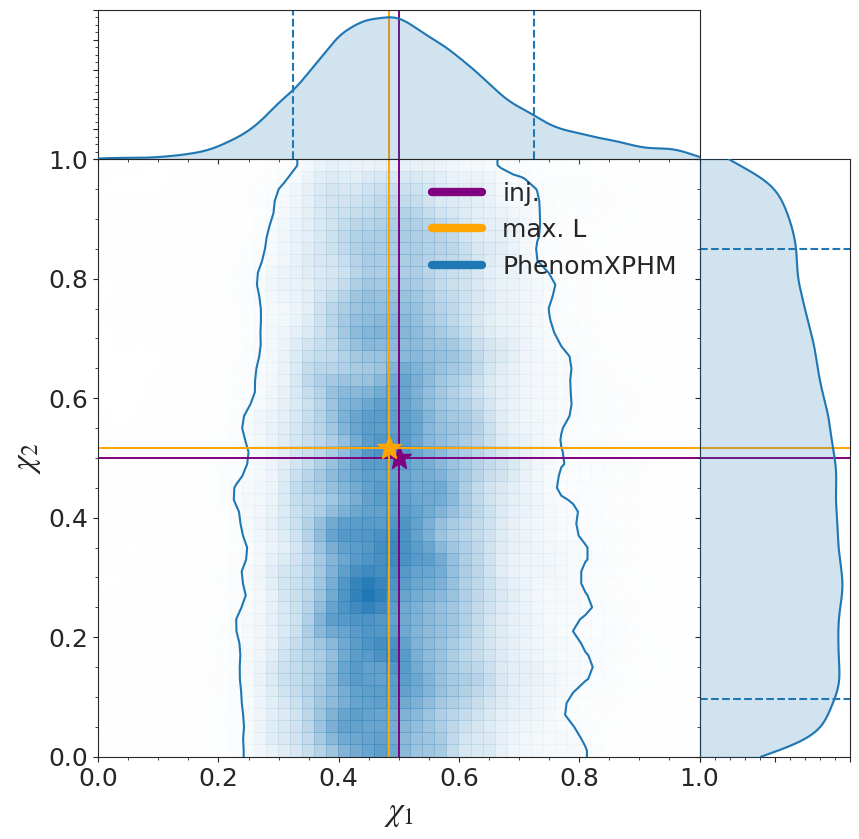}
\caption{Posterior distributions for some of the intrinsic and extrinsic parameters of a synthetic signal obtained starting from the public SXS waveform \textsc{SXS:BBH0049}.
Purple and orange stars mark the values of the injected and maximum-likelihood parameters, respectively.
Dashed lines indicate the $90\%$ credible intervals of the individual posterior distributions.
\label{fig:SXS0049_inj}
}
\end{figure*}

The first injection made use of the public NR waveform \textsc{SXS:BBH:0049}, corresponding to a $q=3$ binary with $\chi_{\mathrm{p}}\approx0.5$.
We simulated a binary with right ascension ra$=3.81$ rad, declination dec$=0.63$ rad, and geocentric time equal to the trigger time reported in GraceDB~\cite{GraceDB}.
We used a minimum frequency of 25.7 Hz, due to the limited length of the NR waveform. 
In Fig.\,\ref{fig:SXS0049_inj} we present posteriors for detector-frame masses, $\chi_{\mathrm{eff}}$, $\chi_{\mathrm{p}}$, luminosity distance, $\theta_{\mathrm{JN}}$, as well as the individual adimensional spin magnitudes $a_1$ and $a_2$. Overall, we observe a very good agreement between the recovered parameters and the injected values, which always lie within the $90\%$ credible intervals of the posterior distributions. In particular, the spin magnitude of the primary is very well constrained, similarly to what happens for GW190412.

\begin{figure*}[htb]
\includegraphics[width=0.8\columnwidth]{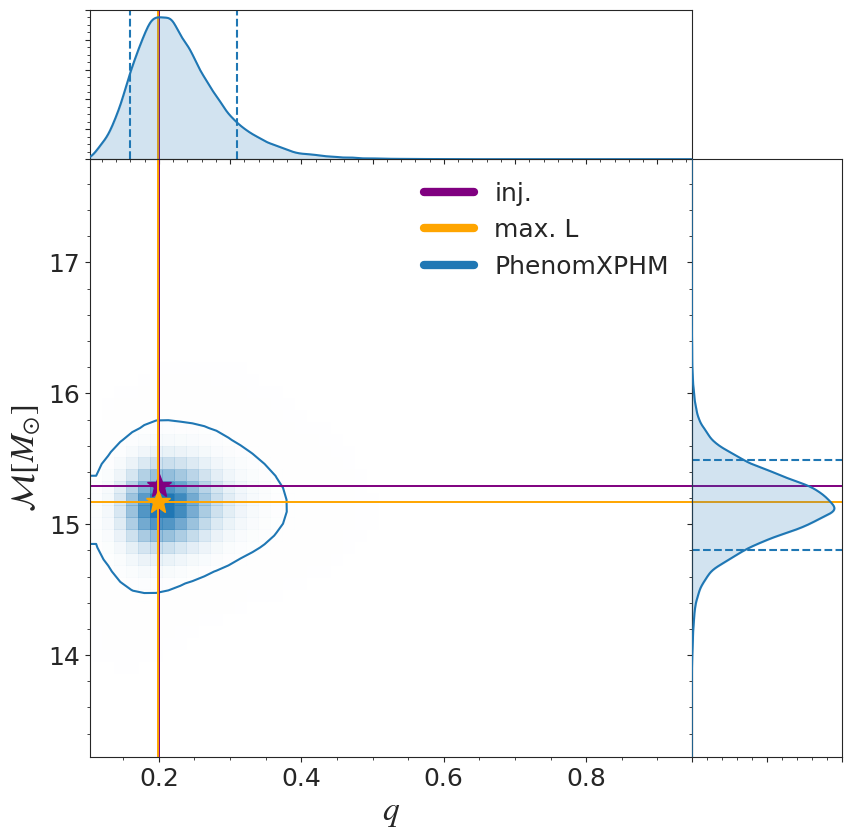}
\includegraphics[width=0.8\columnwidth]{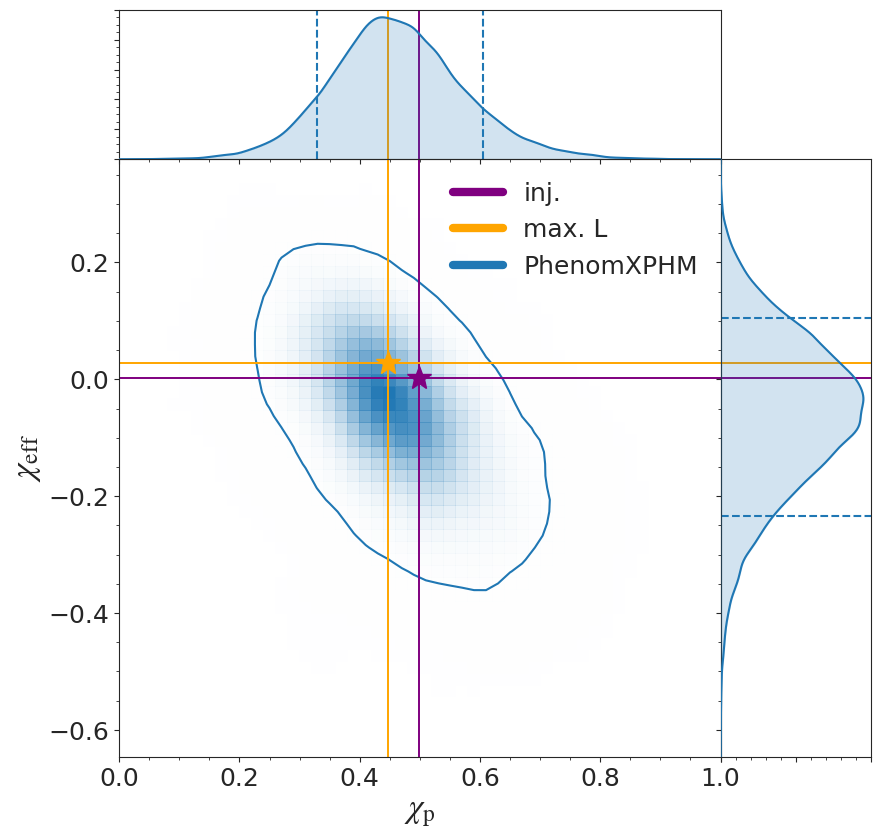}\\
\includegraphics[width=0.8\columnwidth]{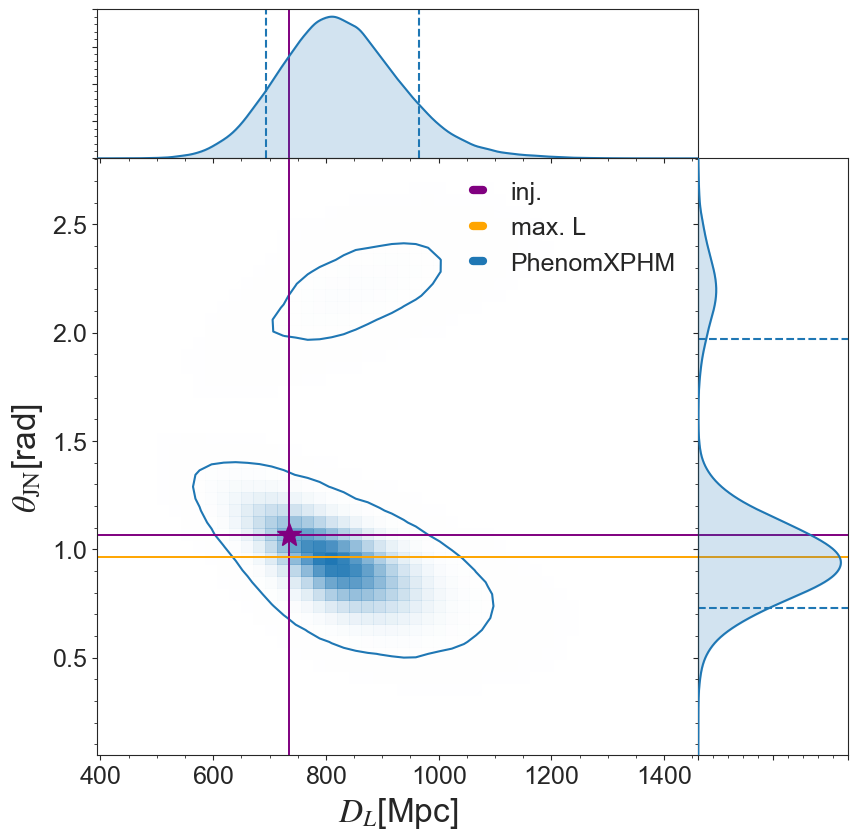}
\includegraphics[width=0.8\columnwidth]{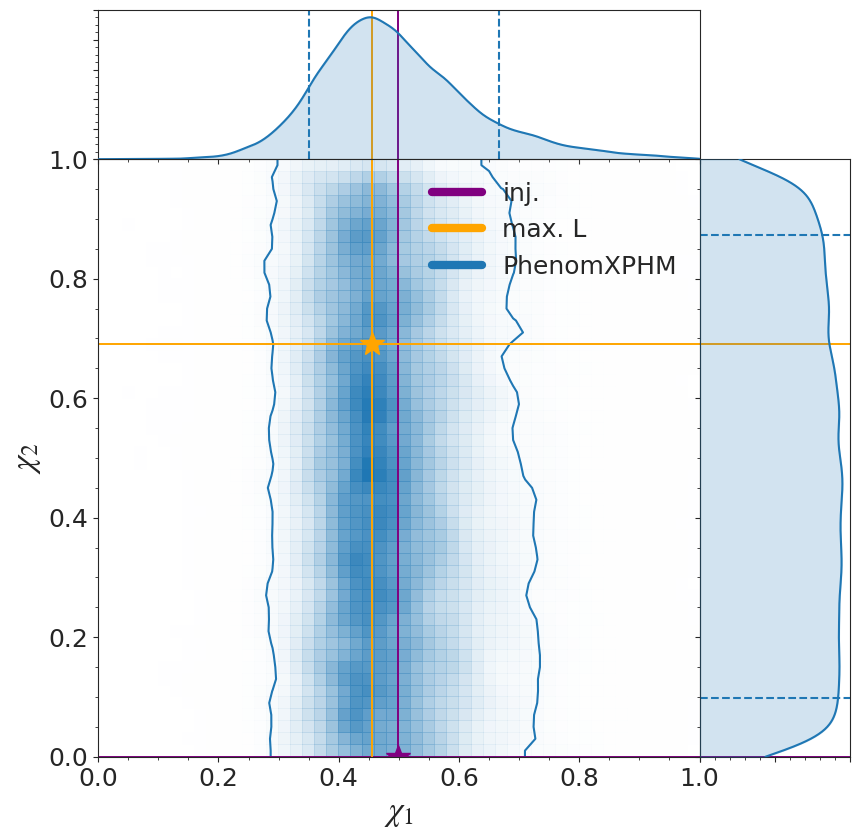}
\caption{Posterior distributions for some of the intrinsic and extrinsic parameters of a synthetic signal obtained starting from the public SXS waveform \textsc{SXS:BBH0058}.
Purple and orange stars mark the values of the injected and maximum-likelihood parameters, respectively.
Dashed lines indicate the $90\%$ credible intervals of the individual posterior distributions.
\label{fig:SXS0058_inj}
}
\end{figure*}

In Fig.\,\ref{fig:SXS0058_inj} we present the results of our second NR injection, that took as input the public SXS waveform \textsc{SXS:BBH0058}, corresponding to a $q=5$ binary with $\chi_{\mathrm{p}}\approx0.5$. The simulated event was assumed to have the same celestial coordinates and geocentric time as the previous mock signal. The total mass of the system was taken to be $50.5\Msun$ in the detector frame, and we used a minimum frequency of 20.5 Hz. We can see that, even in this case, all the recovered parameters are consistent with the ones of the injected signal.

\begin{figure*}[htpb]
\includegraphics[width=0.8\columnwidth]{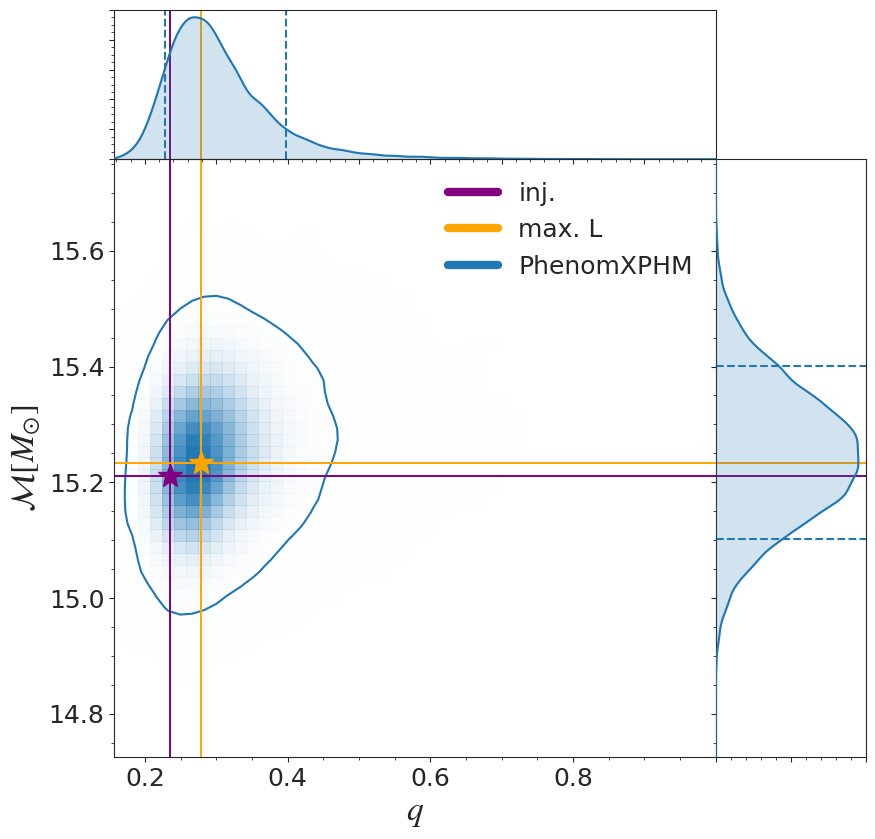}
\includegraphics[width=0.8\columnwidth]{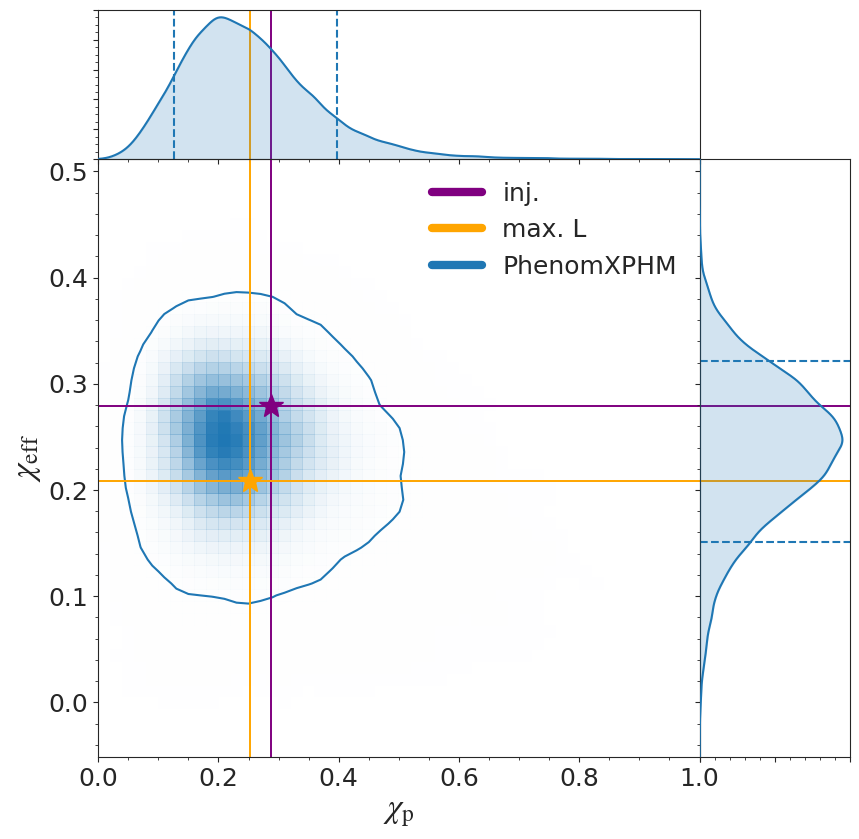}\\
\includegraphics[width=0.8\columnwidth]{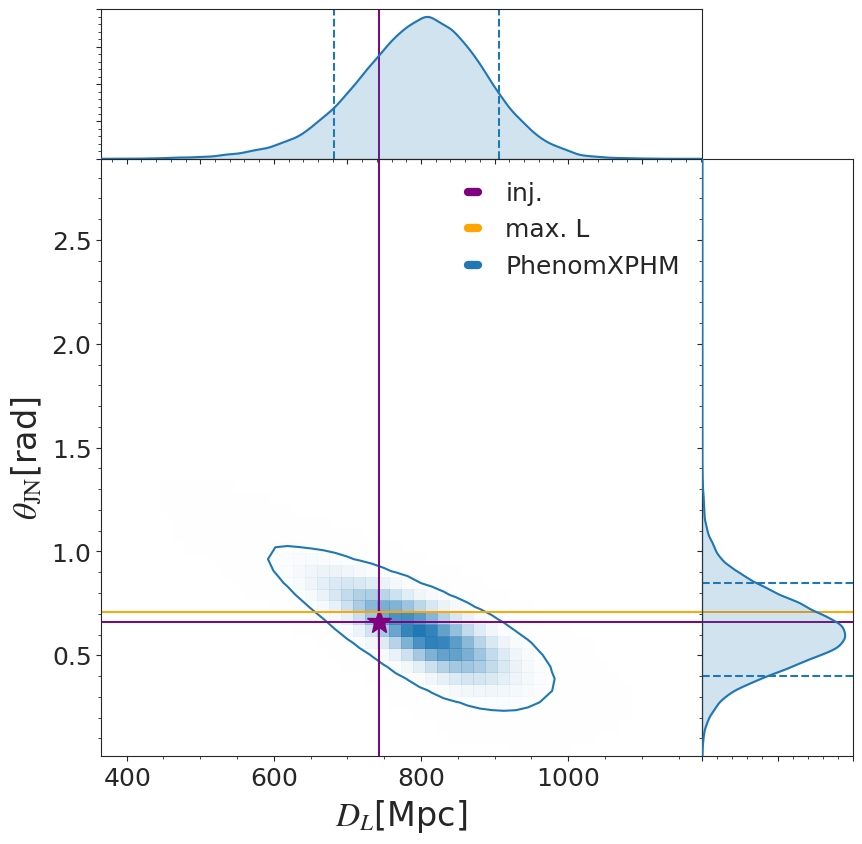}
\includegraphics[width=0.8\columnwidth]{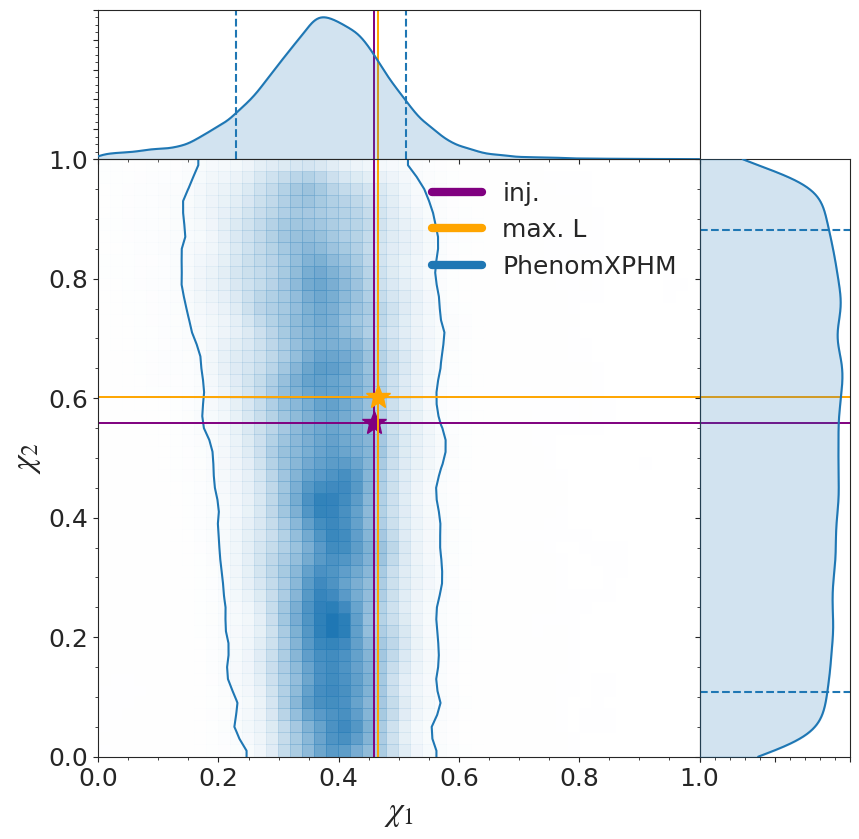}
\caption{Posterior distributions for some of the intrinsic and extrinsic parameters of a synthetic signal obtained from the \seobnrvforphm maximum-likelihood parameters for GW190412.
Purple and orange stars mark the values of the injected and maximum-likelihood parameters, respectively.
Dashed lines indicate the $90\%$ credible intervals of the individual posterior distributions.
\label{fig:maxL_inj}
}
\end{figure*}

Finally, we injected in zero-noise the maximum-likelihood waveform for the \seobnrvforphm sample.
The injected signal had a total mass of $46.66\Msun$ in detector frame, celestial coordinates $\mbox{ra}\approx3.81$ rad and $\mbox{dec}\approx0.64$ rad, and geocentric time of approximately $1239082262.18$ s.
Fig.\,\ref{fig:maxL_inj} summarizes our main results.
\phXPHM recovers well all the parameters of the injected signal.
We therefore conclude that \phXPHM appears to deliver robust estimates of the source properties for events similar to GW190412.   

\section{Conclusions}
\label{sec:conclusions}

We have re-analysed the event GW190412 with the newest generation of phenomenological waveform models:
the frequency-domain model \phXPHM, which includes precession and higher modes,
and the time-domain model \phTHM, which includes higher modes, but not yet precession.
Both models have been constructed with similar techniques and accuracy goals, and we refer to them (and several versions with reduced physics content) jointly as ``generation X''.

The principal code we have used for performing Bayesian inference is parallel Bilby \cite{Smith:2019ucc}, which allows us to work in traditional high performance computing environments and to obtain results on a time scale of hours.
In Sec.~\ref{subsec:cost} we have described some tests varying sampling parameters to convince us that our parallel Bilby runs have converged to reliable results.
In appendix~\ref{subsec:dist_marg} we further test that our main runs, which use distance marginalisation, agree with runs that do not use it, and we study the difference in the number of likelihood evaluations and computational cost. 
Since parallel Bilby is a relatively new code, we also report our cross-checks with the more established LALInference code \cite{Veitch:2014wba} in appendix~\ref{subsec:LI_vs_pbilby}, and find excellent agreement.

For the non-precessing sector, we find excellent agreement between all waveform models where sub-dominant harmonics have been calibrated to numerical waveforms, i.e. \seobnrvforhm, \NRHybSur, \phXHM and \phTHM (see e.g. Fig.~\ref{fig:posteriors_nonprec}).
However we find significant differences for \phHM, where higher modes have been constructed by an approximate map in terms of the dominant quadrupole mode, which was calibrated to a much smaller set of NR waveforms than \phX.
It is to be expected that in the future, EOB and phenomenological models will also be calibrated to numerical relativity waveforms for the precessing sector, and will thus reach similar levels of agreement. 

When adding precession we find that the agreement between phenomenological waveforms and the EOB model improves for some quantities, notably the masses and source location. Not surprisingly, for the effective precession spin this is not the case, although the posterior width for \phXPHM is closer to \seobnrvforphm than \phPvthreehm.
Both \seobnrvforphm  and the frequency-domain phenomenological waveforms use a version of the ``twisting-up'' approximate map between non-precessing and precessing waveforms \cite{Hannam:2013oca}, however the current frequency domain phenomenological waveforms employ the stationary phase approximation (SPA)
in addition to the approximations inherent in the twisting-up (for a recent discussion of these approximations see \cite{Ramos-Buades:2020noq}).
An improved accuracy for precession is thus expected from the extension of the LALSuite implementation of \phT  to include precession following \cite{phenomtp}. Strategies for extending the twisting-up procedure for frequency-domain waveforms beyond the SPA approximation have been discussed in \cite{Marsat:2018oam}.
In appendix~\ref{sec:prec_versions} we compare the two different descriptions of the Euler angles used in the twisting-up procedure implemented in \phXPHM (single-spin and double-spin), and only find small differences, mainly in the effective precession spin parameter $\chi_p$.
Not surprisingly we find that the double-spin description is closer to the results that have been reported for \seobnrvforphm in \cite{Abbott:2020uma}.
In the same section we also compare different prescriptions for the spin of the merger remnant in precessing mergers, and we find no significant differences for GW190412.
In appendix~\ref{sec:phenompv3_comparison} we compare directly against an alternative ``twisting-up'' of the older \phHM aligned-spin model implemented within the \phXPHM framework, finding consistent results with the original \phPvthreehm (which is based on \phHM), thus cleanly demonstrating that our improved results mostly derive from the updated underlying aligned-spin model \phXHM with its calibration of subdominant modes to NR.
We have also shown that the use of the multibanding algorithm leads to a dramatic reduction of the computational cost of parameter estimation runs, nearly halving the cost of non-multibanded runs. 

Additional studies presented in this paper include investigations into possible remaining systematic differences between waveforms, including
waveform match comparisons and injection studies in Sec.\ \ref{sec:extra_checks}.
The tested injections include SXS waveforms with parameters consistent with GW190412 as well as the \seobnrvforphm maximum-likelihood waveform, finding no significant systematic issues with \phXPHM in this part of the parameter space.
We also test the impact of different spin priors in appendix~\ref{sec:spin_priors},
finding that PE results for GW190412 are generally consistent when changing from the standard LVC prior to a volumetric one,
and confirming the results of~\cite{Zevin:2020gxf} that the standard assumption of allowing for a spinning heavier BH component is preferred over runs with a prior that restricts spin to the less massive BH only.

One of the key properties of the \phXF waveform family is its computational efficiency.
We demonstrate that even the most complete \phXPHM model,
when making use of the pBilby sampler and a strong HPC cluster,
allows for extremely fast-turn-around exploratory runs (few hours) and still very fast high-fidelity runs like the ones we report as the main results (see e.g. Fig.~\ref{fig:precessing_posteriors}, with the plotted \phXPHM run taking 2670 CPU hours total and finishing after 28 hours wall-clock time).
The time-domain \phTHM model is also quite competitive in run time despite being a native time domain model, having only 50 more CPU cost per likelihood evaluation than \phXHM. 
More details on computational cost for different model versions and sampler settings are found in Table~\ref{tab:tabRuns} (for the runs with distance marginalization) and Table~,\ref{tab:tab_oldRuns} (for the runs without distance marginalization). As expected, the use of distance marginalization brings significant computational cost savings: this can be readily appreciated by comparing otherwise equivalent runs from the two tables (e.g. run 23 in Table~\ref{tab:tabRuns} and run 20 in Table~\ref{tab:tab_oldRuns}). 

In summary, the studies presented here demonstrate the robustness and efficiency of the ``generation X'' phenomenological waveform models in a real-world application,
and point out how systematic runs with varied settings can be used both to study the physics of an individual detection in detail, and to achieve accuracy requirements at bounded computational cost.
We hope these results will help to advance the routine use of subdominant harmonics in the parameter estimation for compact binary mergers, leading to both more accurate parameter estimates, and the availability of accurate posterior estimates on the time scale of a few hours.

Posterior samples from our preferred run for each waveform model are released in~\cite{datarelease}.

\section*{Acknowledgements}
We gratefully thank the Bilby and parallel Bilby code developers, especially Greg Ashton, Sylvia Biscoveanu, Rory Smith and Colm Talbot for discussions and software fixes.
This work was supported by European Union FEDER funds, the Ministry of Science, 
Innovation and Universities and the Spanish Agencia Estatal de Investigación grants
PID2019-106416GB-I00/AEI/10.13039/501100011033,
FPA2016-76821-P,        
RED2018-102661-T,    
RED2018-102573-E,    
FPA2017-90687-REDC,  
Vicepresidència i Conselleria d’Innovació, Recerca i Turisme, Conselleria d’Educació, i Universitats del Govern de les Illes Balears i Fons Social Europeu, 
Comunitat Autonoma de les Illes Balears through the Direcció General de Política Universitaria i Recerca with funds from the Tourist Stay Tax Law ITS 2017-006 (PRD2018/24),
Generalitat Valenciana (PROMETEO/2019/071),  
EU COST Actions CA18108, CA17137, CA16214, and CA16104, and
the Spanish Ministry of Education, Culture and Sport grants FPU15/03344 and FPU15/01319.
M.C. acknowledges funding from the European Union's Horizon 2020 research and innovation programme, under the Marie Skłodowska-Curie grant agreement No. 751492.
D.K. is supported by the Spanish Ministerio de Ciencia, Innovaci{\'o}n y
Universidades (ref. BEAGAL 18/00148)
and cofinanced by the Universitat de les Illes Balears.
The authors thankfully acknowledge the computer resources at MareNostrum and the technical support provided by Barcelona Supercomputing Center (BSC) through Grants No. AECT-2019-2-0010, AECT-2019-1-0022,  from the Red Española de Supercomputación (RES).
Authors also acknowledge the computational resources at the cluster CIT provided by LIGO Laboratory and supported by National Science Foundation Grants PHY-0757058 and PHY-0823459.
This research has made use of data obtained from the Gravitational Wave Open Science Center~\cite{GWOSC}, a service of LIGO Laboratory, the LIGO Scientific Collaboration and the Virgo Collaboration. LIGO is funded by the U.S. National Science Foundation. Virgo is funded by the French Centre National de Recherche Scientifique (CNRS), the Italian Istituto Nazionale della Fisica Nucleare (INFN) and the Dutch Nikhef, with contributions by Polish and Hungarian institutes.
This work has been assigned LIGO document number \href{https://dcc.ligo.org/P2000402}{P2000402}.

\appendix

\section{Comparison between pBilby and LALInference}
\label{subsec:LI_vs_pbilby}

In order to cross-validate our parallel Bilby results, we perform the runs listed in Table\ \ref{tab:LI_runs} using the Bayesian parameter estimation package LALInference (LI) \cite{Veitch:2014wba}.
The runs that make use of nested sampling employ five different seeds, 2048 live points, and a maximum chain length of 5000.
Those that use MCMC sampling, utilize eight parallel tempering temperatures and 24 parallel chains.

\begin{figure*}[htpb]
 \includegraphics[width=0.6\columnwidth]{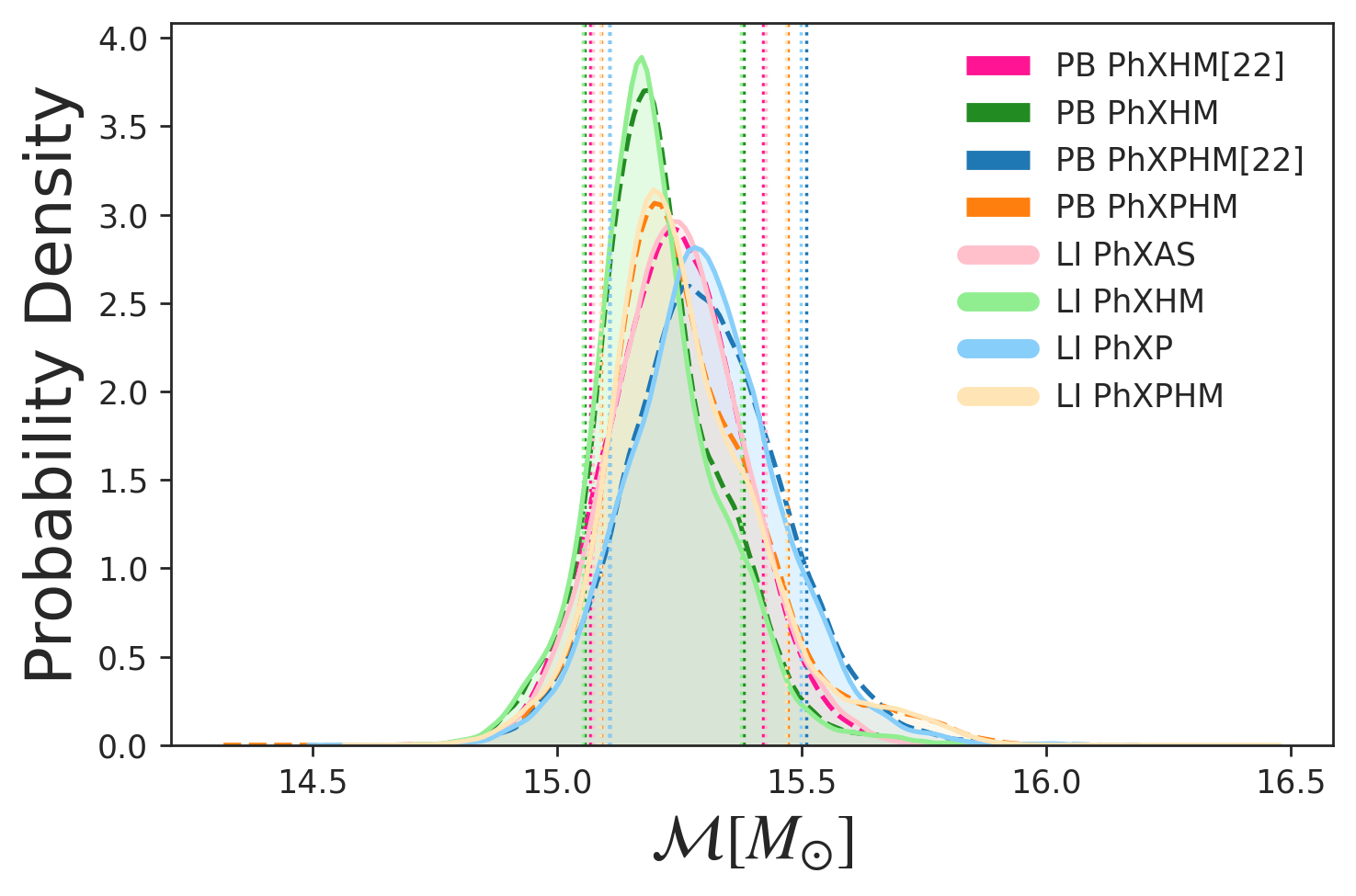}\includegraphics[width=0.6\columnwidth]{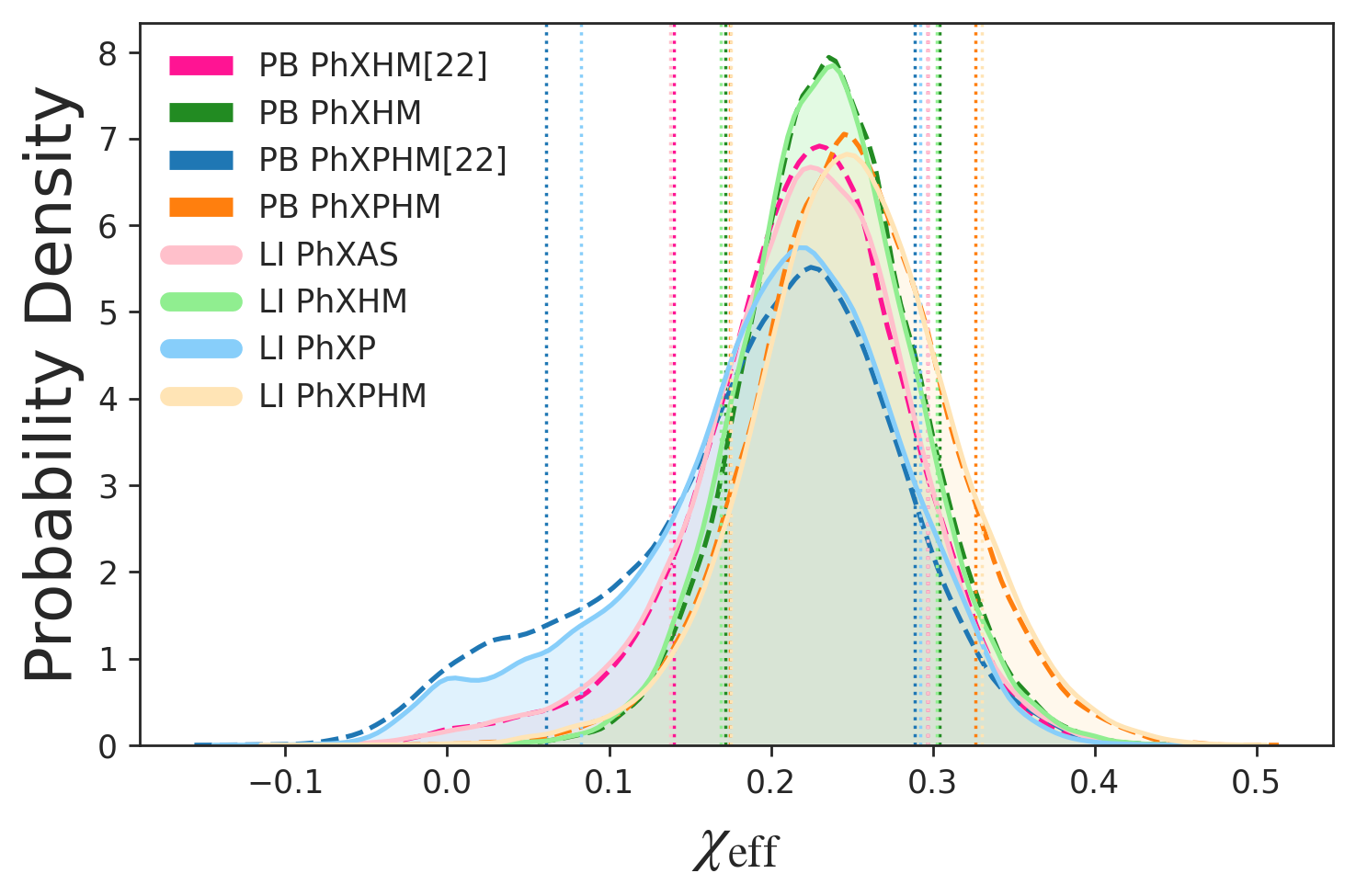}\includegraphics[width=0.6\columnwidth]{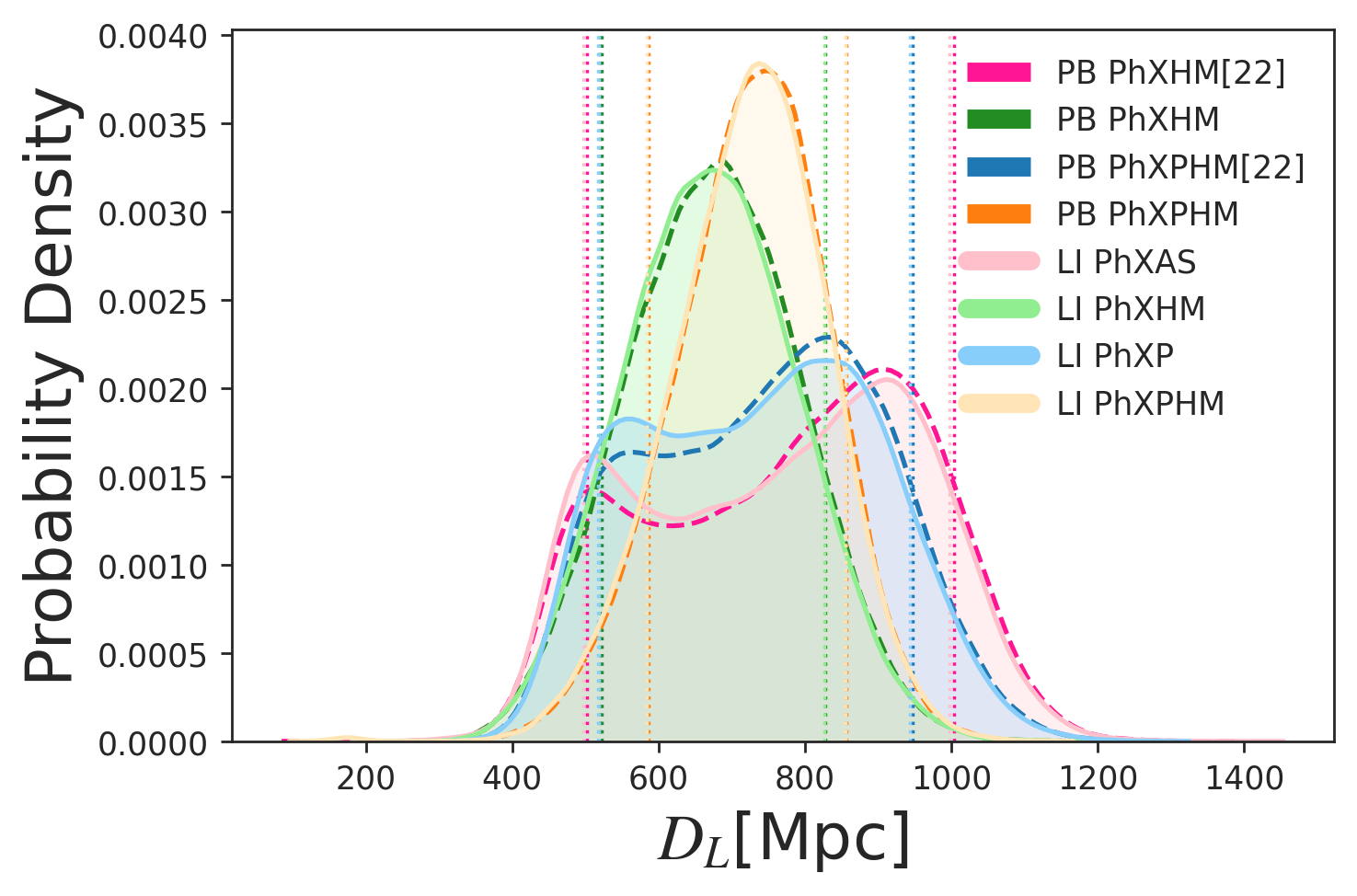}\\
 \caption{Posterior distributions of several source parameters comparing
 the LALInference (LI) runs as listed in Table~\ref{tab:LI_runs} (solid)
 and the standard parallel Bilby runs from Table~\ref{tab:tabRuns} using $n_{\mathrm{live}}=2048$ and $n_\mathrm{act}=10$ (dashed).
  \label{fig:LI_comparison}
 }
\end{figure*}

In Fig.\ \ref{fig:LI_comparison} we compare the standard runs of Table \ref{tab:tabRuns} using $n_{\mathrm{live}} = 2048$ and $n_\mathrm{act}=10$ (dashed) with the default LI runs (solid).
All posteriors are well recovered, and we find generally good 
agreement between pBilby and LI runs.
We also observe that the agreement is better in the presence of higher modes.
A possible explanation is that the breaking of degeneracies due to higher modes reduces the posterior volume and thus benefits better sampling.

\begin{table*}[bhpt]
\begin{center}
\begin{tabular}{|c|c|c|cccc|c|c|cccc|}
\hline 
Approximant &   Run $nº$ &          Modes (l,|m|) & PV &   FS &  PMB & MB &  Prior &  Sampler &  nlive & max. mcmc & nparallel & ntemps \\
\hline \hline 
IMRPhenomXAS & 1&                         (2,2) & - & - & - & - &         Aligned spin  &  Nested &  2048 & 5000 &  5 & - \\
\hline
IMRPhenomXHM & 2& D & - & - & - & D &         Aligned spin  &  Nested &  2048 & 5000 &  5 & -    \\
\hline
\multirow{2}{*}{IMRPhenomXP} & 3&  (2,2) & D & D & D & D &  Precessing &  Nested &  2048 & 5000 &  5 & - \\

 &  4& (2,2) & 102 & D & D & D &  Precessing &  Nested &  2048 & 5000 &  5 & - \\
\hline
\multirow{3}{*}{IMRPhenomXPHM} & 5&  D  & D & D & D & D &  Precessing &  MCMC &  - & - &  24 & 8 \\
& 6&  D  & 102 & D & D & D &  Precessing &  MCMC &  - & - &  24 & 8 \\
& 7&  D  & 102 & D & D & D &  Precessing &  Nested &  2048 & 5000 &  5 & - \\
\hline 
\end{tabular}
\end{center}
\caption{List of all LALInference runs we performed on open data for GW190412 using the \phXF family.
For each run, we indicate the LALSuite waveform approximant called along with along with various waveform settings:
the mode content, precessing prescription (PV), final spin version (FS), and multibanding thresholds applied to the evaluation of the Euler angles (PMB) and of the aligned-spin modes (MB); as well as the prior used.
We also denote the sampler used (Nested or MCMC) and its corresponding settings such as number of live points (nlive),  maximum chain length (max. mcmc), parallel chains (nparallel) and the number of parallel tempering temperatures (ntemps).
\label{tab:LI_runs}
}
\end{table*}

\section{Comparison of runs with and without distance marginalisation}
\label{subsec:dist_marg}

\begin{figure*}[htpb]
\includegraphics[width=0.6\columnwidth]{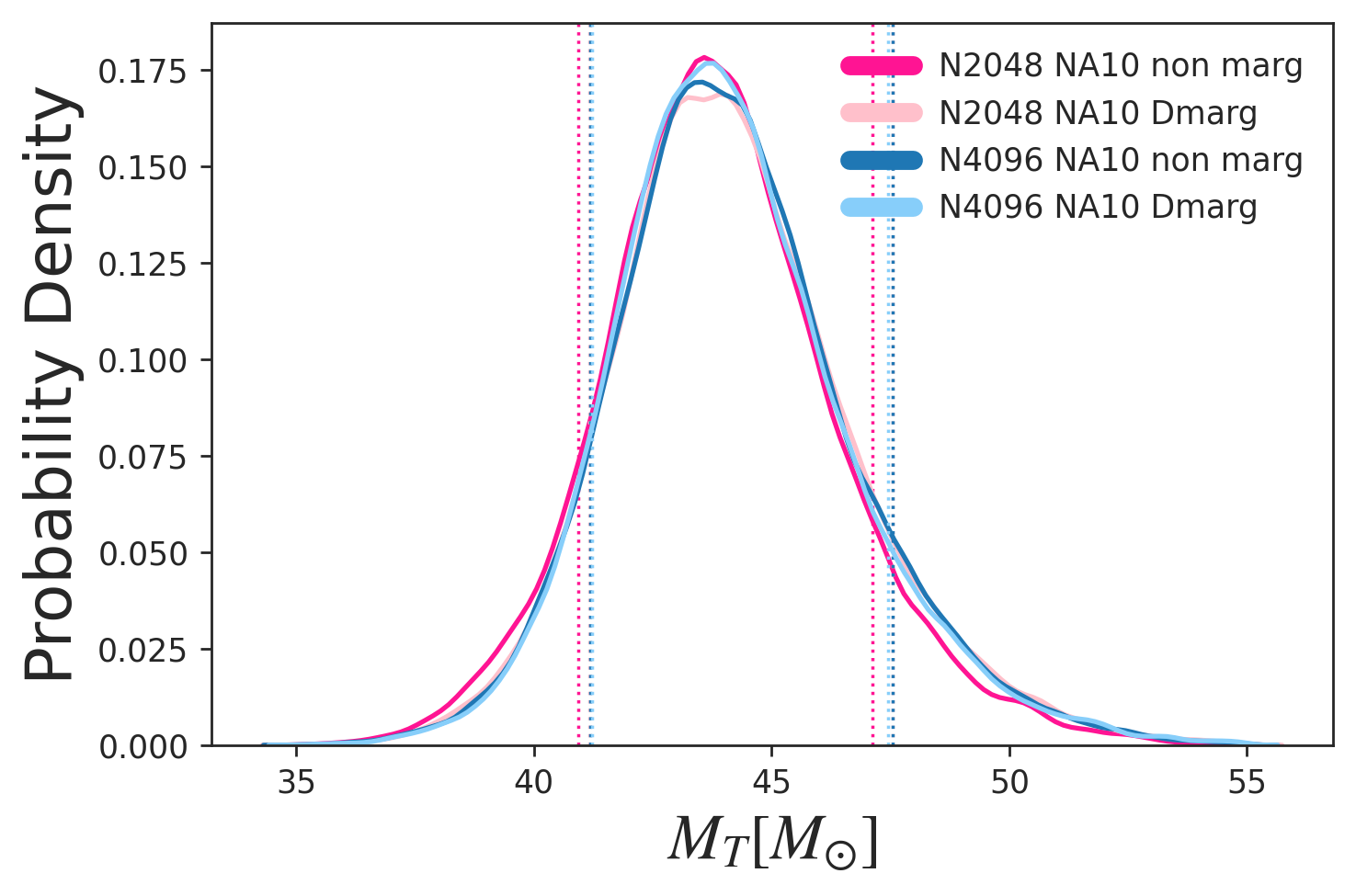}\includegraphics[width=0.6\columnwidth]{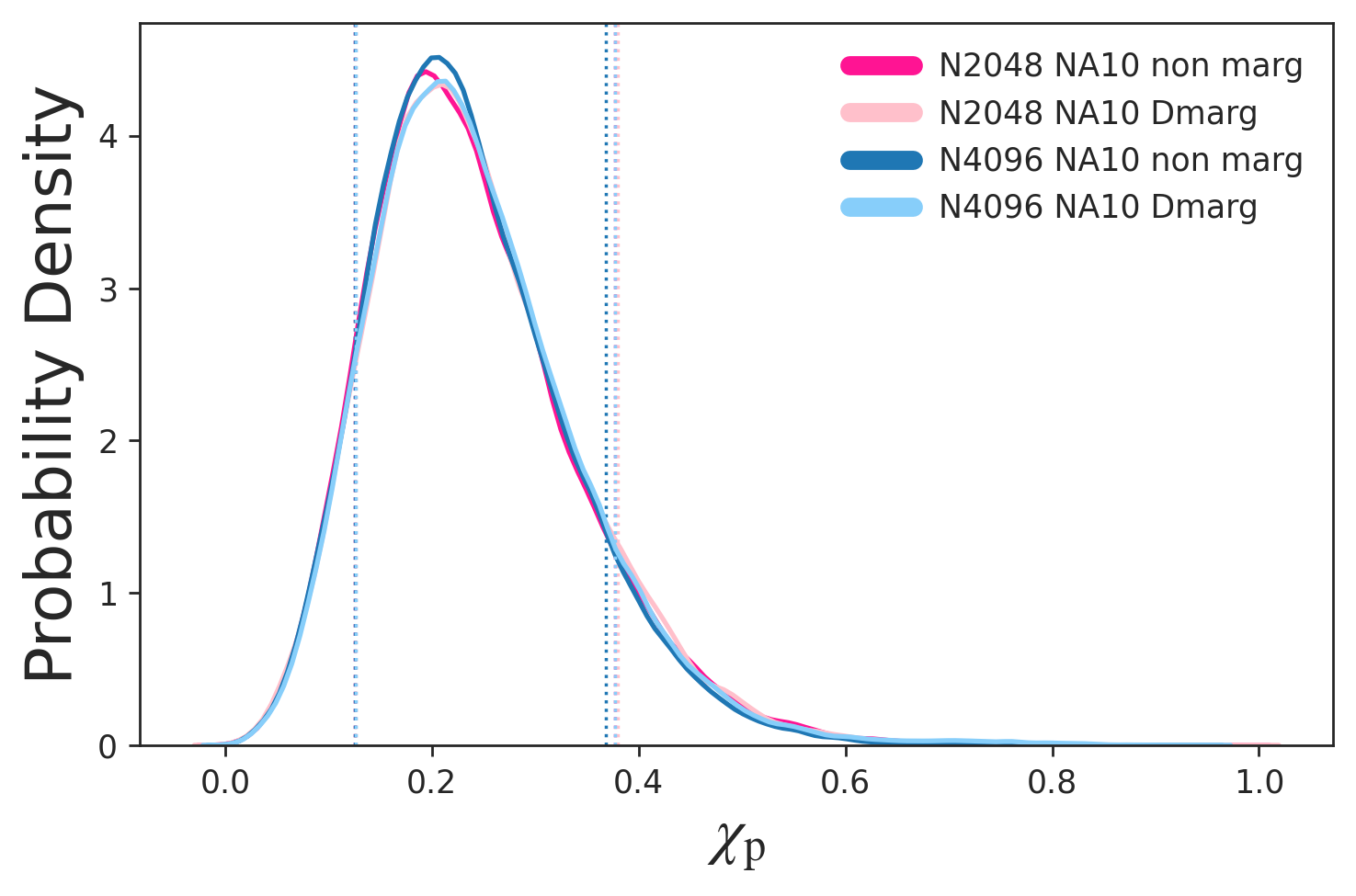}\includegraphics[width=0.6\columnwidth]{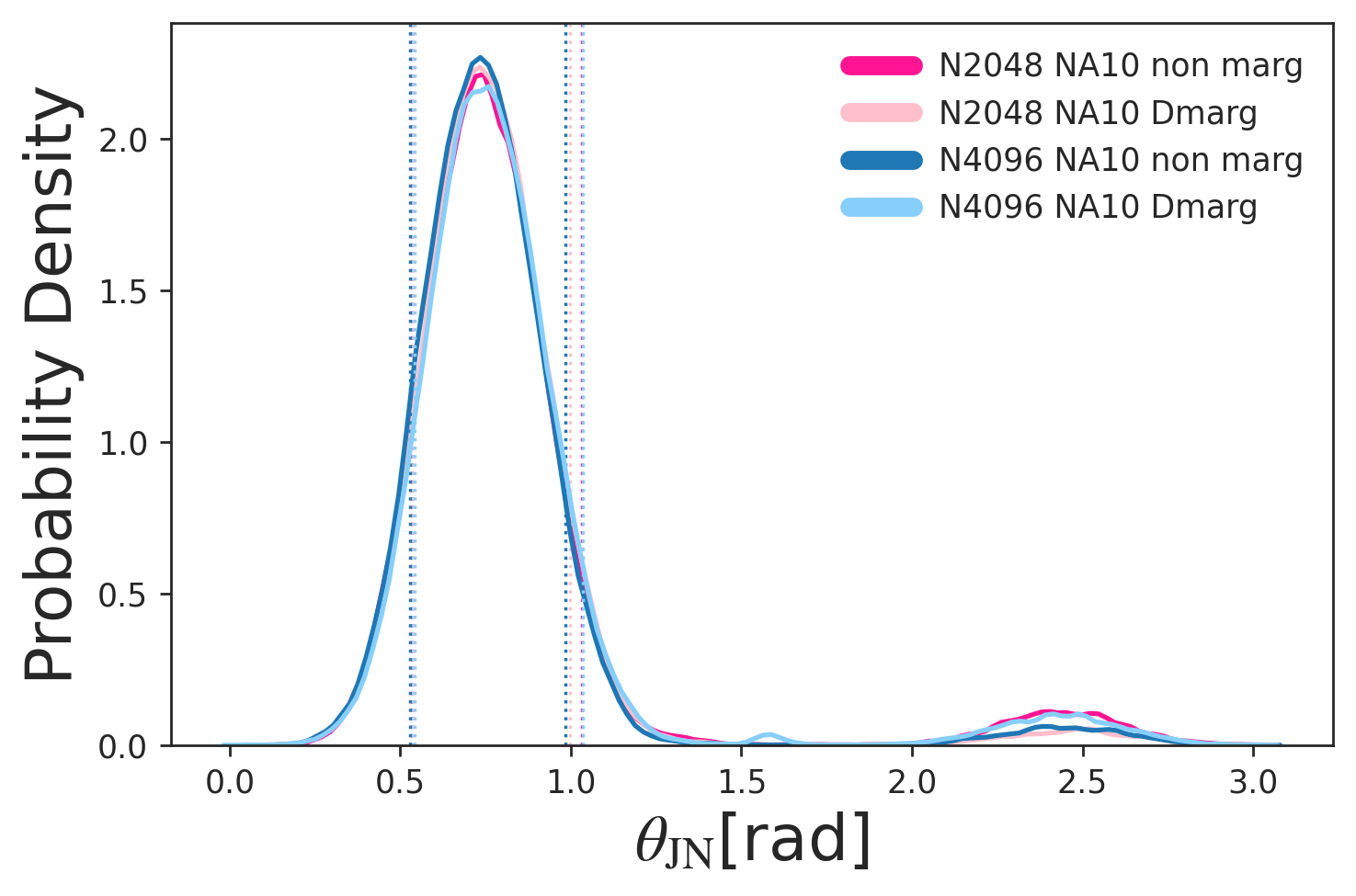}\\
 \caption{Comparison between some key posterior distributions obtained with and without distance marginalisation,
 using two sets of sampler settings
 ($n_{\mathrm{live}}=2048$ and $n_\mathrm{act}=10$ vs. $n_{\mathrm{live}}=4096$ and $n_\mathrm{act}=10$).
 \label{fig:marg_comparison}
}
\end{figure*}

We report in Table\,\ref{tab:tab_oldRuns} the complete list of pBilby runs performed on GW190412 without using distance marginalisation (prior to a bugfix in the distance marginalisation algorithm). We find excellent agreement and show a comparison in Fig.~\ref{fig:marg_comparison} for some key quantities.

\begin{table*}[htpb!]
\begin{center}
\begin{tabular}{|c|c|c|cccc|c|cc|r|c|c|}
\hline 

Approximant &  Run $nº$  &         Modes (l,|m|) & PV &   FS &    PMB & MB & Prior &  $n_\mathrm{live}$  &  $n_\mathrm{act}$ &  CPU h & L. eval. & Cost/L. eval. [ms]\\
\hline \hline 
\phX & 1 & (2,2) & - & - & - & - & Aligned spin  & 2048 & 20 & 2650 & $2.23 \times 10^8$ & 42.77 \\
\hline
\phXHM & 2 &  D &  - & - & - & D &   Aligned spin  &   2048 &  20 &  4734 & $2.90 \times 10^8 $ & 58.67 \\
\hline
\multirow{2}{*}{\phXP}
  & 3 &  (2,2) &  102 & 0 & -&  - &   Precessing &   2048 &   5 &  1042 & $6.57 \times 10^7$ & 57.13 \\
  & 4 & (2,2) &   D & D & - &  - &  Precessing &   2048 &   5 &   1273 & $6.32 \times 10^7$ & 72.45 \\
\hline
\multirow{25}{*}{\phXPHM}

 & 5 & D & D & D &  D &  D &  Precessing & 512 & 5 & 1566 & $1.99 \times 10^7$ & 283.60 \\
 & 6 & D & D & D &  D &  D &  Precessing & 1024 & 5 & 1531 & $3.61 \times 10^7$ & 152.55 \\
 & 7 & D & D & D &  D &  D &  Precessing & 1024 & 20 & 3362 & $1.31 \times 10^8$ & 92.16\\
 & 8 &  D & D & D &  D &  D &  Uniform $m_1$-$m_2$ & 1024 & 30 & 7082 & $2.05 \times 10^8$ & 124.17\\
 & 9 & D & D & D &  D &  D &  Uniform $m_1$-$m_2$ & 1024 & 50 & 8362 & $3.34 \times 10^8$ & 90.10 \\
 & 10 & D & D & D &  D &  D &  Precessing & 1024 & 60 & 3696 & $1.68 \times 10^8$ & 79.33 \\
 & 11 & D &  D &  D &  D &  D &  Precessing & 1500 & 50 & 5041 & $2.31 \times 10^8$ & 78.43 \\
 & 12 &  D &    D &  D &  D &  D  &  Precessing & 2048 & 5 &  1804 & $7.15 \times 10^7$ & 90.85 \\
 & 13 &  D &  D &  D &  $10^{-2}$ &  D &   Precessing &   2048 &  5 &  1295 & $7.19 \times 10^7$ & 64.85 \\
 & 14 & D &  D &  D &  $10^{-2}$ &  D  &  Precessing * & 2048 & 5 & 1690 & $7.18 \times 10^7$ & 84.73 \\
 & 15 & D & 102 &  2 &  D &  D &  Precessing & 2048 & 5 &  1336 & $7.09 \times 10^7$ & 67.83 \\
 & 16 & D & 102 &  0 &  D & D &   Precessing & 2048 & 5 &  1354 & $7.10 \times 10^7$ & 68.61 \\
 & 17 & D &    223 &  2 &  D &  D &  Precessing & 2048 & 5 &  1732 & $7.24 \times 10^7$ & 86.13 \\
 &  18 &    No $(3,2)$ & D & D & D & D &  Precessing & 2048 & 5 & 1465 & $7.21 \times 10^7$ & 73.11 \\
 &  19 & D &   D &  D &  D &  D &  Volumetric spin &  2048 & 5 & 2216 & $7.77 \times 10^7$ & 102.68 \\
 & 20 & D &   D &  D &  D &  D &  Precessing & 2048 & 10 & 4275 &  $1.54 \times 10^8$ & 99.44 \\
 & 21 & D &   D &  D &  D &  D &  Precessing & 2048 & 20 & 6795 & $2.66 \times 10^8$ & 91.79 \\
 & 22 &  D & D & D &  D &  D &  Uniform $m_1$-$m_2$ & 2048 & 30 & 9070 & $4.22 \times 10^8$ & 77.25\\
 & 23 & D &    D &  D &  D &  D &  Precessing & 4096 & 5 & 3434 & $1.46 \times 10^8$ &  84.47 \\
 & 24 & D &    D &  D &  D &  D &  Precessing & 4096 & 10 & 6377 & $2.89 \times 10^8$ & 79.43 \\
 & 25 & D &    102 &  D &  D &  D &  Precessing & 4096 & 10 & 5132 & $2.85 \times 10^8$ & 64.75 \\
  & 26 & D &    D &  D &  D &  D &  Precessing & 4096 & 20 & 11438 & $5.71 \times 10^8$ & 72.08 \\
 & 27 & D & D & D &  D &  D &  Uniform $m_1$-$m_2$ & 4096 & 30 & 20187 &  $9.49 \times 10^8$ & 76.57 \\
 & 28 & D &  D &  D &  D &  D &  Precessing & 8192 & 5 & 8921 & $2.95 \times 10^8$ & 108.94 \\
 & 29 & D &  D &  D &  D &  D &  Precessing, $\chi_1=0$ & 2048 & 5 & 1863 & $6.95 \times 10^7$ & 96.47 \\
 & 30 & D &  D &  D &  D &  D &  Aligned, $\chi_1=0$ & 2048 & 5 & 1726 & $6.31 \times 10^7$ & 98.40 \\

\hline 
\end{tabular}
\end{center}
\caption{List of the runs performed on GW190412 open data with parallel Bilby {\em without} distance marginalisation, using different models of the \phXF family.
For each run, we indicate the LALSuite waveform approximant called along with various waveform settings:
the mode content, precessing prescription (PV), final spin version (FS), and multibanding thresholds applied to the evaluation of the Euler angles (PMB) and of the aligned-spin modes (MB);
as well as the prior used and the chosen sampler settings ($n_\mathrm{live}$ and $n_\mathrm{act}$).
We also provide the computational cost of each run, the number of likelihood evaluations and the mean cost of each evaluation.
We use 4kHz GWOSC data for all runs in this table, except for the one marked with an asterisk, which employs 16kHz data.
\label{tab:tab_oldRuns}
}
\end{table*}

\section{Comparison of \phXF precession and final spin versions}\label{sec:prec_versions}
\begin{figure*}[htpb]
\begin{center}
\includegraphics[width=0.8\columnwidth]{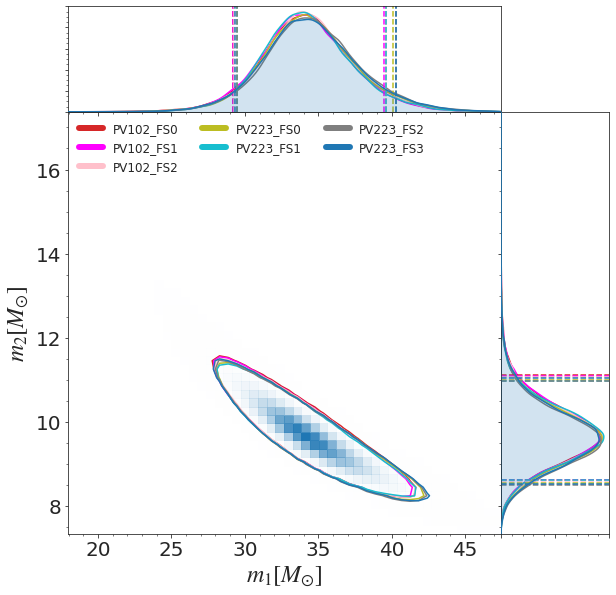}\includegraphics[width=0.8\columnwidth]{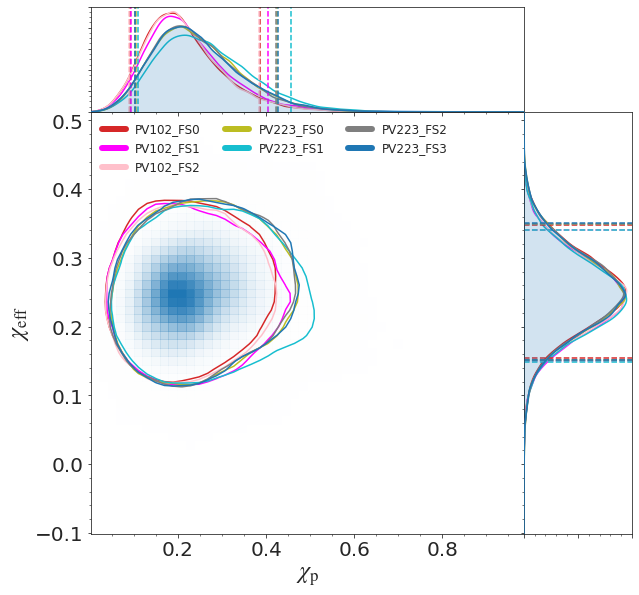}
\caption{1D and 2D posterior distributions for detector-frame masses and spin parameters of GW190412 obtained with three versions of \phXPHM that implement different precessing prescriptions (102 for NNLO angles and 223 for MSA) and final spin formulae (FS0, FS2, FS3, see main text for more details). The two NNLO results are plotted in red (FS0) and purple (FS2), while the default MSA results is plotted in green. Dashed (solid) lines in the 1D (2D) plots indicate 90\% credible intervals (regions).}
\label{fig:compare_settings}
\end{center}
\end{figure*}

We have performed several \phXPHM runs with non-default waveform options to study the robustness of our results under changes of the precession prescription and final spin version.
We compare posterior distributions for the default and non-default runs in Fig.\,\ref{fig:compare_settings}.
Mass parameters appear to be insensitive to changing these settings, while we observe
minor
differences in the spin parameters, with the MSA prescription returning slightly broader posteriors.  

\section{Comparison with \phPvthreehm}
\label{sec:phenompv3_comparison}
\begin{figure*}[htpb]
\includegraphics[width=0.8\columnwidth]{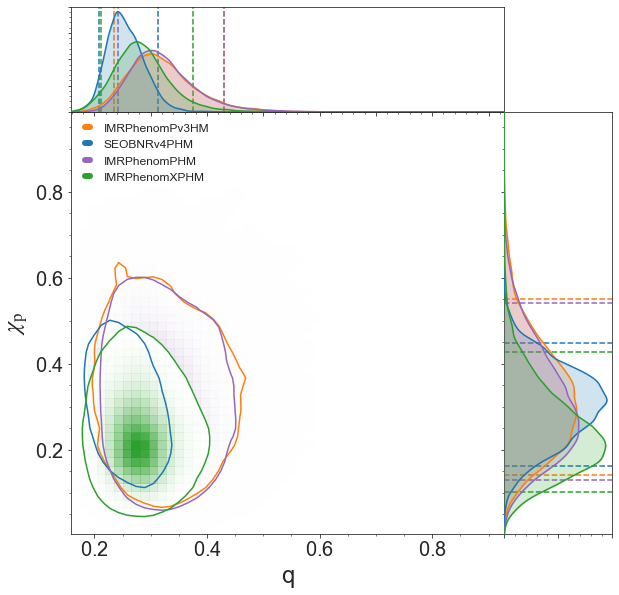}\hspace{2mm}\includegraphics[width=0.9\columnwidth]{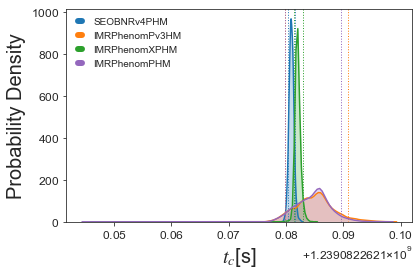}
\caption{Left panel: 1D and 2D posterior distributions for the mass ratio and precession spin of the signal recovered with \phPvthreehm (orange), \seobnrvforphm (blue), \phXPHM in its default version (green) and its ``IMRPhenomPHM'' version (purple, see main text for details).
Dashed (solid) lines indicate 90\% credible intervals (regions).
Right panel: posterior distributions for the geocentric time of the event, as estimated with the same models shown in the top panel.
\label{fig:pv3hm_study}
}
\end{figure*}

As we observed in the previous subsection, there are visible differences in the posterior distributions obtained with \phPvthreehm and \phXPHM.
We mean to quantify here the effect of the underlying aligned-spin model on those results.
This can be easily done with \phXPHM, which has an in-built option allowing to twist-up the GW modes returned by the older model \phHM (we will refer to this configuration as ``IMRPhenomPHM'') instead of \phXHM.
Using this option, one can cleanly separate the effects of the HM calibration from all the details of the precessing extension.
Fig.\,\ref{fig:pv3hm_study} compares previous results with the IMRPhenomPHM run described above. 

It can be seen that the shifts observed in the posteriors can be ascribed to the different underlying aligned-spin models, as, despite the use of independent precessing extensions, \phPvthreehm and IMRPhenomPHM return equivalent results.
Comparing both against the default \phXPHM, we observe in particular a marked difference in the recovery of the geocentric time of the event, due to the improved time-alignment provided by the \phXF models.

\section{Investigating different spin priors}
\label{sec:spin_priors}

For the vast majority of our runs we adopted the same priors employed in the LVC analysis \cite{LIGOScientific:2020stg}. Specifically, we choose uniform priors for detector-frame mass ratio $0.125\leq q \leq 1$ and chirp mass $13\leq \mathcal{M}\leq 18$, with masses constrained to lie in the interval $5\leq m_{1,2}\leq 60$. We use a power law prior with exponent 2 for the luminosity distance, with a lower bound of 100 Mpc and uniform priors for phase and polarization angle. 
We also investigate the effect of different spin priors on the posterior estimates,
in particular on the $\chip$ parameter whose interpretation is subtle.
Our default prior for precessing runs\footnote{For aligned-spin runs we use instead a 'z-prior', corresponding to a projection of the isotropic spin prior along the direction of the total angular momentum.}, in keeping with the LVC standard
(defined as prior ``P1" in appendix C.1 of~\cite{LIGOScientific:2018mvr} and also used in the analysis of GW190412~\cite{LIGOScientific:2020stg}),
employs uniform distributions in the dimensionless component spin magnitudes ($a_i \in [0.0,0.99]$) and isotropically distributed spin tilts (uniform in $\cos(\theta_i)$).
Another common prior choice is a ``volumetric spin'' prior, where the magnitudes follow a power law with exponent 2 (prior ``P2'' in appendix C.1 of~\cite{LIGOScientific:2018mvr}).

\begin{figure}[htpb]
\begin{center}
\includegraphics[width=0.9\columnwidth]{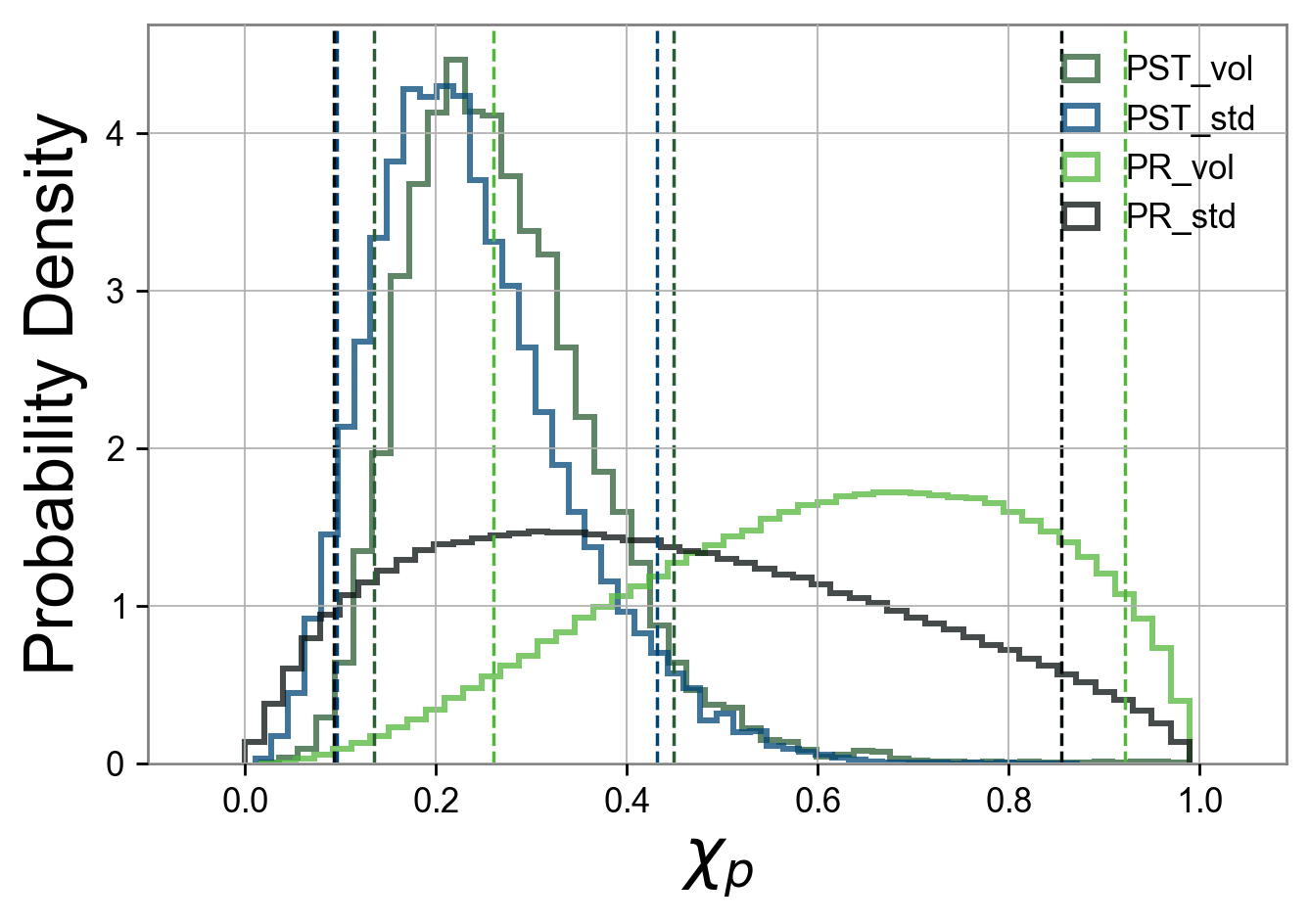}
\caption{Induced prior on the effective precessing spin parameter $\chip$ for two different priors on the component spins:
uniform in magnitudes (default prior for this paper) and volumetric (power law with exponent 2 in the spin magnitudes).
These correspond to priors ``P1" and ``P2" in appendix C.1 of~\cite{LIGOScientific:2018mvr}.
Also shown are the posteriors on $\chip$ from
two \phXPHM runs with the two different prior choices and otherwise identical settings
(runs 12 and 19 in Table~\ref{tab:tab_oldRuns}).
}
\label{fig:spin_priors_chip}
\end{center}
\end{figure}

As shown in Fig.~\ref{fig:spin_priors_chip}, the induced prior on $\chip$ from the volumetric prior prefers higher values than for the standard prior.
We compare two \phXPHM runs with these two prior choices, both with 2048 live points, $n_\mathrm{act}=5$ and without distance marginalisation
(runs 12 and 19 in Table~\ref{tab:tab_oldRuns}).
We find
a slight shift of the inferred $\chip$ towards higher values:
\mbox{$\chip = 0.26_{-0.12}^{+0.19}$}
from the run with a volumetric prior
vs.
\mbox{$\chip = 0.22_{-0.13}^{+0.21}$} from the run with the default prior.
However the resulting estimate is still lower than those reported in~\cite{LIGOScientific:2020stg}
(e.g. \mbox{$\chip = 0.31_{-0.15}^{+0.14}$} for \seobnrvforphm).
Other relevant parameters like $\chieff$ show no noticeable change under this prior.

While the Bayes factor for precession against the corresponding \phXHM run is indecisive for the default prior ($\Delta \log_{10} B = 0.06$), for the volumetric prior precession is actually slightly disfavoured ($\Delta \log_{10} B = -0.84$).
Since the volumetric prior gives more weight to high $\chip$, this result is consistent with the overall picture of no support in the GW190412 data for strongly precessing spins
\footnote{No separate \phXHM run with modified prior was necessary for this Bayes factor comparison, since for the one-dimensional aligned-spin parameter space a ``volumetric" prior is identical to the default uniform prior.}.

In conclusion, the effect of different standard spin prior choices on GW190412 inference is small, with the main conclusions robust against such a change.

We also briefly consider another spin prior as suggested by~\cite{Mandel:2020lhv}.
They argue that if GW190412 formed from the isolated binary evolution channel~\cite{Postnov:2014lrr,Bavera:2019fkg}, the more massive component BH should have low (or even zero) spin but the secondary BH could have high aligned spin.
This suggestion and the effect of spin priors on GW190412 has already been considered in detail by~\cite{Zevin:2020gxf}.
Here we report on two additional \phXPHM runs
with the same settings as above (2048 live points, $n_\mathrm{act}=5$ and without distance marginalisation)
but with priors that
(i) fix $a_1=0$, leaving $a_2$ and the tilt angles unchanged;
(ii) fix $a_1=0$ and the secondary spin to be positive aligned ($a_2$ uniform in [0.0,0.99], tilt $\theta_2=0$).
These are runs 29 and 30 in Table~\ref{tab:tab_oldRuns}.

As suggested by \cite{Mandel:2020lhv} based on reweighting the original LVC posteriors,
and first confirmed by~\cite{Zevin:2020gxf} from full PE runs,
our results also show that it is possible to fit the GW190412 data with zero primary spin,
and that the posteriors then prefer a high value for the secondary spin magnitude:
$a_2>0.52$ or $a_2>0.54$ at 90\% for the two runs respectively,
with both posteriors railing against the upper prior limit.
However, this alternative configuration is actually a somewhat worse fit to the data than the standard interpretation of nonzero primary spin\footnote{Note that, for consistency, we will be comparing here only runs performed without distance marginalization.}: 
while the run with the same settings and default prior (run no. 12 in Table\,\ref{tab:tab_oldRuns}) reaches network matched filter SNRs of $18.80_{-0.3}^{+0.19}$, for the two modified prior runs these are only $18.52_{-0.34}^{+0.23}$ and $18.53_{-0.3}^{+0.19}$.
Correspondingly, we find clear preference against the first and mild preference against the second alternative, with $\log_{10}\mathcal{B}$ of $1.7\pm 0.2$ and $1.0\pm 0.2$ in favour of the standard interpretation.
Hence, our result is consistent with the findings of~\cite{Zevin:2020gxf} for other waveforms:
also with \phXPHM there appears to be no preference for the scenario of \cite{Mandel:2020lhv} from the GW190412 data, since the reduced prior volume cannot make up for the worse fit to the data of waveforms without primary spin.


\vspace{0.1in}

\vfil

\let\c\Originalcdefinition %
\let\d\Originalddefinition %
\let\i\Originalidefinition

\bibliography{phenomx,phenom_refs,eob_refs,nr_refs,postnewtonian,gravitationalwaves}

\providecommand{\noopsort}[1]{}\providecommand{\singleletter}[1]{#1}
\begin{thebibliography}{99}%
\makeatletter
\providecommand \@ifxundefined [1]{%
 \@ifx{#1\undefined}
}%
\providecommand \@ifnum [1]{%
 \ifnum #1\expandafter \@firstoftwo
 \else \expandafter \@secondoftwo
 \fi
}%
\providecommand \@ifx [1]{%
 \ifx #1\expandafter \@firstoftwo
 \else \expandafter \@secondoftwo
 \fi
}%
\providecommand \natexlab [1]{#1}%
\providecommand \enquote  [1]{``#1''}%
\providecommand \bibnamefont  [1]{#1}%
\providecommand \bibfnamefont [1]{#1}%
\providecommand \citenamefont [1]{#1}%
\providecommand \href@noop [0]{\@secondoftwo}%
\providecommand \href [0]{\begingroup \@sanitize@url \@href}%
\providecommand \@href[1]{\@@startlink{#1}\@@href}%
\providecommand \@@href[1]{\endgroup#1\@@endlink}%
\providecommand \@sanitize@url [0]{\catcode `\\12\catcode `\$12\catcode
  `\&12\catcode `\#12\catcode `\^12\catcode `\_12\catcode `\%12\relax}%
\providecommand \@@startlink[1]{}%
\providecommand \@@endlink[0]{}%
\providecommand \url  [0]{\begingroup\@sanitize@url \@url }%
\providecommand \@url [1]{\endgroup\@href {#1}{\urlprefix }}%
\providecommand \urlprefix  [0]{URL }%
\providecommand \Eprint [0]{\href }%
\providecommand \doibase [0]{http://dx.doi.org/}%
\providecommand \selectlanguage [0]{\@gobble}%
\providecommand \bibinfo  [0]{\@secondoftwo}%
\providecommand \bibfield  [0]{\@secondoftwo}%
\providecommand \translation [1]{[#1]}%
\providecommand \BibitemOpen [0]{}%
\providecommand \bibitemStop [0]{}%
\providecommand \bibitemNoStop [0]{.\EOS\space}%
\providecommand \EOS [0]{\spacefactor3000\relax}%
\providecommand \BibitemShut  [1]{\csname bibitem#1\endcsname}%
\let\auto@bib@innerbib\@empty
\bibitem [{\citenamefont {Abbott}\ \emph
  {et~al.}(2020{\natexlab{a}})\citenamefont {Abbott} \emph
  {et~al.}}]{LIGOScientific:2020stg}%
  \BibitemOpen
  \bibfield  {author} {\bibinfo {author} {\bibfnamefont {R.}~\bibnamefont
  {Abbott}} \emph {et~al.} (\bibinfo {collaboration} {LIGO Scientific,
  Virgo}),\ }\href {\doibase 10.1103/PhysRevD.102.043015} {\bibfield  {journal}
  {\bibinfo  {journal} {Phys. Rev. D}\ }\textbf {\bibinfo {volume} {102}},\
  \bibinfo {pages} {043015} (\bibinfo {year} {2020}{\natexlab{a}})},\ \Eprint
  {http://arxiv.org/abs/2004.08342} {arXiv:2004.08342 [astro-ph.HE]}
  \BibitemShut {NoStop}%
\bibitem [{\citenamefont {Abbott}\ \emph
  {et~al.}(2020{\natexlab{b}})\citenamefont {Abbott}, \citenamefont {Abbott},
  \citenamefont {Abraham}, \citenamefont {Acernese}, \citenamefont {Ackley},
  \citenamefont {Adams}, \citenamefont {Adhikari}, \citenamefont {Adya},
  \citenamefont {Affeldt}, \citenamefont {Agathos},\ and\ \citenamefont
  {et~al.}}]{GW190814:journal}%
  \BibitemOpen
  \bibfield  {author} {\bibinfo {author} {\bibfnamefont {R.}~\bibnamefont
  {Abbott}}, \bibinfo {author} {\bibfnamefont {T.~D.}\ \bibnamefont {Abbott}},
  \bibinfo {author} {\bibfnamefont {S.}~\bibnamefont {Abraham}}, \bibinfo
  {author} {\bibfnamefont {F.}~\bibnamefont {Acernese}}, \bibinfo {author}
  {\bibfnamefont {K.}~\bibnamefont {Ackley}}, \bibinfo {author} {\bibfnamefont
  {C.}~\bibnamefont {Adams}}, \bibinfo {author} {\bibfnamefont {R.~X.}\
  \bibnamefont {Adhikari}}, \bibinfo {author} {\bibfnamefont {V.~B.}\
  \bibnamefont {Adya}}, \bibinfo {author} {\bibfnamefont {C.}~\bibnamefont
  {Affeldt}}, \bibinfo {author} {\bibfnamefont {M.}~\bibnamefont {Agathos}}, \
  and\ \bibinfo {author} {\bibnamefont {et~al.}},\ }\href {\doibase
  10.3847/2041-8213/ab960f} {\bibfield  {journal} {\bibinfo  {journal} {The
  Astrophysical Journal}\ }\textbf {\bibinfo {volume} {896}},\ \bibinfo {pages}
  {L44} (\bibinfo {year} {2020}{\natexlab{b}})}\BibitemShut {NoStop}%
\bibitem [{\citenamefont {Abbott}\ and\ \citenamefont {et.
  al.}(2020{\natexlab{a}})}]{GW190521:observ}%
  \BibitemOpen
  \bibfield  {author} {\bibinfo {author} {\bibfnamefont {R.}~\bibnamefont
  {Abbott}}\ and\ \bibinfo {author} {\bibnamefont {et. al.}} (\bibinfo
  {collaboration} {LIGO Scientific Collaboration and Virgo Collaboration}),\
  }\href {\doibase 10.1103/PhysRevLett.125.101102} {\bibfield  {journal}
  {\bibinfo  {journal} {Phys. Rev. Lett.}\ }\textbf {\bibinfo {volume} {125}},\
  \bibinfo {pages} {101102} (\bibinfo {year} {2020}{\natexlab{a}})}\BibitemShut
  {NoStop}%
\bibitem [{\citenamefont {Abbott}\ and\ \citenamefont {et.
  al.}(2020{\natexlab{b}})}]{GW190521:properties}%
  \BibitemOpen
  \bibfield  {author} {\bibinfo {author} {\bibfnamefont {R.}~\bibnamefont
  {Abbott}}\ and\ \bibinfo {author} {\bibnamefont {et. al.}},\ }\href {\doibase
  10.3847/2041-8213/aba493} {\bibfield  {journal} {\bibinfo  {journal} {The
  Astrophysical Journal}\ }\textbf {\bibinfo {volume} {900}},\ \bibinfo {pages}
  {L13} (\bibinfo {year} {2020}{\natexlab{b}})}\BibitemShut {NoStop}%
\bibitem [{\citenamefont {Aasi}\ \emph {et~al.}(2015)\citenamefont {Aasi} \emph
  {et~al.}}]{TheLIGOScientific:2014jea}%
  \BibitemOpen
  \bibfield  {author} {\bibinfo {author} {\bibfnamefont {J.}~\bibnamefont
  {Aasi}} \emph {et~al.} (\bibinfo {collaboration} {LIGO Scientific}),\ }\href
  {\doibase 10.1088/0264-9381/32/7/074001} {\bibfield  {journal} {\bibinfo
  {journal} {Class. Quant. Grav.}\ }\textbf {\bibinfo {volume} {32}},\ \bibinfo
  {pages} {074001} (\bibinfo {year} {2015})},\ \Eprint
  {http://arxiv.org/abs/1411.4547} {arXiv:1411.4547 [gr-qc]} \BibitemShut
  {NoStop}%
\bibitem [{\citenamefont {Acernese}\ \emph {et~al.}(2015)\citenamefont
  {Acernese} \emph {et~al.}}]{TheVirgo:2014hva}%
  \BibitemOpen
  \bibfield  {author} {\bibinfo {author} {\bibfnamefont {F.}~\bibnamefont
  {Acernese}} \emph {et~al.} (\bibinfo {collaboration} {VIRGO}),\ }\href
  {\doibase 10.1088/0264-9381/32/2/024001} {\bibfield  {journal} {\bibinfo
  {journal} {Class. Quant. Grav.}\ }\textbf {\bibinfo {volume} {32}},\ \bibinfo
  {pages} {024001} (\bibinfo {year} {2015})},\ \Eprint
  {http://arxiv.org/abs/1408.3978} {arXiv:1408.3978 [gr-qc]} \BibitemShut
  {NoStop}%
\bibitem [{\citenamefont {Pratten}\ \emph
  {et~al.}(2020{\natexlab{a}})\citenamefont {Pratten}, \citenamefont {Husa},
  \citenamefont {Garc{\'i}a-Quir\'os}, \citenamefont {Colleoni}, \citenamefont
  {Ramos-Buades}, \citenamefont {Estell\'es},\ and\ \citenamefont
  {Jaume}}]{Pratten:2020fqn}%
  \BibitemOpen
  \bibfield  {author} {\bibinfo {author} {\bibfnamefont {G.}~\bibnamefont
  {Pratten}}, \bibinfo {author} {\bibfnamefont {S.}~\bibnamefont {Husa}},
  \bibinfo {author} {\bibfnamefont {C.}~\bibnamefont {Garc{\'i}a-Quir\'os}},
  \bibinfo {author} {\bibfnamefont {M.}~\bibnamefont {Colleoni}}, \bibinfo
  {author} {\bibfnamefont {A.}~\bibnamefont {Ramos-Buades}}, \bibinfo {author}
  {\bibfnamefont {H.}~\bibnamefont {Estell\'es}}, \ and\ \bibinfo {author}
  {\bibfnamefont {R.}~\bibnamefont {Jaume}},\ }\href {\doibase
  10.1103/PhysRevD.102.064001} {\bibfield  {journal} {\bibinfo  {journal}
  {Phys. Rev. D}\ }\textbf {\bibinfo {volume} {102}},\ \bibinfo {pages}
  {064001} (\bibinfo {year} {2020}{\natexlab{a}})}\BibitemShut {NoStop}%
\bibitem [{\citenamefont {Garc{\'i}a-Quir\'os}\ \emph
  {et~al.}(2020)\citenamefont {Garc{\'i}a-Quir\'os}, \citenamefont {Colleoni},
  \citenamefont {Husa}, \citenamefont {Estell\'es}, \citenamefont {Pratten},
  \citenamefont {Ramos-Buades}, \citenamefont {Mateu-Lucena},\ and\
  \citenamefont {Jaume}}]{Garcia-Quiros:2020qpx}%
  \BibitemOpen
  \bibfield  {author} {\bibinfo {author} {\bibfnamefont {C.}~\bibnamefont
  {Garc{\'i}a-Quir\'os}}, \bibinfo {author} {\bibfnamefont {M.}~\bibnamefont
  {Colleoni}}, \bibinfo {author} {\bibfnamefont {S.}~\bibnamefont {Husa}},
  \bibinfo {author} {\bibfnamefont {H.}~\bibnamefont {Estell\'es}}, \bibinfo
  {author} {\bibfnamefont {G.}~\bibnamefont {Pratten}}, \bibinfo {author}
  {\bibfnamefont {A.}~\bibnamefont {Ramos-Buades}}, \bibinfo {author}
  {\bibfnamefont {M.}~\bibnamefont {Mateu-Lucena}}, \ and\ \bibinfo {author}
  {\bibfnamefont {R.}~\bibnamefont {Jaume}},\ }\href {\doibase
  10.1103/PhysRevD.102.064002} {\bibfield  {journal} {\bibinfo  {journal}
  {Phys. Rev. D}\ }\textbf {\bibinfo {volume} {102}},\ \bibinfo {pages}
  {064002} (\bibinfo {year} {2020})}\BibitemShut {NoStop}%
\bibitem [{\citenamefont {Garc{\'i}a-Quir{\'o}s}\ \emph
  {et~al.}(2020)\citenamefont {Garc{\'i}a-Quir{\'o}s}, \citenamefont {Husa},
  \citenamefont {Mateu-Lucena},\ and\ \citenamefont
  {Borchers}}]{Garcia-Quiros:2020qlt}%
  \BibitemOpen
  \bibfield  {author} {\bibinfo {author} {\bibfnamefont {C.}~\bibnamefont
  {Garc{\'i}a-Quir{\'o}s}}, \bibinfo {author} {\bibfnamefont {S.}~\bibnamefont
  {Husa}}, \bibinfo {author} {\bibfnamefont {M.}~\bibnamefont {Mateu-Lucena}},
  \ and\ \bibinfo {author} {\bibfnamefont {A.}~\bibnamefont {Borchers}},\
  }\href@noop {} {\bibfield  {journal} {\bibinfo  {journal} {ArXiv e-prints}\ }
  (\bibinfo {year} {2020})},\ \Eprint {http://arxiv.org/abs/2001.10897}
  {arXiv:2001.10897 [gr-qc]} \BibitemShut {NoStop}%
\bibitem [{\citenamefont {Pratten}\ \emph
  {et~al.}(2020{\natexlab{b}})\citenamefont {Pratten} \emph
  {et~al.}}]{Pratten:2020ceb}%
  \BibitemOpen
  \bibfield  {author} {\bibinfo {author} {\bibfnamefont {G.}~\bibnamefont
  {Pratten}} \emph {et~al.},\ }\href@noop {} {\bibfield  {journal} {\bibinfo
  {journal} {arXiv e-prints}\ } (\bibinfo {year} {2020}{\natexlab{b}})},\
  \Eprint {http://arxiv.org/abs/2004.06503} {arXiv:2004.06503 [gr-qc]}
  \BibitemShut {NoStop}%
\bibitem [{\citenamefont {Estell\'es}\ \emph {et~al.}(2020)\citenamefont
  {Estell\'es}, \citenamefont {Ramos-Buades}, \citenamefont {Husa},
  \citenamefont {Garc\'ia-Quir\'os}, \citenamefont {Colleoni}, \citenamefont
  {Haegel},\ and\ \citenamefont {Jaume}}]{phenomtp}%
  \BibitemOpen
  \bibfield  {author} {\bibinfo {author} {\bibfnamefont {H.}~\bibnamefont
  {Estell\'es}}, \bibinfo {author} {\bibfnamefont {A.}~\bibnamefont
  {Ramos-Buades}}, \bibinfo {author} {\bibfnamefont {S.}~\bibnamefont {Husa}},
  \bibinfo {author} {\bibfnamefont {C.}~\bibnamefont {Garc\'ia-Quir\'os}},
  \bibinfo {author} {\bibfnamefont {M.}~\bibnamefont {Colleoni}}, \bibinfo
  {author} {\bibfnamefont {L.}~\bibnamefont {Haegel}}, \ and\ \bibinfo {author}
  {\bibfnamefont {R.}~\bibnamefont {Jaume}},\ }\href@noop {} {\enquote
  {\bibinfo {title} {{IMRPhenomTP}: A phenomenological time domain model for
  dominant quadrupole gravitational wave signal of coalescing binary black
  holes},}\ } (\bibinfo {year} {2020}),\ \Eprint
  {http://arxiv.org/abs/2004.08302} {arXiv:2004.08302 [gr-qc]} \BibitemShut
  {NoStop}%
\bibitem [{\citenamefont {Estell{\'e}s}\ and\ \citenamefont {et.
  al.}(2020)}]{phenomthm}%
  \BibitemOpen
  \bibfield  {author} {\bibinfo {author} {\bibfnamefont {H.}~\bibnamefont
  {Estell{\'e}s}}\ and\ \bibinfo {author} {\bibnamefont {et. al.}},\
  }\href@noop {} {\enquote {\bibinfo {title} {{IMRPhenomTHM}},}\ } (\bibinfo
  {year} {2020}),\ \bibinfo {note} {{in preparation}}\BibitemShut {NoStop}%
\bibitem [{\citenamefont {Kalaghatgi}\ \emph {et~al.}(2020)\citenamefont
  {Kalaghatgi}, \citenamefont {Hannam},\ and\ \citenamefont
  {Raymond}}]{PE_PhHM}%
  \BibitemOpen
  \bibfield  {author} {\bibinfo {author} {\bibfnamefont {C.}~\bibnamefont
  {Kalaghatgi}}, \bibinfo {author} {\bibfnamefont {M.}~\bibnamefont {Hannam}},
  \ and\ \bibinfo {author} {\bibfnamefont {V.}~\bibnamefont {Raymond}},\ }\href
  {\doibase 10.1103/PhysRevD.101.103004} {\bibfield  {journal} {\bibinfo
  {journal} {Phys. Rev. D}\ }\textbf {\bibinfo {volume} {101}},\ \bibinfo
  {pages} {103004} (\bibinfo {year} {2020})}\BibitemShut {NoStop}%
\bibitem [{\citenamefont {Abbott}\ \emph
  {et~al.}(2019{\natexlab{a}})\citenamefont {Abbott} \emph
  {et~al.}}]{LIGOScientific:2018mvr}%
  \BibitemOpen
  \bibfield  {author} {\bibinfo {author} {\bibfnamefont {B.~P.}\ \bibnamefont
  {Abbott}} \emph {et~al.} (\bibinfo {collaboration} {LIGO Scientific,
  Virgo}),\ }\href {\doibase 10.1103/PhysRevX.9.031040} {\bibfield  {journal}
  {\bibinfo  {journal} {Phys. Rev.}\ }\textbf {\bibinfo {volume} {X9}},\
  \bibinfo {pages} {031040} (\bibinfo {year} {2019}{\natexlab{a}})},\ \Eprint
  {http://arxiv.org/abs/1811.12907} {arXiv:1811.12907 [astro-ph.HE]}
  \BibitemShut {NoStop}%
\bibitem [{\citenamefont {Abbott}\ \emph
  {et~al.}(2020{\natexlab{c}})\citenamefont {Abbott} \emph
  {et~al.}}]{Abbott:2020uma}%
  \BibitemOpen
  \bibfield  {author} {\bibinfo {author} {\bibfnamefont {B.}~\bibnamefont
  {Abbott}} \emph {et~al.} (\bibinfo {collaboration} {LIGO Scientific,
  Virgo}),\ }\href {\doibase 10.3847/2041-8213/ab75f5} {\bibfield  {journal}
  {\bibinfo  {journal} {Astrophys. J. Lett.}\ }\textbf {\bibinfo {volume}
  {892}},\ \bibinfo {pages} {L3} (\bibinfo {year} {2020}{\natexlab{c}})},\
  \Eprint {http://arxiv.org/abs/2001.01761} {arXiv:2001.01761 [astro-ph.HE]}
  \BibitemShut {NoStop}%
\bibitem [{\citenamefont {Abbott}\ \emph
  {et~al.}(2019{\natexlab{b}})\citenamefont {Abbott} \emph
  {et~al.}}]{LIGOScientific:2018jsj}%
  \BibitemOpen
  \bibfield  {author} {\bibinfo {author} {\bibfnamefont {B.}~\bibnamefont
  {Abbott}} \emph {et~al.} (\bibinfo {collaboration} {LIGO Scientific,
  Virgo}),\ }\href {\doibase 10.3847/2041-8213/ab3800} {\bibfield  {journal}
  {\bibinfo  {journal} {Astrophys. J.}\ }\textbf {\bibinfo {volume} {882}},\
  \bibinfo {pages} {L24} (\bibinfo {year} {2019}{\natexlab{b}})},\ \Eprint
  {http://arxiv.org/abs/1811.12940} {arXiv:1811.12940 [astro-ph.HE]}
  \BibitemShut {NoStop}%
\bibitem [{\citenamefont {Mandel}\ and\ \citenamefont
  {Fragos}(2020)}]{Mandel:2020lhv}%
  \BibitemOpen
  \bibfield  {author} {\bibinfo {author} {\bibfnamefont {I.}~\bibnamefont
  {Mandel}}\ and\ \bibinfo {author} {\bibfnamefont {T.}~\bibnamefont
  {Fragos}},\ }\href {\doibase 10.3847/2041-8213/ab8e41} {\bibfield  {journal}
  {\bibinfo  {journal} {Astrophys. J. Lett.}\ }\textbf {\bibinfo {volume}
  {895}},\ \bibinfo {pages} {L28} (\bibinfo {year} {2020})},\ \Eprint
  {http://arxiv.org/abs/2004.09288} {arXiv:2004.09288 [astro-ph.HE]}
  \BibitemShut {NoStop}%
\bibitem [{\citenamefont {Di~Carlo}\ \emph {et~al.}(2020)\citenamefont
  {Di~Carlo} \emph {et~al.}}]{DiCarlo:2020lfa}%
  \BibitemOpen
  \bibfield  {author} {\bibinfo {author} {\bibfnamefont {U.~N.}\ \bibnamefont
  {Di~Carlo}} \emph {et~al.},\ }\href {\doibase 10.1093/mnras/staa2286}
  {\bibfield  {journal} {\bibinfo  {journal} {Monthly Notices of the Royal
  Astronomical Society}\ } (\bibinfo {year} {2020}),\ 10.1093/mnras/staa2286},\
  \Eprint {http://arxiv.org/abs/2004.09525} {arXiv:2004.09525 [astro-ph.HE]}
  \BibitemShut {NoStop}%
\bibitem [{\citenamefont {Olejak}\ \emph {et~al.}(2020)\citenamefont {Olejak},
  \citenamefont {Fishbach}, \citenamefont {Belczynski}, \citenamefont {Holz},
  \citenamefont {Lasota}, \citenamefont {Miller},\ and\ \citenamefont
  {Bulik}}]{Olejak:2020oel}%
  \BibitemOpen
  \bibfield  {author} {\bibinfo {author} {\bibfnamefont {A.}~\bibnamefont
  {Olejak}}, \bibinfo {author} {\bibfnamefont {M.}~\bibnamefont {Fishbach}},
  \bibinfo {author} {\bibfnamefont {K.}~\bibnamefont {Belczynski}}, \bibinfo
  {author} {\bibfnamefont {D.}~\bibnamefont {Holz}}, \bibinfo {author}
  {\bibfnamefont {J.-P.}\ \bibnamefont {Lasota}}, \bibinfo {author}
  {\bibfnamefont {M.}~\bibnamefont {Miller}}, \ and\ \bibinfo {author}
  {\bibfnamefont {T.}~\bibnamefont {Bulik}},\ }\href {\doibase
  10.3847/2041-8213/abb5b5} {\bibfield  {journal} {\bibinfo  {journal}
  {Astrophys. J.}\ }\textbf {\bibinfo {volume} {901}},\ \bibinfo {pages} {L39}
  (\bibinfo {year} {2020})},\ \Eprint {http://arxiv.org/abs/2004.11866}
  {arXiv:2004.11866 [astro-ph.HE]} \BibitemShut {NoStop}%
\bibitem [{\citenamefont {Hamers}\ and\ \citenamefont
  {Safarzadeh}(2020)}]{Hamers:2020huo}%
  \BibitemOpen
  \bibfield  {author} {\bibinfo {author} {\bibfnamefont {A.~S.}\ \bibnamefont
  {Hamers}}\ and\ \bibinfo {author} {\bibfnamefont {M.}~\bibnamefont
  {Safarzadeh}},\ }\href {\doibase 10.3847/1538-4357/ab9b27} {\bibfield
  {journal} {\bibinfo  {journal} {Astrophys. J.}\ }\textbf {\bibinfo {volume}
  {898}},\ \bibinfo {pages} {99} (\bibinfo {year} {2020})},\ \Eprint
  {http://arxiv.org/abs/2005.03045} {arXiv:2005.03045 [astro-ph.HE]}
  \BibitemShut {NoStop}%
\bibitem [{\citenamefont {Rodriguez}\ \emph {et~al.}(2020)\citenamefont
  {Rodriguez} \emph {et~al.}}]{Rodriguez:2020viw}%
  \BibitemOpen
  \bibfield  {author} {\bibinfo {author} {\bibfnamefont {C.~L.}\ \bibnamefont
  {Rodriguez}} \emph {et~al.},\ }\href {\doibase 10.3847/2041-8213/ab961d}
  {\bibfield  {journal} {\bibinfo  {journal} {Astrophys. J. Lett.}\ }\textbf
  {\bibinfo {volume} {896}},\ \bibinfo {pages} {L10} (\bibinfo {year}
  {2020})},\ \Eprint {http://arxiv.org/abs/2005.04239} {arXiv:2005.04239
  [astro-ph.HE]} \BibitemShut {NoStop}%
\bibitem [{\citenamefont {Gerosa}\ \emph {et~al.}(2020)\citenamefont {Gerosa},
  \citenamefont {Vitale},\ and\ \citenamefont {Berti}}]{Gerosa:2020bjb}%
  \BibitemOpen
  \bibfield  {author} {\bibinfo {author} {\bibfnamefont {D.}~\bibnamefont
  {Gerosa}}, \bibinfo {author} {\bibfnamefont {S.}~\bibnamefont {Vitale}}, \
  and\ \bibinfo {author} {\bibfnamefont {E.}~\bibnamefont {Berti}},\ }\href
  {\doibase 10.1103/PhysRevLett.125.101103} {\bibfield  {journal} {\bibinfo
  {journal} {Phys. Rev. Lett.}\ }\textbf {\bibinfo {volume} {125}},\ \bibinfo
  {pages} {101103} (\bibinfo {year} {2020})},\ \Eprint
  {http://arxiv.org/abs/2005.04243} {arXiv:2005.04243 [astro-ph.HE]}
  \BibitemShut {NoStop}%
\bibitem [{\citenamefont {De~Luca}\ \emph {et~al.}(2020)\citenamefont
  {De~Luca}, \citenamefont {Franciolini}, \citenamefont {Pani},\ and\
  \citenamefont {Riotto}}]{DeLuca:2020qqa}%
  \BibitemOpen
  \bibfield  {author} {\bibinfo {author} {\bibfnamefont {V.}~\bibnamefont
  {De~Luca}}, \bibinfo {author} {\bibfnamefont {G.}~\bibnamefont
  {Franciolini}}, \bibinfo {author} {\bibfnamefont {P.}~\bibnamefont {Pani}}, \
  and\ \bibinfo {author} {\bibfnamefont {A.}~\bibnamefont {Riotto}},\ }\href
  {\doibase 10.1088/1475-7516/2020/06/044} {\bibfield  {journal} {\bibinfo
  {journal} {JCAP}\ }\textbf {\bibinfo {volume} {06}},\ \bibinfo {pages} {044}
  (\bibinfo {year} {2020})},\ \Eprint {http://arxiv.org/abs/2005.05641}
  {arXiv:2005.05641 [astro-ph.CO]} \BibitemShut {NoStop}%
\bibitem [{\citenamefont {Safarzadeh}\ and\ \citenamefont
  {Hotokezaka}(2020)}]{Safarzadeh:2020qrc}%
  \BibitemOpen
  \bibfield  {author} {\bibinfo {author} {\bibfnamefont {M.}~\bibnamefont
  {Safarzadeh}}\ and\ \bibinfo {author} {\bibfnamefont {K.}~\bibnamefont
  {Hotokezaka}},\ }\href {\doibase 10.3847/2041-8213/ab9b79} {\bibfield
  {journal} {\bibinfo  {journal} {Astrophys. J. Lett.}\ }\textbf {\bibinfo
  {volume} {897}},\ \bibinfo {pages} {L7} (\bibinfo {year} {2020})},\ \Eprint
  {http://arxiv.org/abs/2005.06519} {arXiv:2005.06519 [astro-ph.HE]}
  \BibitemShut {NoStop}%
\bibitem [{\citenamefont {Kimball}\ \emph {et~al.}(2020)\citenamefont
  {Kimball}, \citenamefont {Talbot}, \citenamefont {Berry}, \citenamefont
  {Carney}, \citenamefont {Zevin}, \citenamefont {Thrane},\ and\ \citenamefont
  {Kalogera}}]{Kimball_2020}%
  \BibitemOpen
  \bibfield  {author} {\bibinfo {author} {\bibfnamefont {C.}~\bibnamefont
  {Kimball}}, \bibinfo {author} {\bibfnamefont {C.}~\bibnamefont {Talbot}},
  \bibinfo {author} {\bibfnamefont {C.~P.~L.}\ \bibnamefont {Berry}}, \bibinfo
  {author} {\bibfnamefont {M.}~\bibnamefont {Carney}}, \bibinfo {author}
  {\bibfnamefont {M.}~\bibnamefont {Zevin}}, \bibinfo {author} {\bibfnamefont
  {E.}~\bibnamefont {Thrane}}, \ and\ \bibinfo {author} {\bibfnamefont
  {V.}~\bibnamefont {Kalogera}},\ }\href {\doibase 10.3847/1538-4357/aba518}
  {\bibfield  {journal} {\bibinfo  {journal} {The Astrophysical Journal}\
  }\textbf {\bibinfo {volume} {900}},\ \bibinfo {pages} {177} (\bibinfo {year}
  {2020})}\BibitemShut {NoStop}%
\bibitem [{\citenamefont {Blanchet}(2006)}]{Blanchet:2006zz}%
  \BibitemOpen
  \bibfield  {author} {\bibinfo {author} {\bibfnamefont {L.}~\bibnamefont
  {Blanchet}},\ }\href@noop {} {\bibfield  {journal} {\bibinfo  {journal}
  {Living Rev.\ Rel.}\ }\textbf {\bibinfo {volume} {9}},\ \bibinfo {pages} {4}
  (\bibinfo {year} {2006})}\BibitemShut {NoStop}%
\bibitem [{\citenamefont {Damour}(2001)}]{Damour:2001tu}%
  \BibitemOpen
  \bibfield  {author} {\bibinfo {author} {\bibfnamefont {T.}~\bibnamefont
  {Damour}},\ }\href {\doibase 10.1103/PhysRevD.64.124013} {\bibfield
  {journal} {\bibinfo  {journal} {Phys. Rev.}\ }\textbf {\bibinfo {volume}
  {D64}},\ \bibinfo {pages} {124013} (\bibinfo {year} {2001})}\BibitemShut
  {NoStop}%
\bibitem [{\citenamefont {Damour}\ \emph {et~al.}(2013)\citenamefont {Damour},
  \citenamefont {Nagar},\ and\ \citenamefont {Bernuzzi}}]{Damour:2012ky}%
  \BibitemOpen
  \bibfield  {author} {\bibinfo {author} {\bibfnamefont {T.}~\bibnamefont
  {Damour}}, \bibinfo {author} {\bibfnamefont {A.}~\bibnamefont {Nagar}}, \
  and\ \bibinfo {author} {\bibfnamefont {S.}~\bibnamefont {Bernuzzi}},\ }\href
  {\doibase 10.1103/PhysRevD.87.084035} {\bibfield  {journal} {\bibinfo
  {journal} {Phys. Rev.}\ }\textbf {\bibinfo {volume} {D87}},\ \bibinfo {pages}
  {084035} (\bibinfo {year} {2013})},\ \Eprint {http://arxiv.org/abs/1212.4357}
  {arXiv:1212.4357 [gr-qc]} \BibitemShut {NoStop}%
\bibitem [{\citenamefont {Cotesta}\ \emph {et~al.}(2018)\citenamefont
  {Cotesta}, \citenamefont {Buonanno}, \citenamefont {Bohé}, \citenamefont
  {Taracchini}, \citenamefont {Hinder},\ and\ \citenamefont
  {Ossokine}}]{Cotesta:2018fcv}%
  \BibitemOpen
  \bibfield  {author} {\bibinfo {author} {\bibfnamefont {R.}~\bibnamefont
  {Cotesta}}, \bibinfo {author} {\bibfnamefont {A.}~\bibnamefont {Buonanno}},
  \bibinfo {author} {\bibfnamefont {A.}~\bibnamefont {Bohé}}, \bibinfo
  {author} {\bibfnamefont {A.}~\bibnamefont {Taracchini}}, \bibinfo {author}
  {\bibfnamefont {I.}~\bibnamefont {Hinder}}, \ and\ \bibinfo {author}
  {\bibfnamefont {S.}~\bibnamefont {Ossokine}},\ }\href {\doibase
  10.1103/PhysRevD.98.084028} {\bibfield  {journal} {\bibinfo  {journal} {Phys.
  Rev.}\ }\textbf {\bibinfo {volume} {D98}},\ \bibinfo {pages} {084028}
  (\bibinfo {year} {2018})},\ \Eprint {http://arxiv.org/abs/1803.10701}
  {arXiv:1803.10701 [gr-qc]} \BibitemShut {NoStop}%
\bibitem [{\citenamefont {Berti}\ \emph {et~al.}(2006)\citenamefont {Berti},
  \citenamefont {Cardoso},\ and\ \citenamefont {Will}}]{Berti_2006}%
  \BibitemOpen
  \bibfield  {author} {\bibinfo {author} {\bibfnamefont {E.}~\bibnamefont
  {Berti}}, \bibinfo {author} {\bibfnamefont {V.}~\bibnamefont {Cardoso}}, \
  and\ \bibinfo {author} {\bibfnamefont {C.~M.}\ \bibnamefont {Will}},\ }\href
  {\doibase 10.1103/physrevd.73.064030} {\bibfield  {journal} {\bibinfo
  {journal} {Physical Review D}\ }\textbf {\bibinfo {volume} {73}} (\bibinfo
  {year} {2006}),\ 10.1103/physrevd.73.064030}\BibitemShut {NoStop}%
\bibitem [{\citenamefont {{SXS Collaboration}}(2019)}]{SXS:catalog}%
  \BibitemOpen
  \bibfield  {author} {\bibinfo {author} {\bibnamefont {{SXS Collaboration}}},\
  }\href@noop {} {\enquote {\bibinfo {title} {{SXS Gravitational Waveform
  Database}},}\ }\bibinfo {howpublished}
  {\url{https://www.black-holes.org/waveforms}} (\bibinfo {year}
  {2019})\BibitemShut {NoStop}%
\bibitem [{\citenamefont {Boyle}\ \emph {et~al.}(2019)\citenamefont {Boyle}
  \emph {et~al.}}]{Boyle:2019kee}%
  \BibitemOpen
  \bibfield  {author} {\bibinfo {author} {\bibfnamefont {M.}~\bibnamefont
  {Boyle}} \emph {et~al.},\ }\href {\doibase 10.1088/1361-6382/ab34e2}
  {\bibfield  {journal} {\bibinfo  {journal} {Class. Quant. Grav.}\ }\textbf
  {\bibinfo {volume} {36}},\ \bibinfo {pages} {195006} (\bibinfo {year}
  {2019})},\ \Eprint {http://arxiv.org/abs/1904.04831} {arXiv:1904.04831
  [gr-qc]} \BibitemShut {NoStop}%
\bibitem [{\citenamefont {Khan}\ \emph {et~al.}(2020)\citenamefont {Khan},
  \citenamefont {Ohme}, \citenamefont {Chatziioannou},\ and\ \citenamefont
  {Hannam}}]{Khan:2019kot}%
  \BibitemOpen
  \bibfield  {author} {\bibinfo {author} {\bibfnamefont {S.}~\bibnamefont
  {Khan}}, \bibinfo {author} {\bibfnamefont {F.}~\bibnamefont {Ohme}}, \bibinfo
  {author} {\bibfnamefont {K.}~\bibnamefont {Chatziioannou}}, \ and\ \bibinfo
  {author} {\bibfnamefont {M.}~\bibnamefont {Hannam}},\ }\href {\doibase
  10.1103/PhysRevD.101.024056} {\bibfield  {journal} {\bibinfo  {journal}
  {Phys. Rev.}\ }\textbf {\bibinfo {volume} {D101}},\ \bibinfo {pages} {024056}
  (\bibinfo {year} {2020})},\ \Eprint {http://arxiv.org/abs/1911.06050}
  {arXiv:1911.06050 [gr-qc]} \BibitemShut {NoStop}%
\bibitem [{\citenamefont {Ossokine}\ \emph {et~al.}(2020)\citenamefont
  {Ossokine} \emph {et~al.}}]{Ossokine:2020kjp}%
  \BibitemOpen
  \bibfield  {author} {\bibinfo {author} {\bibfnamefont {S.}~\bibnamefont
  {Ossokine}} \emph {et~al.},\ }\href {\doibase 10.1103/PhysRevD.102.044055}
  {\bibfield  {journal} {\bibinfo  {journal} {Phys. Rev. D}\ }\textbf {\bibinfo
  {volume} {102}},\ \bibinfo {pages} {044055} (\bibinfo {year} {2020})},\
  \Eprint {http://arxiv.org/abs/2004.09442} {arXiv:2004.09442 [gr-qc]}
  \BibitemShut {NoStop}%
\bibitem [{\citenamefont {Husa}\ \emph {et~al.}(2016)\citenamefont {Husa},
  \citenamefont {Khan}, \citenamefont {Hannam}, \citenamefont {Pürrer},
  \citenamefont {Ohme}, \citenamefont {Jiménez~Forteza},\ and\ \citenamefont
  {Bohé}}]{Husa:2015iqa}%
  \BibitemOpen
  \bibfield  {author} {\bibinfo {author} {\bibfnamefont {S.}~\bibnamefont
  {Husa}}, \bibinfo {author} {\bibfnamefont {S.}~\bibnamefont {Khan}}, \bibinfo
  {author} {\bibfnamefont {M.}~\bibnamefont {Hannam}}, \bibinfo {author}
  {\bibfnamefont {M.}~\bibnamefont {Pürrer}}, \bibinfo {author} {\bibfnamefont
  {F.}~\bibnamefont {Ohme}}, \bibinfo {author} {\bibfnamefont {X.}~\bibnamefont
  {Jiménez~Forteza}}, \ and\ \bibinfo {author} {\bibfnamefont
  {A.}~\bibnamefont {Bohé}},\ }\href {\doibase 10.1103/PhysRevD.93.044006}
  {\bibfield  {journal} {\bibinfo  {journal} {Phys. Rev.}\ }\textbf {\bibinfo
  {volume} {D93}},\ \bibinfo {pages} {044006} (\bibinfo {year} {2016})},\
  \Eprint {http://arxiv.org/abs/1508.07250} {arXiv:1508.07250 [gr-qc]}
  \BibitemShut {NoStop}%
\bibitem [{\citenamefont {Khan}\ \emph {et~al.}(2016)\citenamefont {Khan},
  \citenamefont {Husa}, \citenamefont {Hannam}, \citenamefont {Ohme},
  \citenamefont {Pürrer}, \citenamefont {Jiménez~Forteza},\ and\
  \citenamefont {Bohé}}]{Khan:2015jqa}%
  \BibitemOpen
  \bibfield  {author} {\bibinfo {author} {\bibfnamefont {S.}~\bibnamefont
  {Khan}}, \bibinfo {author} {\bibfnamefont {S.}~\bibnamefont {Husa}}, \bibinfo
  {author} {\bibfnamefont {M.}~\bibnamefont {Hannam}}, \bibinfo {author}
  {\bibfnamefont {F.}~\bibnamefont {Ohme}}, \bibinfo {author} {\bibfnamefont
  {M.}~\bibnamefont {Pürrer}}, \bibinfo {author} {\bibfnamefont
  {X.}~\bibnamefont {Jiménez~Forteza}}, \ and\ \bibinfo {author}
  {\bibfnamefont {A.}~\bibnamefont {Bohé}},\ }\href {\doibase
  10.1103/PhysRevD.93.044007} {\bibfield  {journal} {\bibinfo  {journal} {Phys.
  Rev.}\ }\textbf {\bibinfo {volume} {D93}},\ \bibinfo {pages} {044007}
  (\bibinfo {year} {2016})},\ \Eprint {http://arxiv.org/abs/1508.07253}
  {arXiv:1508.07253 [gr-qc]} \BibitemShut {NoStop}%
\bibitem [{\citenamefont {Hannam}\ \emph {et~al.}(2014)\citenamefont {Hannam},
  \citenamefont {Schmidt}, \citenamefont {Bohé}, \citenamefont {Haegel},
  \citenamefont {Husa} \emph {et~al.}}]{Hannam:2013oca}%
  \BibitemOpen
  \bibfield  {author} {\bibinfo {author} {\bibfnamefont {M.}~\bibnamefont
  {Hannam}}, \bibinfo {author} {\bibfnamefont {P.}~\bibnamefont {Schmidt}},
  \bibinfo {author} {\bibfnamefont {A.}~\bibnamefont {Bohé}}, \bibinfo
  {author} {\bibfnamefont {L.}~\bibnamefont {Haegel}}, \bibinfo {author}
  {\bibfnamefont {S.}~\bibnamefont {Husa}},  \emph {et~al.},\ }\href {\doibase
  10.1103/PhysRevLett.113.151101} {\bibfield  {journal} {\bibinfo  {journal}
  {Phys.Rev.Lett.}\ }\textbf {\bibinfo {volume} {113}},\ \bibinfo {pages}
  {151101} (\bibinfo {year} {2014})},\ \Eprint {http://arxiv.org/abs/1308.3271}
  {arXiv:1308.3271 [gr-qc]} \BibitemShut {NoStop}%
\bibitem [{\citenamefont {Boh{\'e}}\ \emph {et~al.}(2016)\citenamefont
  {Boh{\'e}}, \citenamefont {Hannam}, \citenamefont {Husa}, \citenamefont
  {Ohme}, \citenamefont {Puerrer},\ and\ \citenamefont {Schmidt}}]{Bohe:PPv2}%
  \BibitemOpen
  \bibfield  {author} {\bibinfo {author} {\bibfnamefont {A.}~\bibnamefont
  {Boh{\'e}}}, \bibinfo {author} {\bibfnamefont {M.}~\bibnamefont {Hannam}},
  \bibinfo {author} {\bibfnamefont {S.}~\bibnamefont {Husa}}, \bibinfo {author}
  {\bibfnamefont {F.}~\bibnamefont {Ohme}}, \bibinfo {author} {\bibfnamefont
  {M.}~\bibnamefont {Puerrer}}, \ and\ \bibinfo {author} {\bibfnamefont
  {P.}~\bibnamefont {Schmidt}},\ }\href {https://dcc.ligo.org/LIGO-T1500602}
  {\emph {\bibinfo {title} {PhenomPv2 - Technical Notes for LAL
  Implementation}}},\ \bibinfo {type} {Tech. Rep.}\ \bibinfo {number}
  {{LIGO}-T1500602}\ (\bibinfo  {institution} {{LIGO} Project},\ \bibinfo
  {year} {2016})\BibitemShut {NoStop}%
\bibitem [{\citenamefont {London}\ \emph {et~al.}(2018)\citenamefont {London},
  \citenamefont {Khan}, \citenamefont {Fauchon-Jones}, \citenamefont
  {Garc{\'i}a}, \citenamefont {Hannam}, \citenamefont {Husa}, \citenamefont
  {Jiménez-Forteza}, \citenamefont {Kalaghatgi}, \citenamefont {Ohme},\ and\
  \citenamefont {Pannarale}}]{London:2017bcn}%
  \BibitemOpen
  \bibfield  {author} {\bibinfo {author} {\bibfnamefont {L.}~\bibnamefont
  {London}}, \bibinfo {author} {\bibfnamefont {S.}~\bibnamefont {Khan}},
  \bibinfo {author} {\bibfnamefont {E.}~\bibnamefont {Fauchon-Jones}}, \bibinfo
  {author} {\bibfnamefont {C.}~\bibnamefont {Garc{\'i}a}}, \bibinfo {author}
  {\bibfnamefont {M.}~\bibnamefont {Hannam}}, \bibinfo {author} {\bibfnamefont
  {S.}~\bibnamefont {Husa}}, \bibinfo {author} {\bibfnamefont {X.}~\bibnamefont
  {Jiménez-Forteza}}, \bibinfo {author} {\bibfnamefont {C.}~\bibnamefont
  {Kalaghatgi}}, \bibinfo {author} {\bibfnamefont {F.}~\bibnamefont {Ohme}}, \
  and\ \bibinfo {author} {\bibfnamefont {F.}~\bibnamefont {Pannarale}},\ }\href
  {\doibase 10.1103/PhysRevLett.120.161102} {\bibfield  {journal} {\bibinfo
  {journal} {Phys. Rev. Lett.}\ }\textbf {\bibinfo {volume} {120}},\ \bibinfo
  {pages} {161102} (\bibinfo {year} {2018})},\ \Eprint
  {http://arxiv.org/abs/1708.00404} {arXiv:1708.00404 [gr-qc]} \BibitemShut
  {NoStop}%
\bibitem [{\citenamefont {Khan}\ \emph {et~al.}(2019)\citenamefont {Khan},
  \citenamefont {Chatziioannou}, \citenamefont {Hannam},\ and\ \citenamefont
  {Ohme}}]{Khan:2018fmp}%
  \BibitemOpen
  \bibfield  {author} {\bibinfo {author} {\bibfnamefont {S.}~\bibnamefont
  {Khan}}, \bibinfo {author} {\bibfnamefont {K.}~\bibnamefont {Chatziioannou}},
  \bibinfo {author} {\bibfnamefont {M.}~\bibnamefont {Hannam}}, \ and\ \bibinfo
  {author} {\bibfnamefont {F.}~\bibnamefont {Ohme}},\ }\href {\doibase
  10.1103/PhysRevD.100.024059} {\bibfield  {journal} {\bibinfo  {journal}
  {Phys. Rev.}\ }\textbf {\bibinfo {volume} {D100}},\ \bibinfo {pages} {024059}
  (\bibinfo {year} {2019})},\ \Eprint {http://arxiv.org/abs/1809.10113}
  {arXiv:1809.10113 [gr-qc]} \BibitemShut {NoStop}%
\bibitem [{\citenamefont {{LIGO Scientific Collaboration}}(2020)}]{lalsuite}%
  \BibitemOpen
  \bibfield  {author} {\bibinfo {author} {\bibnamefont {{LIGO Scientific
  Collaboration}}},\ }\href {\doibase 10.7935/GT1W-FZ16} {\enquote {\bibinfo
  {title} {{LIGO} {A}lgorithm {L}ibrary - {LALS}uite},}\ }\bibinfo
  {howpublished} {free software (GPL), \url{https://doi.org/10.7935/GT1W-FZ16}}
  (\bibinfo {year} {2020})\BibitemShut {NoStop}%
\bibitem [{\citenamefont {Varma}\ \emph
  {et~al.}(2019{\natexlab{a}})\citenamefont {Varma}, \citenamefont {Field},
  \citenamefont {Scheel}, \citenamefont {Blackman}, \citenamefont {Gerosa},
  \citenamefont {Stein}, \citenamefont {Kidder},\ and\ \citenamefont
  {Pfeiffer}}]{Varma:2019csw}%
  \BibitemOpen
  \bibfield  {author} {\bibinfo {author} {\bibfnamefont {V.}~\bibnamefont
  {Varma}}, \bibinfo {author} {\bibfnamefont {S.~E.}\ \bibnamefont {Field}},
  \bibinfo {author} {\bibfnamefont {M.~A.}\ \bibnamefont {Scheel}}, \bibinfo
  {author} {\bibfnamefont {J.}~\bibnamefont {Blackman}}, \bibinfo {author}
  {\bibfnamefont {D.}~\bibnamefont {Gerosa}}, \bibinfo {author} {\bibfnamefont
  {L.~C.}\ \bibnamefont {Stein}}, \bibinfo {author} {\bibfnamefont {L.~E.}\
  \bibnamefont {Kidder}}, \ and\ \bibinfo {author} {\bibfnamefont {H.~P.}\
  \bibnamefont {Pfeiffer}},\ }\href {\doibase 10.1103/PhysRevResearch.1.033015}
  {\bibfield  {journal} {\bibinfo  {journal} {Phys. Rev. Research.}\ }\textbf
  {\bibinfo {volume} {1}},\ \bibinfo {pages} {033015} (\bibinfo {year}
  {2019}{\natexlab{a}})},\ \Eprint {http://arxiv.org/abs/1905.09300}
  {arXiv:1905.09300 [gr-qc]} \BibitemShut {NoStop}%
\bibitem [{\citenamefont {Islam}\ \emph {et~al.}(2020)\citenamefont {Islam},
  \citenamefont {Field}, \citenamefont {Haster},\ and\ \citenamefont
  {Smith}}]{GW190412:surro}%
  \BibitemOpen
  \bibfield  {author} {\bibinfo {author} {\bibfnamefont {T.}~\bibnamefont
  {Islam}}, \bibinfo {author} {\bibfnamefont {S.~E.}\ \bibnamefont {Field}},
  \bibinfo {author} {\bibfnamefont {C.-J.}\ \bibnamefont {Haster}}, \ and\
  \bibinfo {author} {\bibfnamefont {R.}~\bibnamefont {Smith}},\ }\href
  {https://dcc.ligo.org/LIGO-P2000384} {\  (\bibinfo {year}
  {2020})}\BibitemShut {NoStop}%
\bibitem [{\citenamefont {Pürrer}(2014)}]{P_rrer_2014}%
  \BibitemOpen
  \bibfield  {author} {\bibinfo {author} {\bibfnamefont {M.}~\bibnamefont
  {Pürrer}},\ }\href {\doibase 10.1088/0264-9381/31/19/195010} {\bibfield
  {journal} {\bibinfo  {journal} {Classical and Quantum Gravity}\ }\textbf
  {\bibinfo {volume} {31}},\ \bibinfo {pages} {195010} (\bibinfo {year}
  {2014})}\BibitemShut {NoStop}%
\bibitem [{\citenamefont {Ashton}\ \emph {et~al.}(2019)\citenamefont {Ashton}
  \emph {et~al.}}]{Ashton:2018jfp}%
  \BibitemOpen
  \bibfield  {author} {\bibinfo {author} {\bibfnamefont {G.}~\bibnamefont
  {Ashton}} \emph {et~al.},\ }\href {\doibase 10.3847/1538-4365/ab06fc}
  {\bibfield  {journal} {\bibinfo  {journal} {Astrophys. J. Suppl.}\ }\textbf
  {\bibinfo {volume} {241}},\ \bibinfo {pages} {27} (\bibinfo {year} {2019})},\
  \Eprint {http://arxiv.org/abs/1811.02042} {arXiv:1811.02042 [astro-ph.IM]}
  \BibitemShut {NoStop}%
\bibitem [{\citenamefont {Smith}\ \emph {et~al.}(2019)\citenamefont {Smith},
  \citenamefont {Ashton}, \citenamefont {Vajpeyi},\ and\ \citenamefont
  {Talbot}}]{Smith:2019ucc}%
  \BibitemOpen
  \bibfield  {author} {\bibinfo {author} {\bibfnamefont {R.}~\bibnamefont
  {Smith}}, \bibinfo {author} {\bibfnamefont {G.}~\bibnamefont {Ashton}},
  \bibinfo {author} {\bibfnamefont {A.}~\bibnamefont {Vajpeyi}}, \ and\
  \bibinfo {author} {\bibfnamefont {C.}~\bibnamefont {Talbot}},\ }\href@noop {}
  {\bibfield  {journal} {\bibinfo  {journal} {arXiv e-prints}\ } (\bibinfo
  {year} {2019})},\ \Eprint {http://arxiv.org/abs/1909.11873} {arXiv:1909.11873
  [gr-qc]} \BibitemShut {NoStop}%
\bibitem [{\citenamefont {Romero-Shaw}\ \emph {et~al.}(2020)\citenamefont
  {Romero-Shaw} \emph {et~al.}}]{Romero-Shaw:2020owr}%
  \BibitemOpen
  \bibfield  {author} {\bibinfo {author} {\bibfnamefont {I.}~\bibnamefont
  {Romero-Shaw}} \emph {et~al.},\ }\href {\doibase 10.1093/mnras/staa2850}
  {\bibfield  {journal} {\bibinfo  {journal} {Monthly Notices of the Royal
  Astronomical Society}\ } (\bibinfo {year} {2020}),\ 10.1093/mnras/staa2850},\
  \Eprint {http://arxiv.org/abs/2006.00714} {arXiv:2006.00714 [astro-ph.IM]}
  \BibitemShut {NoStop}%
\bibitem [{\citenamefont {Veitch}\ \emph {et~al.}(2015)\citenamefont {Veitch}
  \emph {et~al.}}]{Veitch:2014wba}%
  \BibitemOpen
  \bibfield  {author} {\bibinfo {author} {\bibfnamefont {J.}~\bibnamefont
  {Veitch}} \emph {et~al.},\ }\href {\doibase 10.1103/PhysRevD.91.042003}
  {\bibfield  {journal} {\bibinfo  {journal} {Phys. Rev.}\ }\textbf {\bibinfo
  {volume} {D91}},\ \bibinfo {pages} {042003} (\bibinfo {year} {2015})},\
  \Eprint {http://arxiv.org/abs/1409.7215} {arXiv:1409.7215 [gr-qc]}
  \BibitemShut {NoStop}%
\bibitem [{\citenamefont {Colleoni}\ \emph {et~al.}(2020)\citenamefont
  {Colleoni}, \citenamefont {Mateu-Lucena}, \citenamefont {Estellés},
  \citenamefont {García-Quirós}, \citenamefont {Keitel}, \citenamefont
  {Pratten}, \citenamefont {Ramos-Buades},\ and\ \citenamefont
  {Husa}}]{datarelease}%
  \BibitemOpen
  \bibfield  {author} {\bibinfo {author} {\bibfnamefont {M.}~\bibnamefont
  {Colleoni}}, \bibinfo {author} {\bibfnamefont {M.}~\bibnamefont
  {Mateu-Lucena}}, \bibinfo {author} {\bibfnamefont {H.}~\bibnamefont
  {Estellés}}, \bibinfo {author} {\bibfnamefont {C.}~\bibnamefont
  {García-Quirós}}, \bibinfo {author} {\bibfnamefont {D.}~\bibnamefont
  {Keitel}}, \bibinfo {author} {\bibfnamefont {G.}~\bibnamefont {Pratten}},
  \bibinfo {author} {\bibfnamefont {A.}~\bibnamefont {Ramos-Buades}}, \ and\
  \bibinfo {author} {\bibfnamefont {S.}~\bibnamefont {Husa}},\ }\href {\doibase
  10.5281/zenodo.4079188} {\enquote {\bibinfo {title} {{Data release for paper
  `Towards the routine use of subdominant harmonics in gravitational-wave
  inference: re-analysis of GW190412 with generation X waveform models'}},}\
  }\bibinfo {howpublished} {Zenodo.org deposit} (\bibinfo {year}
  {2020})\BibitemShut {NoStop}%
\bibitem [{\citenamefont {{Planck Collaboration}}\ \emph
  {et~al.}(2016)\citenamefont {{Planck Collaboration}}, \citenamefont {{Ade}},
  \citenamefont {{Aghanim}}, \citenamefont {{Arnaud}}, \citenamefont
  {{Ashdown}}, \citenamefont {{Aumont}}, \citenamefont {{Baccigalupi}},
  \citenamefont {{Banday}}, \citenamefont {{Barreiro}}, \citenamefont
  {{Bartlett}} \emph {et~al.}}]{Planck2015}%
  \BibitemOpen
  \bibfield  {author} {\bibinfo {author} {\bibnamefont {{Planck
  Collaboration}}}, \bibinfo {author} {\bibfnamefont {P.~A.~R.}\ \bibnamefont
  {{Ade}}}, \bibinfo {author} {\bibfnamefont {N.}~\bibnamefont {{Aghanim}}},
  \bibinfo {author} {\bibfnamefont {M.}~\bibnamefont {{Arnaud}}}, \bibinfo
  {author} {\bibfnamefont {M.}~\bibnamefont {{Ashdown}}}, \bibinfo {author}
  {\bibfnamefont {J.}~\bibnamefont {{Aumont}}}, \bibinfo {author}
  {\bibfnamefont {C.}~\bibnamefont {{Baccigalupi}}}, \bibinfo {author}
  {\bibfnamefont {A.~J.}\ \bibnamefont {{Banday}}}, \bibinfo {author}
  {\bibfnamefont {R.~B.}\ \bibnamefont {{Barreiro}}}, \bibinfo {author}
  {\bibfnamefont {J.~G.}\ \bibnamefont {{Bartlett}}},  \emph {et~al.},\ }\href
  {\doibase 10.1051/0004-6361/201525830} {\bibfield  {journal} {\bibinfo
  {journal} {A\&A}\ }\textbf {\bibinfo {volume} {594}},\ \bibinfo {eid} {A13}
  (\bibinfo {year} {2016})},\ \Eprint {http://arxiv.org/abs/1502.01589}
  {arXiv:1502.01589 [astro-ph.CO]} \BibitemShut {NoStop}%
\bibitem [{\citenamefont {{Ajith}}\ \emph {et~al.}(2011)\citenamefont
  {{Ajith}}, \citenamefont {{Hannam}}, \citenamefont {{Husa}}, \citenamefont
  {{Chen}}, \citenamefont {{Br{\"u}gmann}}, \citenamefont {{Dorband}},
  \citenamefont {{M{\"u}ller}}, \citenamefont {{Ohme}}, \citenamefont
  {{Pollney}}, \citenamefont {{Reisswig}}, \citenamefont {{Santamar{\'i}a}},\
  and\ \citenamefont {{Seiler}}}]{Ajith:2009bn}%
  \BibitemOpen
  \bibfield  {author} {\bibinfo {author} {\bibfnamefont {P.}~\bibnamefont
  {{Ajith}}}, \bibinfo {author} {\bibfnamefont {M.}~\bibnamefont {{Hannam}}},
  \bibinfo {author} {\bibfnamefont {S.}~\bibnamefont {{Husa}}}, \bibinfo
  {author} {\bibfnamefont {Y.}~\bibnamefont {{Chen}}}, \bibinfo {author}
  {\bibfnamefont {B.}~\bibnamefont {{Br{\"u}gmann}}}, \bibinfo {author}
  {\bibfnamefont {N.}~\bibnamefont {{Dorband}}}, \bibinfo {author}
  {\bibfnamefont {D.}~\bibnamefont {{M{\"u}ller}}}, \bibinfo {author}
  {\bibfnamefont {F.}~\bibnamefont {{Ohme}}}, \bibinfo {author} {\bibfnamefont
  {D.}~\bibnamefont {{Pollney}}}, \bibinfo {author} {\bibfnamefont
  {C.}~\bibnamefont {{Reisswig}}}, \bibinfo {author} {\bibfnamefont
  {L.}~\bibnamefont {{Santamar{\'i}a}}}, \ and\ \bibinfo {author}
  {\bibfnamefont {J.}~\bibnamefont {{Seiler}}},\ }\href {\doibase
  10.1103/PhysRevLett.106.241101} {\bibfield  {journal} {\bibinfo  {journal}
  {Physical Review Letters}\ }\textbf {\bibinfo {volume} {106}},\ \bibinfo
  {eid} {241101} (\bibinfo {year} {2011})}\BibitemShut {NoStop}%
\bibitem [{\citenamefont {{Santamaria}}\ \emph {et~al.}(2010)\citenamefont
  {{Santamaria}}, \citenamefont {{Ohme}}, \citenamefont {{Ajith}},
  \citenamefont {{Bruegmann}}, \citenamefont {{Dorband}}, \citenamefont
  {{Hannam}}, \citenamefont {{Husa}}, \citenamefont {{Moesta}}, \citenamefont
  {{Pollney}}, \citenamefont {{Reisswig}} \emph {et~al.}}]{Santamaria:2010yb}%
  \BibitemOpen
  \bibfield  {author} {\bibinfo {author} {\bibfnamefont {L.}~\bibnamefont
  {{Santamaria}}}, \bibinfo {author} {\bibfnamefont {F.}~\bibnamefont
  {{Ohme}}}, \bibinfo {author} {\bibfnamefont {P.}~\bibnamefont {{Ajith}}},
  \bibinfo {author} {\bibfnamefont {B.}~\bibnamefont {{Bruegmann}}}, \bibinfo
  {author} {\bibfnamefont {N.}~\bibnamefont {{Dorband}}}, \bibinfo {author}
  {\bibfnamefont {M.}~\bibnamefont {{Hannam}}}, \bibinfo {author}
  {\bibfnamefont {S.}~\bibnamefont {{Husa}}}, \bibinfo {author} {\bibfnamefont
  {P.}~\bibnamefont {{Moesta}}}, \bibinfo {author} {\bibfnamefont
  {D.}~\bibnamefont {{Pollney}}}, \bibinfo {author} {\bibfnamefont {E.~L.}\
  \bibnamefont {{Reisswig}}},  \emph {et~al.},\ }\href {\doibase
  10.1103/PhysRevD.82.064016} {\bibfield  {journal} {\bibinfo  {journal} {Phys.
  Rev.}\ }\textbf {\bibinfo {volume} {D82}},\ \bibinfo {pages} {064016}
  (\bibinfo {year} {2010})}\BibitemShut {NoStop}%
\bibitem [{\citenamefont {Schmidt}\ \emph {et~al.}(2015)\citenamefont
  {Schmidt}, \citenamefont {Ohme},\ and\ \citenamefont
  {Hannam}}]{Schmidt:2014iyl}%
  \BibitemOpen
  \bibfield  {author} {\bibinfo {author} {\bibfnamefont {P.}~\bibnamefont
  {Schmidt}}, \bibinfo {author} {\bibfnamefont {F.}~\bibnamefont {Ohme}}, \
  and\ \bibinfo {author} {\bibfnamefont {M.}~\bibnamefont {Hannam}},\ }\href
  {\doibase 10.1103/PhysRevD.91.024043} {\bibfield  {journal} {\bibinfo
  {journal} {Phys. Rev.}\ }\textbf {\bibinfo {volume} {D91}},\ \bibinfo {pages}
  {024043} (\bibinfo {year} {2015})},\ \Eprint {http://arxiv.org/abs/1408.1810}
  {arXiv:1408.1810 [gr-qc]} \BibitemShut {NoStop}%
\bibitem [{\citenamefont {Lange}\ \emph {et~al.}(2018)\citenamefont {Lange},
  \citenamefont {O'Shaughnessy},\ and\ \citenamefont {Rizzo}}]{Lange:2018pyp}%
  \BibitemOpen
  \bibfield  {author} {\bibinfo {author} {\bibfnamefont {J.}~\bibnamefont
  {Lange}}, \bibinfo {author} {\bibfnamefont {R.}~\bibnamefont
  {O'Shaughnessy}}, \ and\ \bibinfo {author} {\bibfnamefont {M.}~\bibnamefont
  {Rizzo}},\ }\href@noop {} {\bibfield  {journal} {\bibinfo  {journal} {arXiv
  e-prints}\ } (\bibinfo {year} {2018})},\ \Eprint
  {http://arxiv.org/abs/1805.10457} {arXiv:1805.10457 [gr-qc]} \BibitemShut
  {NoStop}%
\bibitem [{\citenamefont {Wysocki}\ \emph {et~al.}(2019)\citenamefont
  {Wysocki}, \citenamefont {O'Shaughnessy}, \citenamefont {Lange},\ and\
  \citenamefont {Fang}}]{Wysocki:2019grj}%
  \BibitemOpen
  \bibfield  {author} {\bibinfo {author} {\bibfnamefont {D.}~\bibnamefont
  {Wysocki}}, \bibinfo {author} {\bibfnamefont {R.}~\bibnamefont
  {O'Shaughnessy}}, \bibinfo {author} {\bibfnamefont {J.}~\bibnamefont
  {Lange}}, \ and\ \bibinfo {author} {\bibfnamefont {Y.-L.~L.}\ \bibnamefont
  {Fang}},\ }\href {\doibase 10.1103/PhysRevD.99.084026} {\bibfield  {journal}
  {\bibinfo  {journal} {Phys. Rev. D}\ }\textbf {\bibinfo {volume} {99}},\
  \bibinfo {pages} {084026} (\bibinfo {year} {2019})},\ \Eprint
  {http://arxiv.org/abs/1902.04934} {arXiv:1902.04934 [astro-ph.IM]}
  \BibitemShut {NoStop}%
\bibitem [{\citenamefont {{Postnov}}\ and\ \citenamefont
  {{Yungelson}}(2014)}]{Postnov:2014lrr}%
  \BibitemOpen
  \bibfield  {author} {\bibinfo {author} {\bibfnamefont {K.~A.}\ \bibnamefont
  {{Postnov}}}\ and\ \bibinfo {author} {\bibfnamefont {L.~R.}\ \bibnamefont
  {{Yungelson}}},\ }\href {\doibase 10.12942/lrr-2014-3} {\bibfield  {journal}
  {\bibinfo  {journal} {Living Rel. Rev.}\ }\textbf {\bibinfo {volume} {17}},\
  \bibinfo {eid} {3} (\bibinfo {year} {2014})}\BibitemShut {NoStop}%
\bibitem [{\citenamefont {{Benacquista}}\ and\ \citenamefont
  {{Downing}}(2013)}]{Benacquista:2013lrr}%
  \BibitemOpen
  \bibfield  {author} {\bibinfo {author} {\bibfnamefont {M.~J.}\ \bibnamefont
  {{Benacquista}}}\ and\ \bibinfo {author} {\bibfnamefont {J.~M.~B.}\
  \bibnamefont {{Downing}}},\ }\href {\doibase 10.12942/lrr-2013-4} {\bibfield
  {journal} {\bibinfo  {journal} {Living Rel. Rev.}\ }\textbf {\bibinfo
  {volume} {16}},\ \bibinfo {eid} {4} (\bibinfo {year} {2013})}\BibitemShut
  {NoStop}%
\bibitem [{\citenamefont {Zevin}\ \emph {et~al.}(2020)\citenamefont {Zevin},
  \citenamefont {Berry}, \citenamefont {Coughlin}, \citenamefont
  {Chatziioannou},\ and\ \citenamefont {Vitale}}]{Zevin:2020gxf}%
  \BibitemOpen
  \bibfield  {author} {\bibinfo {author} {\bibfnamefont {M.}~\bibnamefont
  {Zevin}}, \bibinfo {author} {\bibfnamefont {C.~P.}\ \bibnamefont {Berry}},
  \bibinfo {author} {\bibfnamefont {S.}~\bibnamefont {Coughlin}}, \bibinfo
  {author} {\bibfnamefont {K.}~\bibnamefont {Chatziioannou}}, \ and\ \bibinfo
  {author} {\bibfnamefont {S.}~\bibnamefont {Vitale}},\ }\href {\doibase
  10.3847/2041-8213/aba8ef} {\bibfield  {journal} {\bibinfo  {journal}
  {Astrophys. J. Lett.}\ }\textbf {\bibinfo {volume} {899}},\ \bibinfo {pages}
  {L17} (\bibinfo {year} {2020})},\ \Eprint {http://arxiv.org/abs/2006.11293}
  {arXiv:2006.11293 [astro-ph.HE]} \BibitemShut {NoStop}%
\bibitem [{\citenamefont {Boh\'{e}}\ \emph {et~al.}(2017)\citenamefont
  {Boh\'{e}} \emph {et~al.}}]{Bohe:2016gbl}%
  \BibitemOpen
  \bibfield  {author} {\bibinfo {author} {\bibfnamefont {A.}~\bibnamefont
  {Boh\'{e}}} \emph {et~al.},\ }\href {\doibase 10.1103/PhysRevD.95.044028}
  {\bibfield  {journal} {\bibinfo  {journal} {Phys. Rev.}\ }\textbf {\bibinfo
  {volume} {D95}},\ \bibinfo {pages} {044028} (\bibinfo {year} {2017})},\
  \Eprint {http://arxiv.org/abs/1611.03703} {arXiv:1611.03703 [gr-qc]}
  \BibitemShut {NoStop}%
\bibitem [{\citenamefont {Cotesta}\ \emph {et~al.}(2020)\citenamefont
  {Cotesta}, \citenamefont {Marsat},\ and\ \citenamefont
  {P\"urrer}}]{Cotesta:2020qhw}%
  \BibitemOpen
  \bibfield  {author} {\bibinfo {author} {\bibfnamefont {R.}~\bibnamefont
  {Cotesta}}, \bibinfo {author} {\bibfnamefont {S.}~\bibnamefont {Marsat}}, \
  and\ \bibinfo {author} {\bibfnamefont {M.}~\bibnamefont {P\"urrer}},\ }\href
  {\doibase 10.1103/PhysRevD.101.124040} {\bibfield  {journal} {\bibinfo
  {journal} {Phys. Rev. D}\ }\textbf {\bibinfo {volume} {101}},\ \bibinfo
  {pages} {124040} (\bibinfo {year} {2020})},\ \Eprint
  {http://arxiv.org/abs/2003.12079} {arXiv:2003.12079 [gr-qc]} \BibitemShut
  {NoStop}%
\bibitem [{\citenamefont {Babak}\ \emph {et~al.}(2017)\citenamefont {Babak},
  \citenamefont {Taracchini},\ and\ \citenamefont {Buonanno}}]{Babak:2016tgq}%
  \BibitemOpen
  \bibfield  {author} {\bibinfo {author} {\bibfnamefont {S.}~\bibnamefont
  {Babak}}, \bibinfo {author} {\bibfnamefont {A.}~\bibnamefont {Taracchini}}, \
  and\ \bibinfo {author} {\bibfnamefont {A.}~\bibnamefont {Buonanno}},\ }\href
  {\doibase 10.1103/PhysRevD.95.024010} {\bibfield  {journal} {\bibinfo
  {journal} {Phys. Rev.}\ }\textbf {\bibinfo {volume} {D95}},\ \bibinfo {pages}
  {024010} (\bibinfo {year} {2017})},\ \Eprint
  {http://arxiv.org/abs/1607.05661} {arXiv:1607.05661 [gr-qc]} \BibitemShut
  {NoStop}%
\bibitem [{\citenamefont {Pan}\ \emph {et~al.}(2014)\citenamefont {Pan},
  \citenamefont {Buonanno}, \citenamefont {Taracchini}, \citenamefont {Kidder},
  \citenamefont {Mroué}, \citenamefont {Pfeiffer}, \citenamefont {Scheel},\
  and\ \citenamefont {Szilágyi}}]{Pan:2013rra}%
  \BibitemOpen
  \bibfield  {author} {\bibinfo {author} {\bibfnamefont {Y.}~\bibnamefont
  {Pan}}, \bibinfo {author} {\bibfnamefont {A.}~\bibnamefont {Buonanno}},
  \bibinfo {author} {\bibfnamefont {A.}~\bibnamefont {Taracchini}}, \bibinfo
  {author} {\bibfnamefont {L.~E.}\ \bibnamefont {Kidder}}, \bibinfo {author}
  {\bibfnamefont {A.~H.}\ \bibnamefont {Mroué}}, \bibinfo {author}
  {\bibfnamefont {H.~P.}\ \bibnamefont {Pfeiffer}}, \bibinfo {author}
  {\bibfnamefont {M.~A.}\ \bibnamefont {Scheel}}, \ and\ \bibinfo {author}
  {\bibfnamefont {B.}~\bibnamefont {Szilágyi}},\ }\href {\doibase
  10.1103/PhysRevD.89.084006} {\bibfield  {journal} {\bibinfo  {journal} {Phys.
  Rev.}\ }\textbf {\bibinfo {volume} {D89}},\ \bibinfo {pages} {084006}
  (\bibinfo {year} {2014})},\ \Eprint {http://arxiv.org/abs/1307.6232}
  {arXiv:1307.6232 [gr-qc]} \BibitemShut {NoStop}%
\bibitem [{\citenamefont {Varma}\ \emph
  {et~al.}(2019{\natexlab{b}})\citenamefont {Varma}, \citenamefont {Field},
  \citenamefont {Scheel}, \citenamefont {Blackman}, \citenamefont {Kidder},\
  and\ \citenamefont {Pfeiffer}}]{Varma:2018mmi}%
  \BibitemOpen
  \bibfield  {author} {\bibinfo {author} {\bibfnamefont {V.}~\bibnamefont
  {Varma}}, \bibinfo {author} {\bibfnamefont {S.~E.}\ \bibnamefont {Field}},
  \bibinfo {author} {\bibfnamefont {M.~A.}\ \bibnamefont {Scheel}}, \bibinfo
  {author} {\bibfnamefont {J.}~\bibnamefont {Blackman}}, \bibinfo {author}
  {\bibfnamefont {L.~E.}\ \bibnamefont {Kidder}}, \ and\ \bibinfo {author}
  {\bibfnamefont {H.~P.}\ \bibnamefont {Pfeiffer}},\ }\href {\doibase
  10.1103/PhysRevD.99.064045} {\bibfield  {journal} {\bibinfo  {journal} {Phys.
  Rev.}\ }\textbf {\bibinfo {volume} {D99}},\ \bibinfo {pages} {064045}
  (\bibinfo {year} {2019}{\natexlab{b}})},\ \Eprint
  {http://arxiv.org/abs/1812.07865} {arXiv:1812.07865 [gr-qc]} \BibitemShut
  {NoStop}%
\bibitem [{\citenamefont {Ajith}\ \emph {et~al.}(2007)\citenamefont {Ajith},
  \citenamefont {Babak}, \citenamefont {Chen}, \citenamefont {Hewitson},
  \citenamefont {Krishnan}, \citenamefont {Whelan}, \citenamefont {Brügmann},
  \citenamefont {Diener}, \citenamefont {Gonzalez}, \citenamefont {Hannam},\
  and\ \citenamefont {et~al.}}]{Ajith_2007}%
  \BibitemOpen
  \bibfield  {author} {\bibinfo {author} {\bibfnamefont {P.}~\bibnamefont
  {Ajith}}, \bibinfo {author} {\bibfnamefont {S.}~\bibnamefont {Babak}},
  \bibinfo {author} {\bibfnamefont {Y.}~\bibnamefont {Chen}}, \bibinfo {author}
  {\bibfnamefont {M.}~\bibnamefont {Hewitson}}, \bibinfo {author}
  {\bibfnamefont {B.}~\bibnamefont {Krishnan}}, \bibinfo {author}
  {\bibfnamefont {J.~T.}\ \bibnamefont {Whelan}}, \bibinfo {author}
  {\bibfnamefont {B.}~\bibnamefont {Brügmann}}, \bibinfo {author}
  {\bibfnamefont {P.}~\bibnamefont {Diener}}, \bibinfo {author} {\bibfnamefont
  {J.}~\bibnamefont {Gonzalez}}, \bibinfo {author} {\bibfnamefont
  {M.}~\bibnamefont {Hannam}}, \ and\ \bibinfo {author} {\bibnamefont
  {et~al.}},\ }\href {\doibase 10.1088/0264-9381/24/19/s31} {\bibfield
  {journal} {\bibinfo  {journal} {Classical and Quantum Gravity}\ }\textbf
  {\bibinfo {volume} {24}},\ \bibinfo {pages} {S689–S699} (\bibinfo {year}
  {2007})}\BibitemShut {NoStop}%
\bibitem [{\citenamefont {Varma}\ \emph
  {et~al.}(2019{\natexlab{c}})\citenamefont {Varma}, \citenamefont {Field},
  \citenamefont {Scheel}, \citenamefont {Blackman}, \citenamefont {Kidder},\
  and\ \citenamefont {Pfeiffer}}]{Varma_2019}%
  \BibitemOpen
  \bibfield  {author} {\bibinfo {author} {\bibfnamefont {V.}~\bibnamefont
  {Varma}}, \bibinfo {author} {\bibfnamefont {S.~E.}\ \bibnamefont {Field}},
  \bibinfo {author} {\bibfnamefont {M.~A.}\ \bibnamefont {Scheel}}, \bibinfo
  {author} {\bibfnamefont {J.}~\bibnamefont {Blackman}}, \bibinfo {author}
  {\bibfnamefont {L.~E.}\ \bibnamefont {Kidder}}, \ and\ \bibinfo {author}
  {\bibfnamefont {H.~P.}\ \bibnamefont {Pfeiffer}},\ }\href {\doibase
  10.1103/physrevd.99.064045} {\bibfield  {journal} {\bibinfo  {journal}
  {Physical Review D}\ }\textbf {\bibinfo {volume} {99}} (\bibinfo {year}
  {2019}{\natexlab{c}}),\ 10.1103/physrevd.99.064045}\BibitemShut {NoStop}%
\bibitem [{\citenamefont {Schmidt}\ \emph {et~al.}(2011)\citenamefont
  {Schmidt}, \citenamefont {Hannam}, \citenamefont {Husa},\ and\ \citenamefont
  {Ajith}}]{Schmidt:2010it}%
  \BibitemOpen
  \bibfield  {author} {\bibinfo {author} {\bibfnamefont {P.}~\bibnamefont
  {Schmidt}}, \bibinfo {author} {\bibfnamefont {M.}~\bibnamefont {Hannam}},
  \bibinfo {author} {\bibfnamefont {S.}~\bibnamefont {Husa}}, \ and\ \bibinfo
  {author} {\bibfnamefont {P.}~\bibnamefont {Ajith}},\ }\href {\doibase
  10.1103/PhysRevD.84.024046} {\bibfield  {journal} {\bibinfo  {journal} {Phys.
  Rev.}\ }\textbf {\bibinfo {volume} {D84}},\ \bibinfo {pages} {024046}
  (\bibinfo {year} {2011})},\ \Eprint {http://arxiv.org/abs/1012.2879}
  {arXiv:1012.2879 [gr-qc]} \BibitemShut {NoStop}%
\bibitem [{\citenamefont {Schmidt}\ \emph {et~al.}(2012)\citenamefont
  {Schmidt}, \citenamefont {Hannam},\ and\ \citenamefont
  {Husa}}]{Schmidt:2012rh}%
  \BibitemOpen
  \bibfield  {author} {\bibinfo {author} {\bibfnamefont {P.}~\bibnamefont
  {Schmidt}}, \bibinfo {author} {\bibfnamefont {M.}~\bibnamefont {Hannam}}, \
  and\ \bibinfo {author} {\bibfnamefont {S.}~\bibnamefont {Husa}},\ }\href
  {\doibase 10.1103/PhysRevD.86.104063} {\bibfield  {journal} {\bibinfo
  {journal} {Phys. Rev.}\ }\textbf {\bibinfo {volume} {D86}},\ \bibinfo {pages}
  {104063} (\bibinfo {year} {2012})},\ \Eprint {http://arxiv.org/abs/1207.3088}
  {arXiv:1207.3088 [gr-qc]} \BibitemShut {NoStop}%
\bibitem [{\citenamefont {Chatziioannou}\ \emph {et~al.}(2017)\citenamefont
  {Chatziioannou}, \citenamefont {Klein}, \citenamefont {Yunes},\ and\
  \citenamefont {Cornish}}]{Chatziioannou:2017tdw}%
  \BibitemOpen
  \bibfield  {author} {\bibinfo {author} {\bibfnamefont {K.}~\bibnamefont
  {Chatziioannou}}, \bibinfo {author} {\bibfnamefont {A.}~\bibnamefont
  {Klein}}, \bibinfo {author} {\bibfnamefont {N.}~\bibnamefont {Yunes}}, \ and\
  \bibinfo {author} {\bibfnamefont {N.}~\bibnamefont {Cornish}},\ }\href
  {\doibase 10.1103/PhysRevD.95.104004} {\bibfield  {journal} {\bibinfo
  {journal} {Phys. Rev.}\ }\textbf {\bibinfo {volume} {D95}},\ \bibinfo {pages}
  {104004} (\bibinfo {year} {2017})},\ \Eprint
  {http://arxiv.org/abs/1703.03967} {arXiv:1703.03967 [gr-qc]} \BibitemShut
  {NoStop}%
\bibitem [{\citenamefont {Vinciguerra}\ \emph {et~al.}(2017)\citenamefont
  {Vinciguerra}, \citenamefont {Veitch},\ and\ \citenamefont
  {Mandel}}]{Vinciguerra:2017ngf}%
  \BibitemOpen
  \bibfield  {author} {\bibinfo {author} {\bibfnamefont {S.}~\bibnamefont
  {Vinciguerra}}, \bibinfo {author} {\bibfnamefont {J.}~\bibnamefont {Veitch}},
  \ and\ \bibinfo {author} {\bibfnamefont {I.}~\bibnamefont {Mandel}},\ }\href
  {\doibase 10.1088/1361-6382/aa6d44} {\bibfield  {journal} {\bibinfo
  {journal} {Class. Quant. Grav.}\ }\textbf {\bibinfo {volume} {34}},\ \bibinfo
  {pages} {115006} (\bibinfo {year} {2017})},\ \Eprint
  {http://arxiv.org/abs/1703.02062} {arXiv:1703.02062 [gr-qc]} \BibitemShut
  {NoStop}%
\bibitem [{\citenamefont {Buonanno}\ \emph {et~al.}(2009)\citenamefont
  {Buonanno}, \citenamefont {Iyer}, \citenamefont {Ochsner}, \citenamefont
  {Pan},\ and\ \citenamefont {Sathyaprakash}}]{Buonanno_2009}%
  \BibitemOpen
  \bibfield  {author} {\bibinfo {author} {\bibfnamefont {A.}~\bibnamefont
  {Buonanno}}, \bibinfo {author} {\bibfnamefont {B.~R.}\ \bibnamefont {Iyer}},
  \bibinfo {author} {\bibfnamefont {E.}~\bibnamefont {Ochsner}}, \bibinfo
  {author} {\bibfnamefont {Y.}~\bibnamefont {Pan}}, \ and\ \bibinfo {author}
  {\bibfnamefont {B.~S.}\ \bibnamefont {Sathyaprakash}},\ }\href {\doibase
  10.1103/physrevd.80.084043} {\bibfield  {journal} {\bibinfo  {journal}
  {Physical Review D}\ }\textbf {\bibinfo {volume} {80}} (\bibinfo {year}
  {2009}),\ 10.1103/physrevd.80.084043}\BibitemShut {NoStop}%
\bibitem [{\citenamefont {Damour}\ and\ \citenamefont
  {Nagar}(2014)}]{Damour_2014}%
  \BibitemOpen
  \bibfield  {author} {\bibinfo {author} {\bibfnamefont {T.}~\bibnamefont
  {Damour}}\ and\ \bibinfo {author} {\bibfnamefont {A.}~\bibnamefont {Nagar}},\
  }\href {\doibase 10.1103/physrevd.90.024054} {\bibfield  {journal} {\bibinfo
  {journal} {Physical Review D}\ }\textbf {\bibinfo {volume} {90}} (\bibinfo
  {year} {2014}),\ 10.1103/physrevd.90.024054}\BibitemShut {NoStop}%
\bibitem [{\citenamefont {{GW Open Science Center}}(2020)}]{GW190412:gwosc-v2}%
  \BibitemOpen
  \bibfield  {author} {\bibinfo {author} {\bibnamefont {{GW Open Science
  Center}}},\ }\href {\doibase 10.7935/20yv-ka61} {\enquote {\bibinfo {title}
  {{GW190412 strain data release v2}},}\ }\bibinfo {howpublished}
  {\url{https://doi.org/10.7935/20yv-ka61}} (\bibinfo {year}
  {2020})\BibitemShut {NoStop}%
\bibitem [{\citenamefont {Vallisneri}\ \emph {et~al.}(2015)\citenamefont
  {Vallisneri}, \citenamefont {Kanner}, \citenamefont {Williams}, \citenamefont
  {Weinstein},\ and\ \citenamefont {Stephens}}]{Vallisneri:2014vxa}%
  \BibitemOpen
  \bibfield  {author} {\bibinfo {author} {\bibfnamefont {M.}~\bibnamefont
  {Vallisneri}}, \bibinfo {author} {\bibfnamefont {J.}~\bibnamefont {Kanner}},
  \bibinfo {author} {\bibfnamefont {R.}~\bibnamefont {Williams}}, \bibinfo
  {author} {\bibfnamefont {A.}~\bibnamefont {Weinstein}}, \ and\ \bibinfo
  {author} {\bibfnamefont {B.}~\bibnamefont {Stephens}},\ }\bibfield
  {booktitle} {\emph {\bibinfo {booktitle} {{Proceedings, 10th International
  LISA Symposium: Gainesville, Florida, USA, May 18-23, 2014}}},\ }\href
  {\doibase 10.1088/1742-6596/610/1/012021} {\bibfield  {journal} {\bibinfo
  {journal} {J. Phys. Conf. Ser.}\ }\textbf {\bibinfo {volume} {610}},\
  \bibinfo {pages} {012021} (\bibinfo {year} {2015})},\ \Eprint
  {http://arxiv.org/abs/1410.4839} {arXiv:1410.4839 [gr-qc]} \BibitemShut
  {NoStop}%
\bibitem [{\citenamefont {Abbott}\ \emph
  {et~al.}(2019{\natexlab{c}})\citenamefont {Abbott} \emph
  {et~al.}}]{Abbott:2019ebz}%
  \BibitemOpen
  \bibfield  {author} {\bibinfo {author} {\bibfnamefont {R.}~\bibnamefont
  {Abbott}} \emph {et~al.} (\bibinfo {collaboration} {LIGO Scientific,
  Virgo}),\ }\href@noop {} {\bibfield  {journal} {\bibinfo  {journal} {ArXiv
  e-prints}\ } (\bibinfo {year} {2019}{\natexlab{c}})},\ \Eprint
  {http://arxiv.org/abs/1912.11716} {arXiv:1912.11716 [gr-qc]} \BibitemShut
  {NoStop}%
\bibitem [{\citenamefont {Vajente}\ \emph {et~al.}(2020)\citenamefont
  {Vajente}, \citenamefont {Huang}, \citenamefont {Isi}, \citenamefont
  {Driggers}, \citenamefont {Kissel}, \citenamefont {Szczepanczyk},\ and\
  \citenamefont {Vitale}}]{Vajente:2019ycy}%
  \BibitemOpen
  \bibfield  {author} {\bibinfo {author} {\bibfnamefont {G.}~\bibnamefont
  {Vajente}}, \bibinfo {author} {\bibfnamefont {Y.}~\bibnamefont {Huang}},
  \bibinfo {author} {\bibfnamefont {M.}~\bibnamefont {Isi}}, \bibinfo {author}
  {\bibfnamefont {J.~C.}\ \bibnamefont {Driggers}}, \bibinfo {author}
  {\bibfnamefont {J.~S.}\ \bibnamefont {Kissel}}, \bibinfo {author}
  {\bibfnamefont {M.~J.}\ \bibnamefont {Szczepanczyk}}, \ and\ \bibinfo
  {author} {\bibfnamefont {S.}~\bibnamefont {Vitale}},\ }\href {\doibase
  10.1103/PhysRevD.101.042003} {\bibfield  {journal} {\bibinfo  {journal}
  {Phys. Rev. D}\ }\textbf {\bibinfo {volume} {101}},\ \bibinfo {pages}
  {042003} (\bibinfo {year} {2020})},\ \Eprint
  {http://arxiv.org/abs/1911.09083} {arXiv:1911.09083 [gr-qc]} \BibitemShut
  {NoStop}%
\bibitem [{\citenamefont {Littenberg}\ and\ \citenamefont
  {Cornish}(2015)}]{Littenberg:2014oda}%
  \BibitemOpen
  \bibfield  {author} {\bibinfo {author} {\bibfnamefont {T.~B.}\ \bibnamefont
  {Littenberg}}\ and\ \bibinfo {author} {\bibfnamefont {N.~J.}\ \bibnamefont
  {Cornish}},\ }\href {\doibase 10.1103/PhysRevD.91.084034} {\bibfield
  {journal} {\bibinfo  {journal} {Phys. Rev. D}\ }\textbf {\bibinfo {volume}
  {91}},\ \bibinfo {pages} {084034} (\bibinfo {year} {2015})}\BibitemShut
  {NoStop}%
\bibitem [{\citenamefont {Chatziioannou}\ \emph {et~al.}(2019)\citenamefont
  {Chatziioannou}, \citenamefont {Haster}, \citenamefont {Littenberg},
  \citenamefont {Farr}, \citenamefont {Ghonge}, \citenamefont {Millhouse},
  \citenamefont {Clark},\ and\ \citenamefont
  {Cornish}}]{Chatziioannou:2019zvs}%
  \BibitemOpen
  \bibfield  {author} {\bibinfo {author} {\bibfnamefont {K.}~\bibnamefont
  {Chatziioannou}}, \bibinfo {author} {\bibfnamefont {C.-J.}\ \bibnamefont
  {Haster}}, \bibinfo {author} {\bibfnamefont {T.~B.}\ \bibnamefont
  {Littenberg}}, \bibinfo {author} {\bibfnamefont {W.~M.}\ \bibnamefont
  {Farr}}, \bibinfo {author} {\bibfnamefont {S.}~\bibnamefont {Ghonge}},
  \bibinfo {author} {\bibfnamefont {M.}~\bibnamefont {Millhouse}}, \bibinfo
  {author} {\bibfnamefont {J.~A.}\ \bibnamefont {Clark}}, \ and\ \bibinfo
  {author} {\bibfnamefont {N.}~\bibnamefont {Cornish}},\ }\href {\doibase
  10.1103/PhysRevD.100.104004} {\bibfield  {journal} {\bibinfo  {journal}
  {Phys. Rev. D}\ }\textbf {\bibinfo {volume} {100}},\ \bibinfo {pages}
  {104004} (\bibinfo {year} {2019})}\BibitemShut {NoStop}%
\bibitem [{\citenamefont {Sun}\ \emph {et~al.}(2020)\citenamefont {Sun} \emph
  {et~al.}}]{Sun:2020wke}%
  \BibitemOpen
  \bibfield  {author} {\bibinfo {author} {\bibfnamefont {L.}~\bibnamefont
  {Sun}} \emph {et~al.},\ }\href {\doibase 10.1088/1361-6382/abb14e} {\bibfield
   {journal} {\bibinfo  {journal} {Classical and Quantum Gravity}\ } (\bibinfo
  {year} {2020}),\ 10.1088/1361-6382/abb14e},\ \Eprint
  {http://arxiv.org/abs/2005.02531} {arXiv:2005.02531 [astro-ph.IM]}
  \BibitemShut {NoStop}%
\bibitem [{\citenamefont {{LIGO Scientific and Virgo
  Collaborations}}(2020)}]{GW190412:dcc}%
  \BibitemOpen
  \bibfield  {author} {\bibinfo {author} {\bibnamefont {{LIGO Scientific and
  Virgo Collaborations}}},\ }\href@noop {} {\enquote {\bibinfo {title}
  {{GW190412 posterior samples data release}},}\ }\bibinfo {howpublished}
  {\url{https://dcc.ligo.org/P190412/public}} (\bibinfo {year}
  {2020})\BibitemShut {NoStop}%
\bibitem [{\citenamefont {{LIGO Scientific Collaboration and Virgo
  Collaboration}}(2020)}]{GraceDB}%
  \BibitemOpen
  \bibfield  {author} {\bibinfo {author} {\bibnamefont {{LIGO Scientific
  Collaboration and Virgo Collaboration}}},\ }\href {https://gracedb.ligo.org/}
  {\enquote {\bibinfo {title} {{Gravitational-Wave Candidate Event
  Database}},}\ } (\bibinfo {year} {2020})\BibitemShut {NoStop}%
\bibitem [{\citenamefont {Smith}\ \emph {et~al.}(2020)\citenamefont {Smith},
  \citenamefont {Ashton}, \citenamefont {Vajpeyi},\ and\ \citenamefont
  {Talbot}}]{10.1093/mnras/staa2483}%
  \BibitemOpen
  \bibfield  {author} {\bibinfo {author} {\bibfnamefont {R.~J.~E.}\
  \bibnamefont {Smith}}, \bibinfo {author} {\bibfnamefont {G.}~\bibnamefont
  {Ashton}}, \bibinfo {author} {\bibfnamefont {A.}~\bibnamefont {Vajpeyi}}, \
  and\ \bibinfo {author} {\bibfnamefont {C.}~\bibnamefont {Talbot}},\ }\href
  {\doibase 10.1093/mnras/staa2483} {\bibfield  {journal} {\bibinfo  {journal}
  {Monthly Notices of the Royal Astronomical Society}\ }\textbf {\bibinfo
  {volume} {498}},\ \bibinfo {pages} {4492} (\bibinfo {year} {2020})},\ \Eprint
  {http://arxiv.org/abs/https://academic.oup.com/mnras/article-pdf/498/3/4492/33798799/staa2483.pdf}
  {https://academic.oup.com/mnras/article-pdf/498/3/4492/33798799/staa2483.pdf}
  \BibitemShut {NoStop}%
\bibitem [{\citenamefont {Smith}\ \emph {et~al.}(2016)\citenamefont {Smith},
  \citenamefont {Field}, \citenamefont {Blackburn}, \citenamefont {Haster},
  \citenamefont {P\"urrer}, \citenamefont {Raymond},\ and\ \citenamefont
  {Schmidt}}]{Smith:2016qas}%
  \BibitemOpen
  \bibfield  {author} {\bibinfo {author} {\bibfnamefont {R.}~\bibnamefont
  {Smith}}, \bibinfo {author} {\bibfnamefont {S.~E.}\ \bibnamefont {Field}},
  \bibinfo {author} {\bibfnamefont {K.}~\bibnamefont {Blackburn}}, \bibinfo
  {author} {\bibfnamefont {C.-J.}\ \bibnamefont {Haster}}, \bibinfo {author}
  {\bibfnamefont {M.}~\bibnamefont {P\"urrer}}, \bibinfo {author}
  {\bibfnamefont {V.}~\bibnamefont {Raymond}}, \ and\ \bibinfo {author}
  {\bibfnamefont {P.}~\bibnamefont {Schmidt}},\ }\href {\doibase
  10.1103/PhysRevD.94.044031} {\bibfield  {journal} {\bibinfo  {journal} {Phys.
  Rev. D}\ }\textbf {\bibinfo {volume} {94}},\ \bibinfo {pages} {044031}
  (\bibinfo {year} {2016})}\BibitemShut {NoStop}%
\bibitem [{\citenamefont {{Skilling}}(2004)}]{Skilling:2004ns}%
  \BibitemOpen
  \bibfield  {author} {\bibinfo {author} {\bibfnamefont {J.}~\bibnamefont
  {{Skilling}}},\ }in\ \href {\doibase 10.1063/1.1835238} {\emph {\bibinfo
  {booktitle} {24th International Workshop on Bayesian Inference and Maximum
  Entropy Methods in Science and Engineering}}},\ \bibinfo {series} {AIP Conf.
  Proc.}, Vol.\ \bibinfo {volume} {735},\ \bibinfo {editor} {edited by\
  \bibinfo {editor} {\bibfnamefont {R.}~\bibnamefont {{Fischer}}}, \bibinfo
  {editor} {\bibfnamefont {R.}~\bibnamefont {{Preuss}}}, \ and\ \bibinfo
  {editor} {\bibfnamefont {U.~V.}\ \bibnamefont {{Toussaint}}}}\ (\bibinfo
  {year} {2004})\ pp.\ \bibinfo {pages} {395--405}\BibitemShut {NoStop}%
\bibitem [{\citenamefont {Speagle}(2020)}]{10.1093/mnras/staa278}%
  \BibitemOpen
  \bibfield  {author} {\bibinfo {author} {\bibfnamefont {J.~S.}\ \bibnamefont
  {Speagle}},\ }\href {\doibase 10.1093/mnras/staa278} {\bibfield  {journal}
  {\bibinfo  {journal} {Monthly Notices of the Royal Astronomical Society}\
  }\textbf {\bibinfo {volume} {493}},\ \bibinfo {pages} {3132} (\bibinfo {year}
  {2020})},\ \Eprint
  {http://arxiv.org/abs/https://academic.oup.com/mnras/article-pdf/493/3/3132/32890730/staa278.pdf}
  {https://academic.oup.com/mnras/article-pdf/493/3/3132/32890730/staa278.pdf}
  \BibitemShut {NoStop}%
\bibitem [{\citenamefont {Hoy}\ and\ \citenamefont
  {Raymond}(2020)}]{Hoy:2020vys}%
  \BibitemOpen
  \bibfield  {author} {\bibinfo {author} {\bibfnamefont {C.}~\bibnamefont
  {Hoy}}\ and\ \bibinfo {author} {\bibfnamefont {V.}~\bibnamefont {Raymond}},\
  }\href@noop {} {\bibfield  {journal} {\bibinfo  {journal} {ArXiv e-prints}\ }
  (\bibinfo {year} {2020})},\ \Eprint {http://arxiv.org/abs/2006.06639}
  {arXiv:2006.06639 [astro-ph.IM]} \BibitemShut {NoStop}%
\bibitem [{\citenamefont {Thrane}\ and\ \citenamefont
  {Talbot}(2019)}]{Thrane:2018qnx}%
  \BibitemOpen
  \bibfield  {author} {\bibinfo {author} {\bibfnamefont {E.}~\bibnamefont
  {Thrane}}\ and\ \bibinfo {author} {\bibfnamefont {C.}~\bibnamefont
  {Talbot}},\ }\href {\doibase 10.1017/pasa.2019.2} {\bibfield  {journal}
  {\bibinfo  {journal} {Publ. Astron. Soc. Austral.}\ }\textbf {\bibinfo
  {volume} {36}},\ \bibinfo {pages} {e010} (\bibinfo {year} {2019})},\ \Eprint
  {http://arxiv.org/abs/1809.02293} {arXiv:1809.02293 [astro-ph.IM]}
  \BibitemShut {NoStop}%
\bibitem [{\citenamefont {Majtey}\ \emph {et~al.}(2005)\citenamefont {Majtey},
  \citenamefont {Lamberti}, \citenamefont {Martin},\ and\ \citenamefont
  {Plastino}}]{Majtey2005}%
  \BibitemOpen
  \bibfield  {author} {\bibinfo {author} {\bibfnamefont {A.}~\bibnamefont
  {Majtey}}, \bibinfo {author} {\bibfnamefont {P.~W.}\ \bibnamefont
  {Lamberti}}, \bibinfo {author} {\bibfnamefont {M.~T.}\ \bibnamefont
  {Martin}}, \ and\ \bibinfo {author} {\bibfnamefont {A.}~\bibnamefont
  {Plastino}},\ }\href {\doibase 10.1140/epjd/e2005-00005-1} {\bibfield
  {journal} {\bibinfo  {journal} {The European Physical Journal D - Atomic,
  Molecular, Optical and Plasma Physics}\ }\textbf {\bibinfo {volume} {32}},\
  \bibinfo {pages} {413} (\bibinfo {year} {2005})}\BibitemShut {NoStop}%
\bibitem [{\citenamefont {Kullback}\ and\ \citenamefont
  {Leibler}(1951)}]{kullback1951}%
  \BibitemOpen
  \bibfield  {author} {\bibinfo {author} {\bibfnamefont {S.}~\bibnamefont
  {Kullback}}\ and\ \bibinfo {author} {\bibfnamefont {R.~A.}\ \bibnamefont
  {Leibler}},\ }\href {\doibase 10.1214/aoms/1177729694} {\bibfield  {journal}
  {\bibinfo  {journal} {Ann. Math. Statist.}\ }\textbf {\bibinfo {volume}
  {22}},\ \bibinfo {pages} {79} (\bibinfo {year} {1951})}\BibitemShut {NoStop}%
\bibitem [{\citenamefont {Shaik}\ \emph {et~al.}(2020)\citenamefont {Shaik},
  \citenamefont {Lange}, \citenamefont {Field}, \citenamefont {O'Shaughnessy},
  \citenamefont {Varma}, \citenamefont {Kidder}, \citenamefont {Pfeiffer},\
  and\ \citenamefont {Wysocki}}]{PhysRevD.101.124054}%
  \BibitemOpen
  \bibfield  {author} {\bibinfo {author} {\bibfnamefont {F.~H.}\ \bibnamefont
  {Shaik}}, \bibinfo {author} {\bibfnamefont {J.}~\bibnamefont {Lange}},
  \bibinfo {author} {\bibfnamefont {S.~E.}\ \bibnamefont {Field}}, \bibinfo
  {author} {\bibfnamefont {R.}~\bibnamefont {O'Shaughnessy}}, \bibinfo {author}
  {\bibfnamefont {V.}~\bibnamefont {Varma}}, \bibinfo {author} {\bibfnamefont
  {L.~E.}\ \bibnamefont {Kidder}}, \bibinfo {author} {\bibfnamefont {H.~P.}\
  \bibnamefont {Pfeiffer}}, \ and\ \bibinfo {author} {\bibfnamefont
  {D.}~\bibnamefont {Wysocki}},\ }\href {\doibase 10.1103/PhysRevD.101.124054}
  {\bibfield  {journal} {\bibinfo  {journal} {Phys. Rev. D}\ }\textbf {\bibinfo
  {volume} {101}},\ \bibinfo {pages} {124054} (\bibinfo {year}
  {2020})}\BibitemShut {NoStop}%
\bibitem [{\citenamefont {Kumar}\ \emph {et~al.}(2019)\citenamefont {Kumar},
  \citenamefont {Blackman}, \citenamefont {Field}, \citenamefont {Scheel},
  \citenamefont {Galley}, \citenamefont {Boyle}, \citenamefont {Kidder},
  \citenamefont {Pfeiffer}, \citenamefont {Szilagyi},\ and\ \citenamefont
  {Teukolsky}}]{PhysRevD.99.124005}%
  \BibitemOpen
  \bibfield  {author} {\bibinfo {author} {\bibfnamefont {P.}~\bibnamefont
  {Kumar}}, \bibinfo {author} {\bibfnamefont {J.}~\bibnamefont {Blackman}},
  \bibinfo {author} {\bibfnamefont {S.~E.}\ \bibnamefont {Field}}, \bibinfo
  {author} {\bibfnamefont {M.}~\bibnamefont {Scheel}}, \bibinfo {author}
  {\bibfnamefont {C.~R.}\ \bibnamefont {Galley}}, \bibinfo {author}
  {\bibfnamefont {M.}~\bibnamefont {Boyle}}, \bibinfo {author} {\bibfnamefont
  {L.~E.}\ \bibnamefont {Kidder}}, \bibinfo {author} {\bibfnamefont {H.~P.}\
  \bibnamefont {Pfeiffer}}, \bibinfo {author} {\bibfnamefont {B.}~\bibnamefont
  {Szilagyi}}, \ and\ \bibinfo {author} {\bibfnamefont {S.~A.}\ \bibnamefont
  {Teukolsky}},\ }\href {\doibase 10.1103/PhysRevD.99.124005} {\bibfield
  {journal} {\bibinfo  {journal} {Phys. Rev. D}\ }\textbf {\bibinfo {volume}
  {99}},\ \bibinfo {pages} {124005} (\bibinfo {year} {2019})}\BibitemShut
  {NoStop}%
\bibitem [{\citenamefont {Usman}\ \emph {et~al.}(2019)\citenamefont {Usman},
  \citenamefont {Mills},\ and\ \citenamefont {Fairhurst}}]{Usman_2019}%
  \BibitemOpen
  \bibfield  {author} {\bibinfo {author} {\bibfnamefont {S.~A.}\ \bibnamefont
  {Usman}}, \bibinfo {author} {\bibfnamefont {J.~C.}\ \bibnamefont {Mills}}, \
  and\ \bibinfo {author} {\bibfnamefont {S.}~\bibnamefont {Fairhurst}},\ }\href
  {\doibase 10.3847/1538-4357/ab0b3e} {\bibfield  {journal} {\bibinfo
  {journal} {The Astrophysical Journal}\ }\textbf {\bibinfo {volume} {877}},\
  \bibinfo {pages} {82} (\bibinfo {year} {2019})}\BibitemShut {NoStop}%
\bibitem [{\citenamefont {Abbott}\ \emph {et~al.}(2018)\citenamefont {Abbott}
  \emph {et~al.}}]{Aasi:2013wya}%
  \BibitemOpen
  \bibfield  {author} {\bibinfo {author} {\bibfnamefont {B.}~\bibnamefont
  {Abbott}} \emph {et~al.} (\bibinfo {collaboration} {KAGRA, LIGO Scientific,
  VIRGO}),\ }\href {\doibase 10.1007/s41114-018-0012-9} {\bibfield  {journal}
  {\bibinfo  {journal} {Living Rev. Rel.}\ }\textbf {\bibinfo {volume} {21}},\
  \bibinfo {pages} {3} (\bibinfo {year} {2018})},\ \Eprint
  {http://arxiv.org/abs/1304.0670} {arXiv:1304.0670 [gr-qc]} \BibitemShut
  {NoStop}%
\bibitem [{\citenamefont {Barsotti}\ \emph {et~al.}(2018)\citenamefont
  {Barsotti}, \citenamefont {Fritschel}, \citenamefont {Evans},\ and\
  \citenamefont {Gras}}]{adligopsd}%
  \BibitemOpen
  \bibfield  {author} {\bibinfo {author} {\bibfnamefont {L.}~\bibnamefont
  {Barsotti}}, \bibinfo {author} {\bibfnamefont {P.}~\bibnamefont {Fritschel}},
  \bibinfo {author} {\bibfnamefont {M.}~\bibnamefont {Evans}}, \ and\ \bibinfo
  {author} {\bibfnamefont {S.}~\bibnamefont {Gras}},\ }\href
  {https://dcc.ligo.org/T1800044/public} {\emph {\bibinfo {title} {{The updated
  Advanced LIGO design curve}}}},\ \bibinfo {type} {Tech. Rep.}\ \bibinfo
  {number} {LIGO-T1800044}\ (\bibinfo  {institution} {{LIGO Project}},\
  \bibinfo {year} {2018})\BibitemShut {NoStop}%
\bibitem [{\citenamefont {Schmidt}\ \emph {et~al.}(2017)\citenamefont
  {Schmidt}, \citenamefont {Harry},\ and\ \citenamefont
  {Pfeiffer}}]{Schmidt:2017btt}%
  \BibitemOpen
  \bibfield  {author} {\bibinfo {author} {\bibfnamefont {P.}~\bibnamefont
  {Schmidt}}, \bibinfo {author} {\bibfnamefont {I.~W.}\ \bibnamefont {Harry}},
  \ and\ \bibinfo {author} {\bibfnamefont {H.~P.}\ \bibnamefont {Pfeiffer}},\
  }\href@noop {} {\bibfield  {journal} {\bibinfo  {journal} {ArXiv e-prints}\ }
  (\bibinfo {year} {2017})},\ \Eprint {http://arxiv.org/abs/1703.01076}
  {arXiv:1703.01076 [gr-qc]} \BibitemShut {NoStop}%
\bibitem [{\citenamefont {Harry}\ \emph {et~al.}(2016)\citenamefont {Harry},
  \citenamefont {Privitera}, \citenamefont {Boh\'e},\ and\ \citenamefont
  {Buonanno}}]{Harry:2016aa}%
  \BibitemOpen
  \bibfield  {author} {\bibinfo {author} {\bibfnamefont {I.}~\bibnamefont
  {Harry}}, \bibinfo {author} {\bibfnamefont {S.}~\bibnamefont {Privitera}},
  \bibinfo {author} {\bibfnamefont {A.}~\bibnamefont {Boh\'e}}, \ and\ \bibinfo
  {author} {\bibfnamefont {A.}~\bibnamefont {Buonanno}},\ }\href {\doibase
  10.1103/PhysRevD.94.024012} {\bibfield  {journal} {\bibinfo  {journal} {Phys.
  Rev. D}\ }\textbf {\bibinfo {volume} {94}},\ \bibinfo {pages} {024012}
  (\bibinfo {year} {2016})}\BibitemShut {NoStop}%
\bibitem [{\citenamefont {Ramos-Buades}\ \emph {et~al.}(2020)\citenamefont
  {Ramos-Buades}, \citenamefont {Schmidt}, \citenamefont {Pratten},\ and\
  \citenamefont {Husa}}]{Ramos-Buades:2020noq}%
  \BibitemOpen
  \bibfield  {author} {\bibinfo {author} {\bibfnamefont {A.}~\bibnamefont
  {Ramos-Buades}}, \bibinfo {author} {\bibfnamefont {P.}~\bibnamefont
  {Schmidt}}, \bibinfo {author} {\bibfnamefont {G.}~\bibnamefont {Pratten}}, \
  and\ \bibinfo {author} {\bibfnamefont {S.}~\bibnamefont {Husa}},\ }\href
  {\doibase 10.1103/PhysRevD.101.103014} {\bibfield  {journal} {\bibinfo
  {journal} {Phys. Rev. D}\ }\textbf {\bibinfo {volume} {101}},\ \bibinfo
  {pages} {103014} (\bibinfo {year} {2020})},\ \Eprint
  {http://arxiv.org/abs/2001.10936} {arXiv:2001.10936 [gr-qc]} \BibitemShut
  {NoStop}%
\bibitem [{\citenamefont {Marsat}\ and\ \citenamefont
  {Baker}(2018)}]{Marsat:2018oam}%
  \BibitemOpen
  \bibfield  {author} {\bibinfo {author} {\bibfnamefont {S.}~\bibnamefont
  {Marsat}}\ and\ \bibinfo {author} {\bibfnamefont {J.~G.}\ \bibnamefont
  {Baker}},\ }\href@noop {} {\bibfield  {journal} {\bibinfo  {journal} {ArXiv
  e-prints}\ } (\bibinfo {year} {2018})},\ \Eprint
  {http://arxiv.org/abs/1806.10734} {arXiv:1806.10734 [gr-qc]} \BibitemShut
  {NoStop}%
\bibitem [{\citenamefont {{LIGO Scientific Collaboration, Virgo
  Collaboration}}(2019)}]{GWOSC}%
  \BibitemOpen
  \bibfield  {author} {\bibinfo {author} {\bibnamefont {{LIGO Scientific
  Collaboration, Virgo Collaboration}}},\ }\href@noop {} {\enquote {\bibinfo
  {title} {{Gravitational Wave Open Science Center}},}\ }\bibinfo
  {howpublished}
  {\href{https://www.gw-openscience.org}{https://www.gw-openscience.org}}
  (\bibinfo {year} {2019})\BibitemShut {NoStop}%
\bibitem [{\citenamefont {Bavera}\ \emph {et~al.}(2020)\citenamefont {Bavera},
  \citenamefont {Fragos}, \citenamefont {Qin}, \citenamefont {Zapartas},
  \citenamefont {Neijssel}, \citenamefont {Mandel}, \citenamefont {Batta},
  \citenamefont {Gaebel}, \citenamefont {Kimball},\ and\ \citenamefont
  {Stevenson}}]{Bavera:2019fkg}%
  \BibitemOpen
  \bibfield  {author} {\bibinfo {author} {\bibfnamefont {S.~S.}\ \bibnamefont
  {Bavera}}, \bibinfo {author} {\bibfnamefont {T.}~\bibnamefont {Fragos}},
  \bibinfo {author} {\bibfnamefont {Y.}~\bibnamefont {Qin}}, \bibinfo {author}
  {\bibfnamefont {E.}~\bibnamefont {Zapartas}}, \bibinfo {author}
  {\bibfnamefont {C.~J.}\ \bibnamefont {Neijssel}}, \bibinfo {author}
  {\bibfnamefont {I.}~\bibnamefont {Mandel}}, \bibinfo {author} {\bibfnamefont
  {A.}~\bibnamefont {Batta}}, \bibinfo {author} {\bibfnamefont {S.~M.}\
  \bibnamefont {Gaebel}}, \bibinfo {author} {\bibfnamefont {C.}~\bibnamefont
  {Kimball}}, \ and\ \bibinfo {author} {\bibfnamefont {S.}~\bibnamefont
  {Stevenson}},\ }\href {\doibase 10.1051/0004-6361/201936204} {\bibfield
  {journal} {\bibinfo  {journal} {Astron. Astrophys.}\ }\textbf {\bibinfo
  {volume} {635}},\ \bibinfo {pages} {A97} (\bibinfo {year} {2020})},\ \Eprint
  {http://arxiv.org/abs/1906.12257} {arXiv:1906.12257 [astro-ph.HE]}
  \BibitemShut {NoStop}%
\end{thebibliography}%


\end{document}